\documentclass[lettersize,journal]{IEEEtran}
\AtBeginDocument{%
  }

\usepackage{orcidlink}
\usepackage{amssymb}
\usepackage{booktabs}
\newsavebox\CBox

\usepackage{svg}

\usepackage{xspace}
\newcommand\approach{\textsc{MANGO}\xspace}
\newcommand\approachb{\textsc{MANGO$_o$}\xspace} 
\usepackage{multirow}
\usepackage{graphicx}
\usepackage[table]{xcolor}     
\usepackage[linesnumbered,algoruled,boxed,lined]{algorithm2e}
\SetKwRepeat{Do}{do}{while}
\usepackage{algpseudocode}
\newcommand{\best}[1]{{\underline{\textbf{#1}}}}
\usepackage[most]{tcolorbox}
\newcommand{\revision}[1]{{\color{black}#1}}
\usepackage{listings}
\usepackage{xcolor}

\definecolor{dslblue}{RGB}{40,75,115}
\definecolor{dslgray}{RGB}{248,249,251}
\definecolor{dslframe}{RGB}{190,198,208}

\lstdefinestyle{oracledsl}{
    basicstyle=\ttfamily\scriptsize,
    backgroundcolor=\color{dslgray},
    frame=single,
    framerule=0.4pt,
    rulecolor=\color{dslframe},
    framesep=6pt,
    xleftmargin=0.5em,
    xrightmargin=0.5em,
    framexleftmargin=0.5em,
    framexrightmargin=0.5em,
    columns=fullflexible,
    keepspaces=true,
    showstringspaces=false,
    breaklines=true,
    captionpos=b,
    aboveskip=0.8em,
    belowskip=0.8em,
    emph={FineGrainedOracle,Step,Action,Monitor,
          ExecutionMode,FunctionCall, Value, Parameter},
    emphstyle=\bfseries\color{dslblue}
}

\begin{document}



\title{MANGO: Automated Multi-Agent Test Oracle Generation for Vision-Language-Action Models}





\author{Pablo~Valle~\orcidlink{0000-0002-0588-316X}, Shaukat~Ali~\orcidlink{0000-0002-9979-3519}, Aitor~Arrieta~\orcidlink{0000-0001-7507-5080} and Lionel~Briand~~\orcidlink{0000-0002-1393-1010}, \textit{Fellow, IEEE}%
\thanks{Pablo Valle and Aitor Arrieta are with Mondragon University, Guipuzcoa, Spain. E-mail: \url{pvalle@mondragon.edu}, \url{aarrieta@mondragon.edu}. 

Shaukat Ali is with Simula Research Laboratory, Oslo, Norway. E-mail: \url{shaukat@simula.no}.

Lionel Briand is with the University of Ottawa, Ottawa, ON K1N 6N5, Canada, and also with the Research Ireland Lero Centre for Software, University of Limerick, V94 T9PX, Limerick, Ireland. E-mail: \url{lbriand@uottawa.ca}; \url{lionel.briand@lero.ie} }%
}
       
\markboth{}%
{Shell \MakeLowercase{\textit{et al.}}: A Sample Article Using IEEEtran.cls for IEEE Journals}

\maketitle

\begin{abstract}
Vision-Language-Action (VLA) models are emerging robotic control systems that integrate perception, language understanding, and action generation in a unified architecture. Existing testing approaches for \revision{VLAs} rely on manually constructed symbolic \revision{end-state} test oracles that determine task success from final environment states. These oracles are costly to construct, require domain expertise, and are often tightly coupled to specific tasks and environments, limiting scalability and reuse. Furthermore, they provide only end-state assessments of task outcomes, offering limited insight into intermediate behavior and fault localization. To address these limitations, we introduce \approach, a multi-agent framework that automatically generates fine-grained oracles from natural-language descriptions of robotic tasks. \approach first generates a reusable library of atomic tasks, then generates simulator-grounded oracle definitions for each atomic task, and finally produces executable fine-grained oracles by decomposing complex instructions into ordered sequences of atomic actions and corresponding oracles. The framework uses collaborative Generator, Assessor, and Judge agents that iteratively refine generated artifacts through structured feedback. We evaluate \approach on the LIBERO\_10 and RoboCasa Humanoid Tabletop benchmarks. Results show that \approach generates executable, fine-grained oracles that detect a similar number of failures as symbolic \revision{end-state} oracles while accurately localizing them and providing richer diagnostic information. Through ablation studies, we further analyzed component contributions and the effect of the initial task set, while preserving oracle quality. Overall, the results show the feasibility and effectiveness of test oracle generation for \revision{VLA} testing.
\end{abstract}





\section{Introduction}\label{sec:introduction}
Vision-Language-Action (VLA) models are emerging as a new generation of control algorithms for robots~\cite{zitkovich2023rt,kim2024openvla, nvidia2025gr00tn1openfoundation}. Unlike traditional robotic architectures that separate perception, planning, and control into independent modules, VLA models integrate visual observations, natural-language instructions, and robot state information into a unified multimodal representation from which executable actions are generated. This enables robots to interpret high-level instructions such as \textit{``store the bottle of water in the fridge''} and autonomously translate them into sequences of low-level control commands. Recent works have demonstrated remarkable capabilities across many manipulation tasks~\cite{black2024pi0visionlanguageactionflowmodel, nvidia2025gr00tn1openfoundation, qu2025eo1}, enabling general-purpose robots operating in complex environments. 

As VLAs become more capable, ensuring their correctness and reliability becomes critical. Current testing approaches~\cite{wang2025vlatest, zhang2025vlabench} primarily rely on symbolic \revision{end-state} test oracles provided by robotic benchmarks~\cite{liu2023libero, li24simpler}. Given task instructions and simulator execution, these oracles determine task success by checking whether the environment's final state matches the expected goal state. For example, in a task requiring a robot to place an object inside a container, the oracle verifies whether the object is inside the target container at the end of the execution. This strategy is widely used in existing benchmarks due to its simplicity and automation. 

However, symbolic \revision{end-state} oracles have limitations~\cite{valle2025evaluating}. First, they reduce task correctness to a binary decision based solely on the final environment state, ignoring the action sequence that produced it, and thus cannot distinguish between failures arising from different reasons~\cite{valle2025evaluating}. For example, a robot tasked with placing a bottle in a fridge may fail because it cannot open the fridge, drops the bottle during transport, or attempts to place it before opening the door. From the perspective of a symbolic \revision{end-state} oracle, all such executions are indistinguishable, as they collapse to the same outcome: task failure. Second, symbolic \revision{end-state} oracles provide little support for debugging, as they cannot identify which sub-task failed within a complex instruction that failed during execution. As VLAs increasingly execute long-horizon tasks with multiple objects and intermediate goals, this lack of diagnostic information becomes a major obstacle to development and testing. These limitations reflect the test oracle problem in software testing, i.e., the difficulty of determining output correctness automatically~\cite{barr2014oracle}. While symbolic \revision{end-state} oracles provide a partial solution for robotic benchmarks, they are limited to predefined goal states and fail to capture the structure and quality of task execution. This limitation is particularly problematic in long-horizon robotic manipulation, where success depends not only on achieving a final configuration but also on correctly executing a sequence of interdependent sub-tasks.

A natural direction to address this issue is finer-grained evaluation. Rather than treating a complex instruction as a single objective, it can be decomposed into atomic tasks corresponding to reusable capabilities such as opening, grasping, placing, or closing. Associating an oracle with each atomic task reframes evaluation as structured verification that assesses both final goal completion and the correctness of intermediate step and their order. Such fine-grained oracles provide richer feedback, improve failure localization, and enable more interpretable assessment of robot behavior. However, manually defining them is impractical at scale, as benchmarks often comprise dozens to hundreds of tasks~\cite{zhang2025vlabench, fei2026libero, nasiriany2024robocasa}, each with different objects, environments, and execution constraints. Designing atomic task decompositions and corresponding oracles would require substantial human effort and domain expertise. Moreover, maintaining consistency across manually authored fine-grained oracles becomes increasingly difficult as benchmark complexity grows. In addition, existing benchmarks typically implement symbolic \revision{end-state} oracles in heterogeneous formats and programming languages; for example, Python~\cite{nasiriany2024robocasa} or .bddl files~\cite{liu2023libero}. This fragmentation further complicates reuse, interoperability, and scalability across benchmarks. As a result, scalable fine-grained evaluation requires automated methods that translate natural-language instructions into executable, fine-grained oracle specifications.

In this paper, we present \approach, a \textbf{M}ulti-\textbf{A}gent Fi\textbf{N}e-\textbf{G}rained \textbf{O}racle Generator for automatically generating fine-grained test oracles for VLAs from natural-language task descriptions. \approach operates in three stages. First, it automatically generates a reusable library of atomic tasks from user-provided robotic instructions. Second, it generates oracle definitions for each atomic task using available target simulator functions. Finally, it generates executable fine-grained oracles for individual tasks by decomposing instructions into ordered atomic tasks and linking each step to its corresponding oracle. \revision{To identify and correct generation errors, \approach employs a collaborative multi-agent architecture with \textit{Generator}, \textit{Assessor}, and \textit{Judge} agents that iteratively refine generated artifacts through structured feedback. We assume that the supplied instructions convey the intended goal and relevant task-specific constraints, including spatial relationships when needed to distinguish ambiguous outcomes. Detailed motion specifications are not required, as \approach derives atomic tasks and grounds their oracles using the available environment information and simulator functions. Our evaluation focuses on oracle generation under these assumptions; thus, systematic robustness testing with ambiguous, incomplete, or adversarial instructions is outside the paper's scope.}

We evaluate \approach on two representative robotic benchmarks, LIBERO\_10~\cite{liu2023libero} and RoboCasa Humanoid Tabletop~\cite{nvidia2025gr00tn1openfoundation}, assessing the ability of \approach to generate fine-grained oracles, comparing against symbolic \revision{end-state} oracles for failure detection and localization, and conducting ablation studies on key framework components. We also study how much the initial task set can be minimized while preserving the quality of the generated fine-grained oracles, providing insights for practitioners.

The main contributions of the paper are:

\begin{itemize}
    \item A novel formulation of fine-grained test oracles for VLAs based on task decomposition into reusable atomic tasks.
    \item \approach, the first multi-agent framework for automatically generating fine-grained test oracles from natural language task descriptions for VLAs.
    \item An extensive empirical evaluation on two robotic benchmarks assessing generation quality, failure detection capability, and failure localization effectiveness.
    \item A replication package, which will be made available upon acceptance, containing the implementation of \approach, generated oracle libraries, experimental artifacts, and instructions to facilitate reproducibility and future research.
\end{itemize}


\section{Background}\label{sec:background}

In this section, we present the background on VLAs, followed by their testing.

\subsection{\revision{Vision Language Action Models}}
\begin{figure}[ht]
    \centering
    \includegraphics[width=\linewidth]{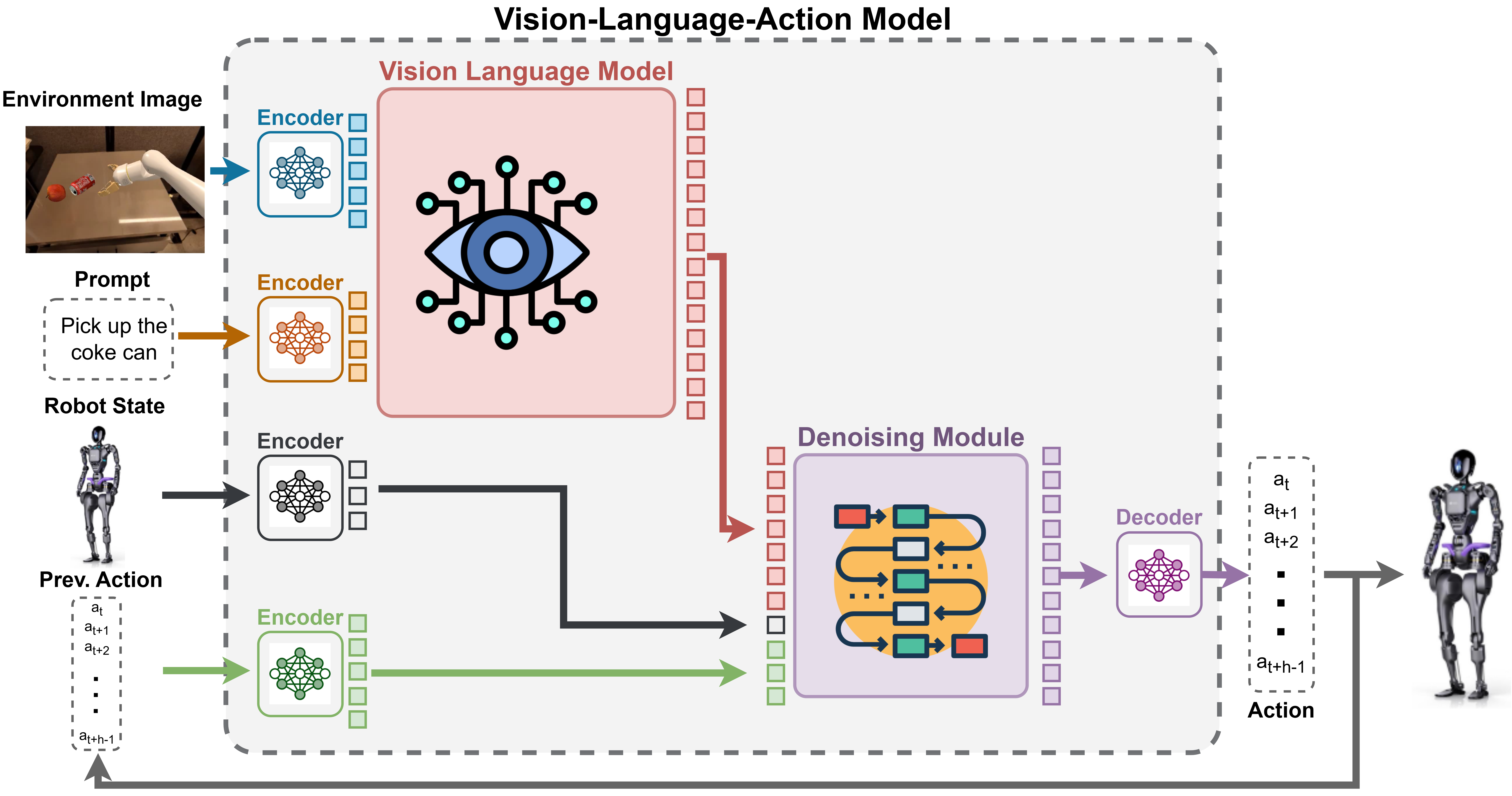}
    \caption{\revision{Overview of the action generation process in a VLA-enabled robot, mapping multimodal inputs to executable action chunks}}
    \label{fig:vla_arch}
\end{figure}

\revision{Vision-Language-Action (VLA) models are multimodal foundation models designed to translate visual observations, natural-language instructions, and robot states into executable actions. Robotic systems that integrate these models as runtime decision-making components constitute an emerging class of Cyber-Physical Systems (CPSs).} As Figure~\ref{fig:vla_arch} depicts, unlike traditional autonomous CPSs, which decompose perception, planning, and control into modular components, \revision{VLA models encode} visual observations, language instructions, and proprioceptive information into a shared representation space~\cite{nvidia2025gr00tn1openfoundation,black2024pi0visionlanguageactionflowmodel,kim2024openvla, qu2025eo1}. This design departs from conventional Deep Neural Network (DNN) specialization, such as Convolutional Neural Networks (CNNs) for visual feature extraction~\cite{lecun2002gradient,krizhevsky2012imagenet,simonyan2014very,szegedy2015going} or transformer-based Large Language Models (LLMs) for textual reasoning~\cite{vaswani2017attention,achiam2023gpt,team2023gemini,grattafiori2024llama,jiang2023mistral7b}, by enabling a single model to bridge perception and action. Within a robotic system, this capability allows the agent to interpret instructions such as ``\textit{Pick up the coke can}'' in its environment, resolve referential ambiguities through visual understanding, and generate executable control commands.

Formally, at each control step $t$, the \revision{VLA model} receives an observation $o_t = \left[I^t_1, I^t_2, \dots, I^t_n, \ell_t, \theta_t \right]$, where $I^t_i$ is an RGB image captured by the camera of the robot (i.e., Environment image in Figure~\ref{fig:vla_arch}), $\ell_t$ is the task instruction in natural language (i.e., Prompt in Figure~\ref{fig:vla_arch}), and $\theta_t$ represents the robot’s proprioceptive state (i.e., Robot state in Figure~\ref{fig:vla_arch}). These inputs are encoded and represented in a shared latent space, from which the VLA model predicts a temporally extended control command, referred to as an action chunk $AC_t$. Each action chunk $AC_t = \left[a_t, a_{t+1}, \dots, a_{t+h-1}\right]$ corresponds to a sequence of actions over a finite time horizon $h$. By iteratively producing such chunks, the VLA model constructs an action sequence $A = \left[ AC_{t}, AC_{t+h}, \dots, AC_{t+(N-1)h} \right]$ of length $N$, which when executed, produces a trajectory $\mathcal{T} = \left[ S_t, S_{t+1}, \dots, S_{t+T} \right]$ of $T$ states. Each state $S_t$ captures the configuration of the robot's end-effector (i.e., the part of the robot that interacts with the environment, such as a gripper), including its position $(x, y, z)$ and orientation in quaternions $(q_1, q_2, q_3, q_4)$. When the corresponding actions are fed to the robot, the robot’s control system automatically computes and applies the necessary joint movements to move the end-effector to the specified positions and orientations. This way, the VLA model serves as the central decision-making component of the robot, continuously translating natural-language instructions into physically grounded behavior.

Recent VLA models further enhance action generation by incorporating diffusion-based action refinement mechanisms~\cite{nvidia2025gr00tn1openfoundation, black2024pi0visionlanguageactionflowmodel}, as shown in Figure~\ref{fig:vla_arch}, which act as iterative post-processing layers over initially predicted actions. Inside the denoising module, an initial action-chunk prediction is iteratively refined through a denoising diffusion process implemented via a diffusion transformer. At each refinement step, the denoising diffusion module iterates over the same multimodal context, visual observations, language instructions, and proprioceptive state to progressively improve action consistency and feasibility. Instead of generating actions in a single pass, this iterative process acts as a structured optimization procedure over the action space. This approach has been shown to reduce error accumulation and improve robustness in long-horizon tasks~\cite{nvidia2025gr00tn1openfoundation, black2024pi0visionlanguageactionflowmodel,  zhao2023learning}.

\subsection{\revision{Testing Vision-Language-Action Models}}

\begin{figure*}[ht]
    \centering
    \includegraphics[width=\linewidth]{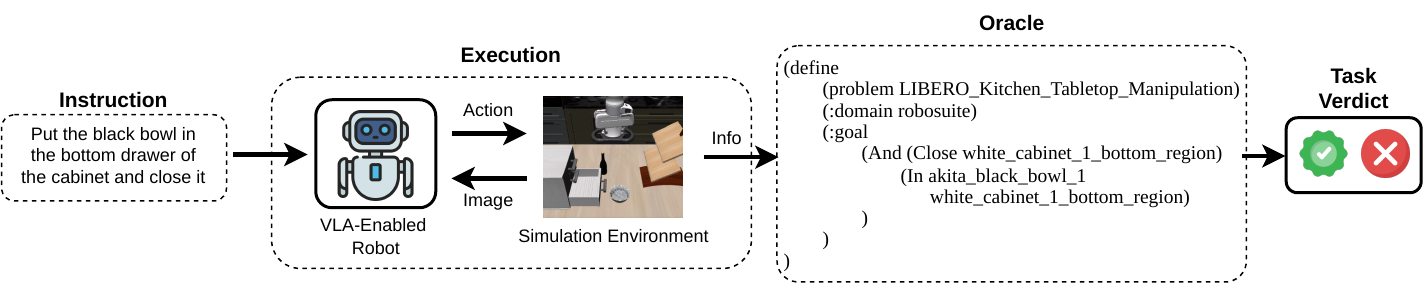}
    \caption{Task execution pipeline for the ``\textit{Put the black bowl in the bottom drawer of the cabinet and close it}'' task}
    \label{fig:symbolic_oracle}
\end{figure*}

Figure~\ref{fig:symbolic_oracle} depicts the full pipeline for testing a \revision{VLA model} within a given scenario and task instruction. For each task evaluation, we assume the availability of three key components: (1) a natural language instruction, (2) a simulation environment in which the task is executed, and (3) an oracle that determines whether the task is completed.

\subsubsection{Instruction} The instruction corresponds to the task description provided to the \revision{VLA model} in natural language. It specifies the intended behavior, including the goal to be achieved and, implicitly, the interactions required with the environment. As shown in Figure~\ref{fig:symbolic_oracle}, the instruction ``\textit{Put the black bowl in the bottom drawer of the cabinet and close it}'' defines the desired behavior of the robot and serves as the primary interface between the human intent and the robot execution.

\subsubsection{Execution} The execution component corresponds to the instantiation of the simulation environment in which the robot operates. This includes setting up the scene, initializing relevant objects, such as those the robot must interact with, and defining their spatial relationships so that the robot can interact with the environment to perform the specified task. The robot interacts with this environment to perform the task specified by the instruction. During execution, at each time step, the robot receives visual inputs (i.e., RGB images) from the simulator and acts on the environment by executing a sequence of actions generated by its controller (i.e., the VLA model), thereby generating a trajectory that accomplishes the specified task.

\subsubsection{Oracle}\label{sec:background_oracle} The oracle defines the task success criteria in an executable form. It depends on both the instruction and the environment configuration and is typically defined per task and scenario. These oracles are manually defined by each benchmark's developers, which results in implementation differences across benchmarks. For instance, in the LIBERO benchmark~\cite{liu2023libero} (i.e., the one shown in Figure~\ref{fig:symbolic_oracle}), oracles are defined as separate specification files in \textit{.bddl} format instantiated per instruction. In contrast, the RoboCasa Humanoid Tabletop benchmark~\cite{nvidia2025gr00tn1openfoundation} embeds the oracle within the scenario definition and implements it directly in Python. Despite these differences, both approaches rely on predefined simulator functions, such as object contact detection and spatial relations (e.g., whether an object is inside or on top of another), to evaluate task completion. These oracles are symbolic, rely on the aforementioned simulator functions, and produce a binary success/failure result based on whether the environment's goal state is achieved. 

In the example shown in Figure~\ref{fig:symbolic_oracle}, the robot is instructed to store a bowl inside the bottom drawer of the cabinet and to close it. Therefore, the oracle evaluates task success by checking whether the bottom drawer is closed, i.e., \texttt{Close white\_cabinet\_1\_bottom\_drawer}, and whether the bowl is inside the bottom drawer of the cabinet, i.e., \texttt{In akita\_black\_bowl\_1 white\_cabinet\_1\_bottom\_region}. While this captures the task's final state, it does not account for the intermediate steps required to complete it, such as grasping the bowl or checking whether the drawer is open before placing it inside. As a result, the oracle is limited to the task's end state, leading to several limitations. On the one hand, test execution assessment does not consider intermediate steps. For instance, in prior studies~\cite{valle2025evaluating}, we found high variability in quality across successful executions (e.g., in many successful executions, \revision{the object was accidentally dropped outside the cabinet next to the door; therefore, closing the door pushed it inside, resulting in a successful task execution.); these oracles are unable to capture such issues. Another limitation is that, for long-horizon unsuccessful tasks, it is not possible to know which sub-task caused the failure; this information could eventually be used for debugging and repair.}

\section{Fine-Grained Oracles} \label{sec:fg_oracles}

Existing evaluation approaches for \revision{VLAs}~\cite{wang2025vlatest, fei2026libero, peng2025nebula, zhang2025vlabench} typically rely on symbolic \revision{end-state} oracles that determine task success by verifying whether the final state of the environment is the same as the expected one, as explained in Section~\ref{sec:background_oracle}. Formally, given an instruction ($I$), a symbolic \revision{end-state} oracle evaluates the final state ($s_{n-1}$) of an execution trace ($\tau = \langle s_0, a_0, \dots, s_{n-1} \rangle$) and returns a binary outcome ($O_s(s_{n-1}) \in {\texttt{Pass}, \texttt{Fail}}$). For instance, for the instruction \textit{``store the bottle of water in the fridge''}, a symbolic \revision{end-state} oracle may simply check whether the bottle is located inside the fridge at the end of the execution. While this formulation is effective for simple goal assessment, it fundamentally reduces task correctness to a single end-state condition. As a result, it provides no insight into how the task was executed, nor can it detect which step in the task execution caused the failure. In the given example, a robot may fail to open the fridge or place the bottle inside it; from the perspective of symbolic \revision{end-state} oracles, these two failures appear indistinguishable, even though they likely correspond to different underlying causes.

To address this limitation, we propose the use of \emph{fine-grained oracles} for task assessment in the context of \revision{VLAs}. The central idea is to decompose the evaluation of a natural-language instruction into a structured set of smaller, reusable instructions that align with the task's underlying structure. Instead of associating a complete instruction with a single symbolic \revision{end-state} oracle, we first decompose the instruction into a sequence of \emph{atomic tasks}, and then associate each atomic task with its own symbolic oracle. Finally, the fine-grained oracle is defined as the composition of these atomic symbolic oracles. Any given instruction $I$ can be decomposed as a sequence of parameterized atomic tasks as:
\begin{equation}
D(I) = \langle a_0(\theta_0), a_1(\theta_1), \ldots, a_{n-1}(\theta_{n-1}) \rangle,
\end{equation}

\noindent where each atomic task $a_i(\cdot)$ is a reusable functional template and $\theta_i$ denotes its task-specific arguments, such as objects or target regions. \revision{For instance, in the \texttt{Place}(\texttt{Bottle}, \texttt{Fridge}) function, the first argument specifies the object to be placed, while the second specifies the destination where the object should be placed.} The resulting sequence may be strictly ordered, unordered, or partially ordered, depending on the structure and ambiguity of the input instruction. \revision{Task ordering is defined through two basic relations: \emph{ordered} and \emph{unordered}. An \emph{ordered} relation imposes a precedence constraint between two task components, requiring one to be completed before the other can be executed. In contrast, an \emph{unordered} relation imposes no precedence constraint between the corresponding components, meaning that either may be executed first and, when the execution platform permits it, they may also be executed concurrently. Through the hierarchical composition of these relations, the representation can encode partially ordered task structures in which some atomic tasks are sequentially constrained while others remain independent. This specification is robot-independent. The fine-grained oracle defines the admissible task-level orderings, while the VLA determines how to execute them given the robot's capabilities. Unordered tasks may therefore be completed in either order or concurrently when feasible. In such cases, simultaneous execution is not required, and the oracle verifies them without enforcing a specific order.} 

For example, in a pick-and-place task involving a single object, the sequence is strictly ordered because the actions must follow a fixed order (i.e., the object must first be picked and then placed). When two independent objects are involved, the overall task may instead be partially ordered: the pick-and-place actions associated with each object remain internally ordered, while no precedence constraint is imposed between the two object-level subtasks. Consequently, either object may be processed first. Under this formulation, atomic tasks are not tied to a specific instruction instance but instead represent general-purpose manipulation tasks instantiated with context-dependent parameters. For example, the instruction \textit{``store the bottle of water in the fridge''} can be decomposed as:
\begin{equation}
\begin{split}
D(I) = \langle &\texttt{Open}(\texttt{Fridge}), \texttt{Pick}(\texttt{Bottle}), \\
&\texttt{Place}(\texttt{Bottle}, \texttt{Fridge}), \texttt{Close}(\texttt{Fridge}) \rangle.
\end{split}
\end{equation}

Each atomic task $a(\theta)$ is associated with a reusable symbolic oracle that evaluates whether the intended effect of that parameterized operation has been achieved in the environment, \revision{i.e., a specific state in the environment is achieved}. These atomic symbolic oracles are also parameterized functions over the state of the environment and task arguments, allowing them to generalize across different objects and contexts. Formally, each atomic oracle is defined as:
$
O_{a(\theta)} : s \rightarrow \{\texttt{Pass}, \texttt{Fail}\},
$
where $s$ denotes the current state of the environment. Importantly, the evaluation performed by $O_{a(\theta)}$ is conditioned on the task parameters $\theta$. For instance, the oracle for the task $\texttt{Open}(\texttt{Fridge})$ evaluates a task such as:
\begin{equation}
O_{\texttt{Open}(\texttt{Fridge})}(s) = \texttt{isOpen}(\texttt{Fridge}, s),
\end{equation}

\noindent which returns \texttt{Pass} if the fridge door is open in state $s$, and \texttt{Fail} otherwise. Similarly, $\texttt{Pick}(\texttt{Bottle})$ evaluates whether the robot is currently holding the specified object, and $\texttt{Place}(\texttt{Bottle}, \texttt{Fridge})$ checks whether the bottle is located inside the target container. This parameterized formulation is crucial because it decouples the oracle's structure from specific object instances, enabling the same atomic oracle template to be reused across a wide range of instructions involving different objects and environments.

A fine-grained oracle is then constructed by composing the symbolic oracles associated with the decomposed atomic tasks. Given an instruction ($I$) and its decomposition ($D(I) = \langle a_1, \dots, a_n \rangle$), the corresponding fine-grained oracle is defined as:
$
O_F(I) = \bigwedge_{i=1}^{n} O_{a_i}.
$
This formulation transforms the evaluation from a monolithic decision problem into a structured verification process over atomic tasks. In the example of \textit{``store the bottle of water in the fridge''}, the fine-grained oracle can be defined as follows:
\begin{equation}
\resizebox{\columnwidth}{!}{$
O_F =
\big( O_{\texttt{OpenFridge}} \parallel O_{\texttt{PickBottle}} \big)
\prec O_{\texttt{PlaceBottleInFridge}}
\prec O_{\texttt{CloseFridge}}
$}
\end{equation}

\noindent where $ \parallel $ denotes that the two initial actions can be executed unordered, and $ \prec $ enforces a strict temporal ordering constraint requiring one step to be completed before the next begins. This ensures that both opening the fridge and picking up the bottle are completed, regardless of order, before placing the bottle inside the fridge, and that closing the fridge occurs only after placement. \revision{Note that, as aforementioned, this oracle is robot-independent. The unordered initial actions may be executed concurrently by a multi-arm robot or sequentially by a single-arm robot; the VLA controlling the robot decides. Importantly, the absence of an ordering constraint permits, but does not require, parallel execution.}

This decomposition provides a more informative notion of correctness by explicitly incorporating the instruction's structure and ordering constraints. Rather than producing a single binary outcome, the oracle identifies not only which atomic task was not satisfied but also whether violations occurred in the required temporal sequence. For instance, a failure may arise because the bottle was never successfully grasped, or because it was placed outside the fridge despite correctly executing all prior steps. Importantly, it can also distinguish ordering violations, such as attempting to place the bottle before the fridge has been opened or before the object has been grasped, which represent fundamentally different failure modes from missing or incorrect actions. As a result, fine-grained oracles provide a structured, compositional, and reusable mechanism for evaluating complex robotic tasks, enabling scalable and interpretable assessment of \revision{VLAs}. However, manually defining such fine-grained oracles for every task would be extremely expensive and would not scale to large, diverse task collections. This motivates the need for automatically generating fine-grained oracles from natural language instructions, which is the goal of \approach, presented in Section~\ref{sec:mango}.

\subsection{\revision{Fine-Grained Oracle DSL}} \label{sec:fg_oracles_2}
\revision{Listing~\ref{lst} presents the typed abstract grammar of the JSON-based DSL for representing fine-grained oracles. The grammar defines the required fields, their types, and the hierarchical relationships among task steps and monitoring conditions.}

\revision{In Listing~\ref{lst}, \texttt{NonEmptyList\textless T\textgreater} refers to a JSON array containing one or more elements of type \texttt{T}. Thus, an oracle may contain any positive number of steps, each step may contain any positive number of actions, and each action may contain multiple parameters and monitoring conditions.  The root-level \texttt{execution\_mode} determines the ordering among steps, whereas the step-level mode determines the ordering among the actions within a step. Each \texttt{action\_id} refers to an atomic-task template, and its parameters bind that template to entities in the environment. Each \texttt{Monitor} associates the action with an atomic-oracle template and its instantiated simulator function.}

\begin{lstlisting}[
style=oracledsl,
caption={Typed abstract grammar of the fine-grained oracle DSL},
label={lst}
]
FineGrainedOracle ::= {
instruction : String,
execution_mode : ExecutionMode,
plan : NonEmptyList<Step>,
}


Step ::= {
step : PositiveInteger,
instruction : String,
execution_mode : ExecutionMode,
actions : NonEmptyList<Action>
}

Action ::= {
action_id : AtomicTaskIdentifier,
parameters : NonEmptyList<Parameter>,
monitoring : NonEmptyList<Monitor>
}

Parameter ::= ParameterName | ObjectIdentifier

Monitor ::= {
oracle_id : AtomicOracleIdentifier,
handle_as : FunctionCall
}

ExecutionMode ::= "sequential" | "any_order"

FunctionCall ::= {
function : Identifier,
arguments : List<Value>
}

Value ::= ObjectIdentifier | String | Number | Boolean
\end{lstlisting}
\section{MANGO: Multi-Agent-Based Fine-Grained Oracle Generator}\label{sec:mango}

\begin{figure*}[ht]
\centering
\includegraphics[width=\linewidth]{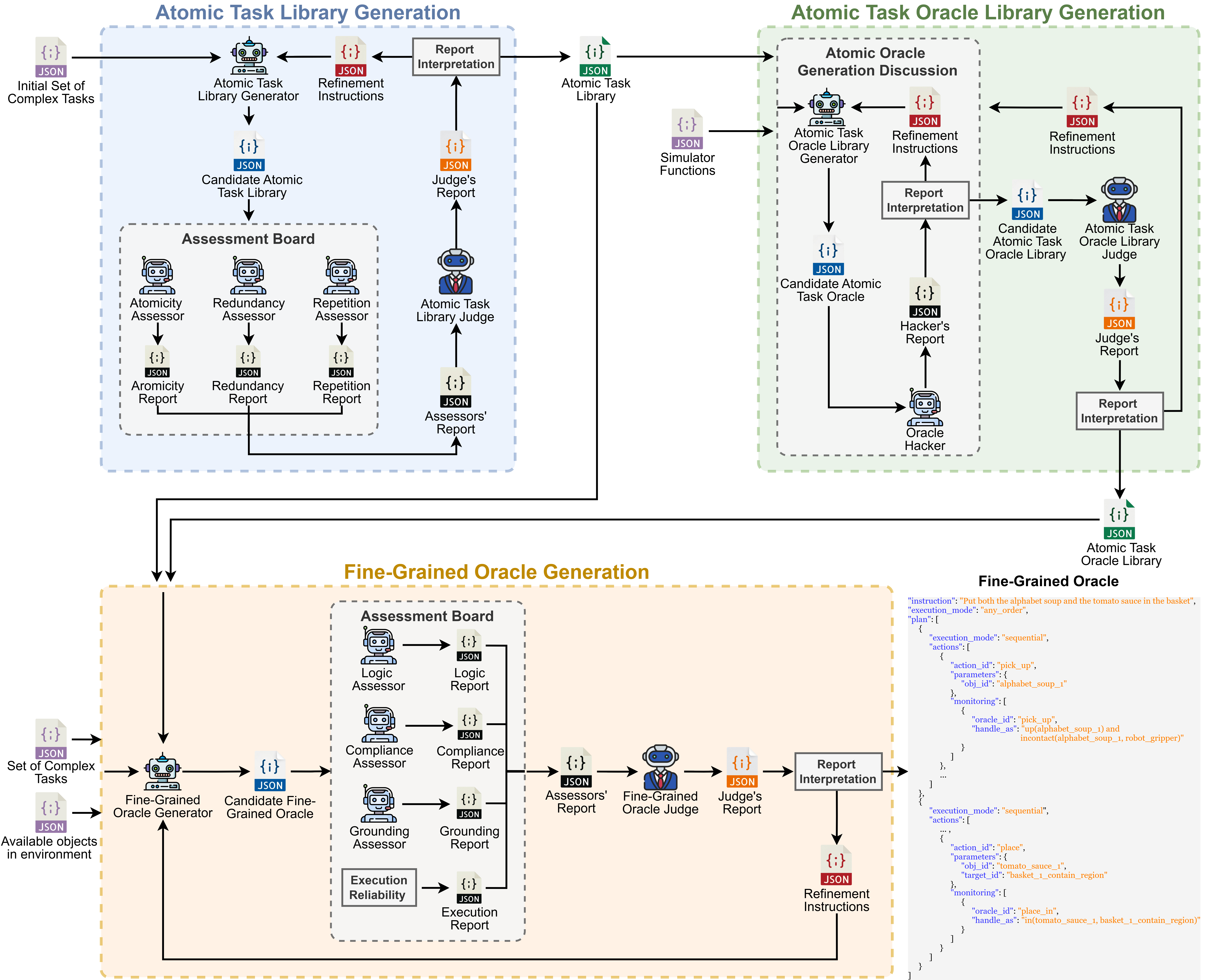}
\caption{Overview of \approach and a fine-grained oracle for the task ``Put both the alphabet soup and the tomato sauce in the basket''.}
\label{fig:approach}
\end{figure*}

\revision{This section presents \approach, a \textbf{M}ulti-\textbf{A}gent fi\textbf{N}e-\textbf{G}rained \textbf{O}racle generator for evaluating VLAs from natural-language task descriptions. As illustrated in Figure~\ref{fig:approach}, the approach comprises three modules. Module I extracts reusable atomic tasks from an initial natural-language instruction set. Module II generates an oracle for each atomic task in the generated library using simulator functions. Module III combines these libraries with a task instruction and the available environment objects to produce a fine-grained oracle.

The libraries generated by Modules I and II  can be reused within the same environment. They need not be repeated for subsequent instructions whose required atomic capabilities are already represented. Module III uses these libraries to generate the fine-grained oracles which specify the expected atomic tasks, their ordering constraints, and the conditions used to assess their completion. The VLA independently generates the actions to control the robot, whereas the generated fine-grained oracle evaluates the resulting execution.}

\subsection{\revision{Agent Roles and Refinement}}

\begin{figure*}[ht]
\centering
\includegraphics[width=\linewidth]{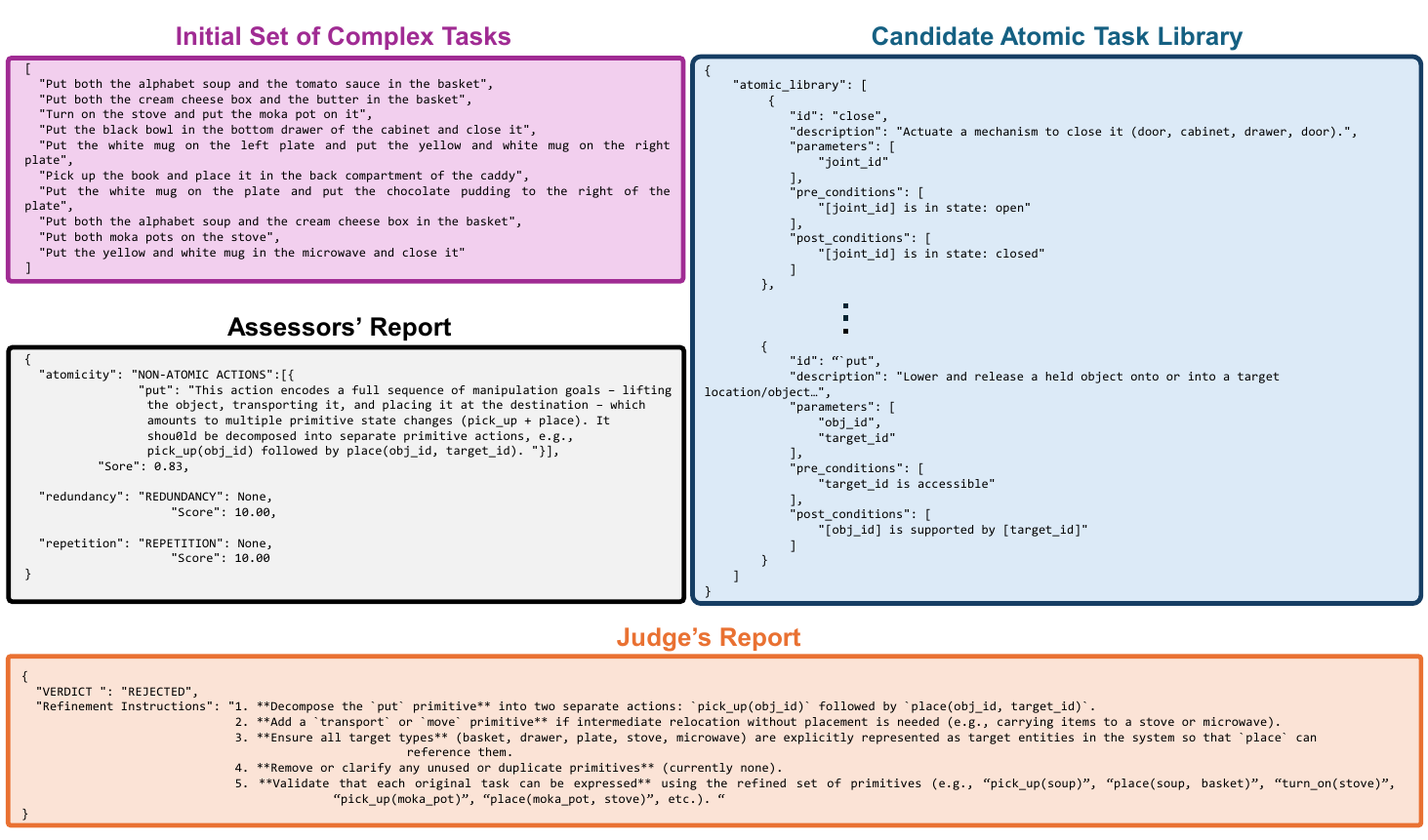}
\caption{Example inputs and outputs of the agents in Module I.}
\label{fig:input_output_module_1}
\end{figure*}

\revision{All three modules follow a common process involving candidate generation, assessment, and refinement. A generator proposes a candidate, specialized assessors examine its properties, and a judge determines whether to accept it or request changes. Feedback from rejected candidates guides subsequent generation. Each module stops upon acceptance or when it exhausts its iteration budget. This process provides opportunities to identify and correct unreliable foundation-model outputs before they propagate through the subsequent modules. However, it does not eliminate non-determinism or guarantee correctness, as different runs may produce different candidates, and errors may remain after refinement. The three agent roles use different model types to balance generation quality, assessment cost, and reasoning capability. Their responsibilities are described below, and detailed agent specifications are provided in the Appendix~\ref{sec:appendix}.

\subsubsection{Generator}\label{Sec:Generator}
The generator produces candidate task libraries, atomic oracles, or fine-grained oracles, depending on the module. It uses a world model because generation requires reasoning about object properties, spatial relationships, and the effects of manipulation tasks~\cite{ha2018world,zhang2025step,team2025gemini,agarwal2025cosmos}. This choice was also supported by the generation quality observed in our preliminary model comparison (Section~\ref{sec:baselines}).

\subsubsection{Assessor}\label{Sec:Assessor}
Assessors use lightweight language models to evaluate specific requirements, such as atomicity, object grounding, or function compliance, within a reasonable execution time and with low resource usage. Multiple assessors form an assessment board and operate independently in parallel. Each provides focused feedback, allowing the approach to examine several properties without assigning every check to a single reasoning agent.

\subsubsection{Judge}\label{Sec:Judge}
The judge uses a reasoning LLM to review the candidate together with the assessment reports. It resolves conflicting or incomplete feedback and decides whether to accept or reject the candidate. When refinement is required, it identifies the necessary corrections and generates a correction report for the Generator. A report-processing component converts this feedback into instructions for the next generation cycle.

\subsection{Module I: Atomic Task Library Generation}

Module I generates a reusable atomic task library from an initial set of complex natural-language instructions. The generated library captures the atomic tasks needed to decompose these complex instructions and provides the atomic tasks used by Modules II and III. Figure~\ref{fig:input_output_module_1} illustrates the generation and assessment process.}

\begin{figure*}[ht]
\centering
\includegraphics[width=\linewidth]{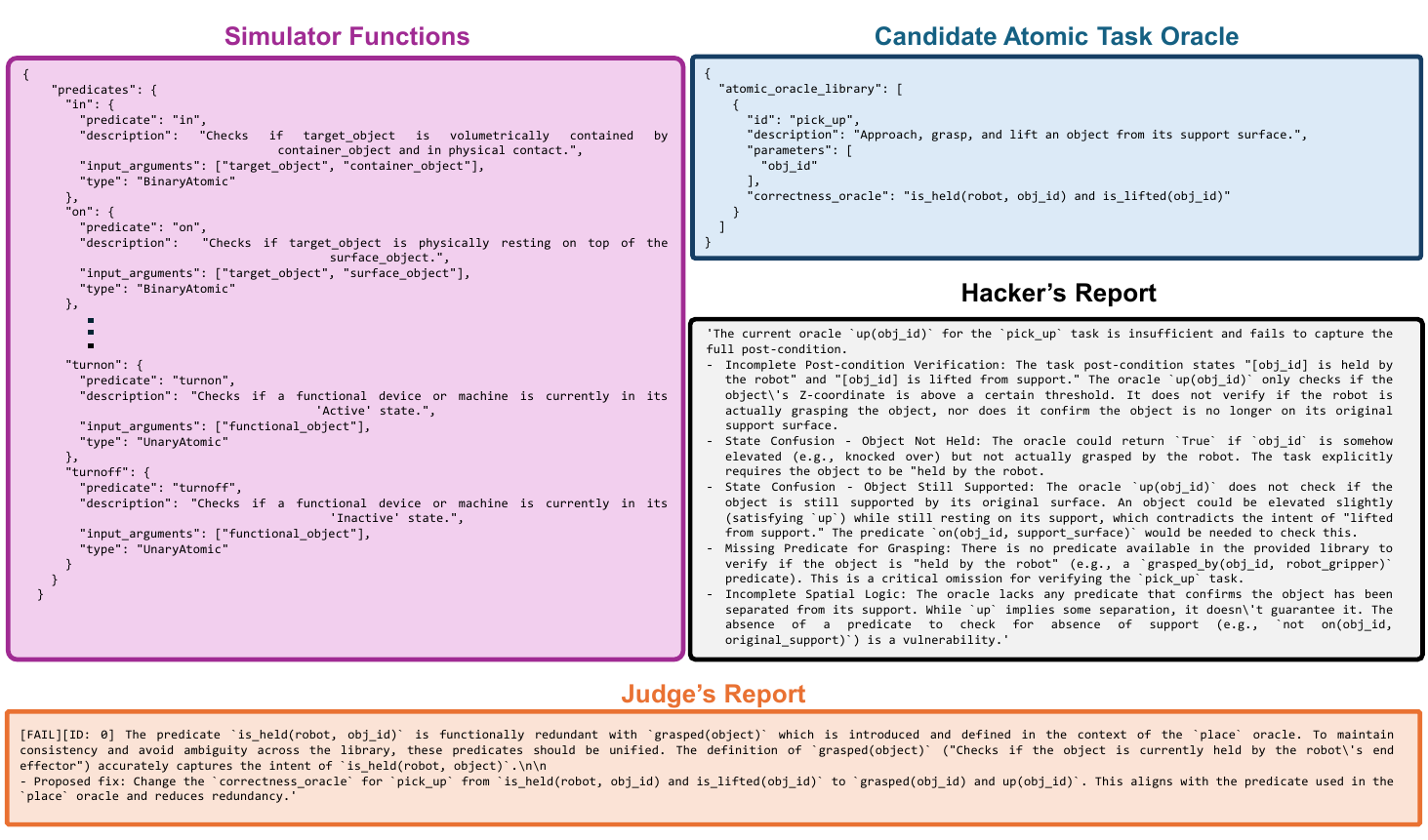}
\caption{Example inputs and outputs of the agents in Module II.}
\label{fig:input_output_module_2}
\end{figure*}

\revision{\subsubsection{Candidate Generation}
The generator decomposes the instructions into parameterized atomic tasks. Each entry includes a semantic description, required parameters, and preconditions and postconditions. These entries describe task-level operations, such as picking or placing an object, rather than the low-level movements needed to execute them. The generator is instructed to generate reusable definitions over object-specific variants. For example, opening a drawer and opening a closet may share a parameterized opening task when their conditions support that abstraction. The candidate library should cover the required capabilities without including unnecessary specializations or duplicate definitions.

\subsubsection{Library Assessment}
Three assessors compose the Assessment Board and examine complementary properties of the candidate. The atomicity assessor identifies tasks that combine operations that should be represented separately. For example, ``pick up an object and place it in a container'' combines grasping and placement. The redundancy assessor identifies specialized definitions already covered by a more general task. For instance, ``place on'' is redundant if an existing parameterized ``place '' task accepts the object and target surface and expresses the same conditions. The repetition assessor instead detects equivalent definitions, such as ``pick'' and ``pick up'' when their parameters, preconditions, and postconditions coincide. Each assessor returns a JSON report containing the detected issues, a justification, and a score out of 10. The judge reviews these reports alongside the candidate library and either accepts it or requests changes. Refinement instructions identify the affected tasks and the required operations, such as decomposition, removal, merging, or generalization.

\subsubsection{Refinement Budget}
Module I permits up to 10 refinement iterations. This practical budget bounds computational cost while allowing repeated correction. A smaller limit, such as that used by CANDOR~\cite{xu2026hallucination}, may terminate refinement before a candidate satisfies the acceptance criteria, whereas a substantially larger limit permits prolonged and costly refinement. Related iterative approaches have also used a ten-iteration strategy when foundation models are used~\cite{vallecillos2025art}.

\subsection{Module II: Atomic Oracle Library Generation}

Module II generates oracles for each atomic task in the atomic task library generated by Module I using the functions available in the simulator. It receives the atomic task library and the available simulator functions, and generates a parameterized oracle for each task. The resulting library supplies the oracles that Module III uses to generate the fine-grained oracles for particular instructions.

In this module, the assessment takes place at two levels. First, the system reviews individual oracle definitions during generation; second, a judge evaluates the complete library. Figure~\ref{fig:input_output_module_2} depicts both stages.}

\revision{\subsubsection{Individual Oracle Generation and Review}
For each atomic task, the generator receives its definition and the simulator's available functions. It produces a candidate oracle containing the task identifier, a description, the oracle parameters, and a logical expression defining the condition to be checked. Parameterization lets the same expression apply to different objects or spatial regions. The assessor, labeled Oracle Hacker in Figure~\ref{fig:input_output_module_2}, examines the candidate for missing or wrongly used simulator functions, inconsistent parameter bindings, and invalid logical expressions. The assessor either accepts the candidate or provides feedback for another generation cycle. This process continues until acceptance or until a limit of 10 iterations is reached, and it is repeated for each atomic task.

\subsubsection{Complete-Library Assessment}
The resulting atomic oracles form a candidate oracle library. The judge reviews their logical coherence, their alignment with the corresponding atomic tasks, and the compatibility of their parameters with the referenced simulator functions. If the library is rejected, the feedback identifies the atomic oracles requiring further changes. Library-level refinement also uses up to 10 iterations. Reviewing both individual definitions and the complete collection addresses local defects while retaining a final assessment of the library as a whole.

\subsection{Module III: Fine-Grained Oracle Generation}}

\begin{figure*}[ht]
\centering
\includegraphics[width=\linewidth]{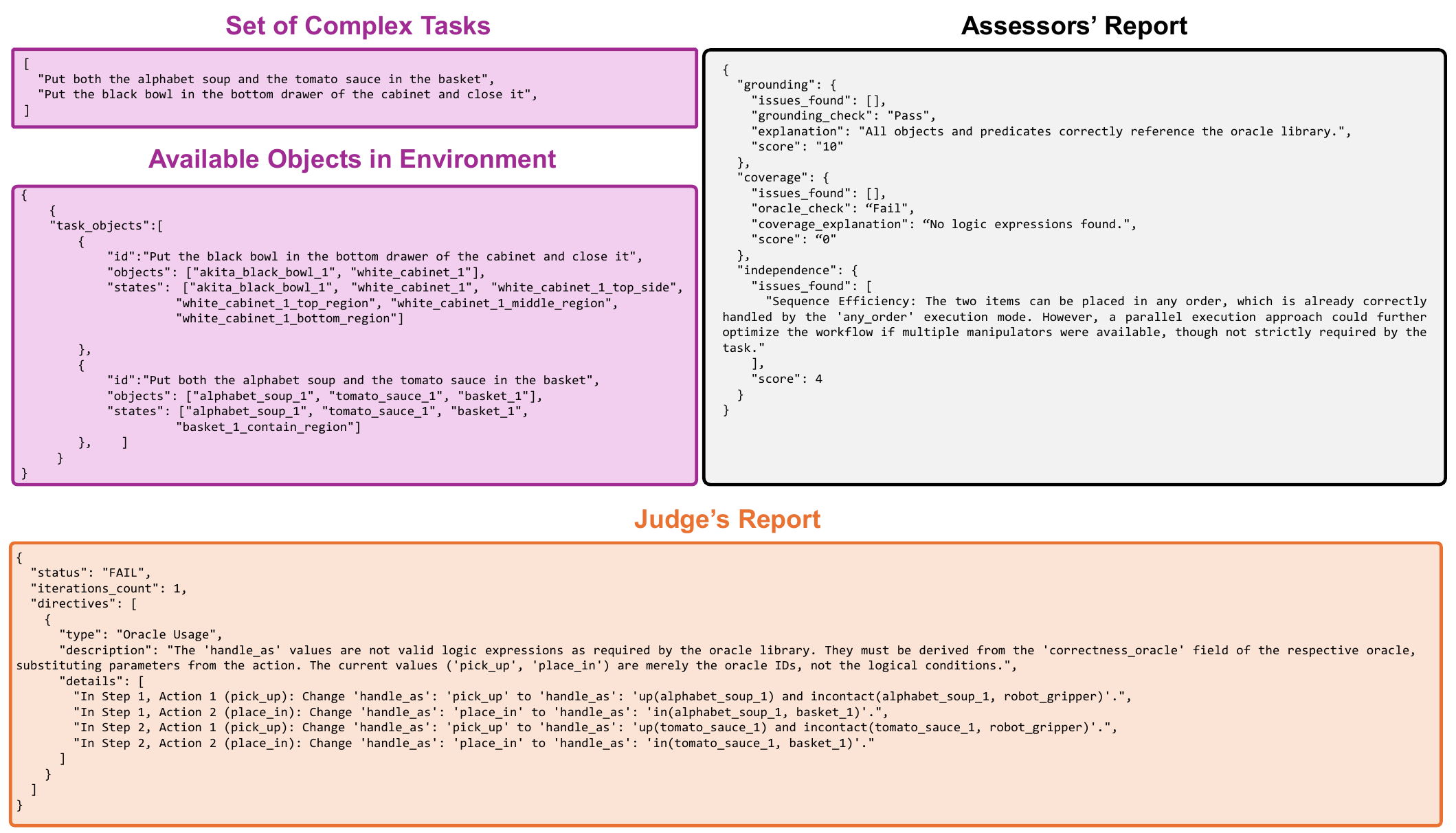}
\caption{Example inputs and outputs of the agents in Module III.}
\label{fig:input_output_module_3}
\end{figure*}

\begin{figure*}[ht]
\centering
\includegraphics[width=\linewidth]{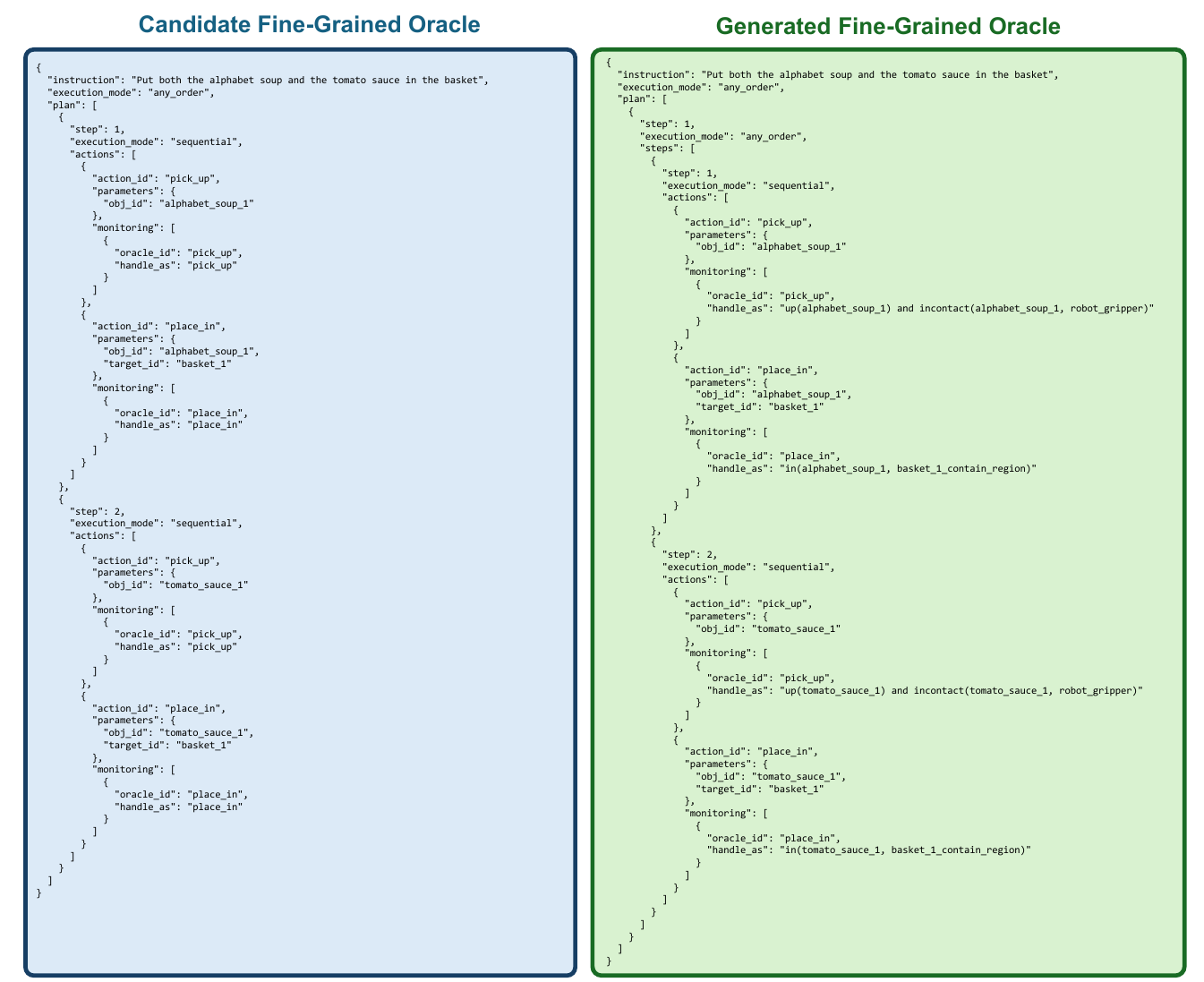}
\caption{A fine-grained oracle before and after refinement.}
\label{fig:oracle_examples}
\end{figure*}

\revision{Module III generates a task-specific fine-grained oracle from a natural-language instruction, the two libraries generated by Modules I and II, and the objects available in the environment. It combines task decomposition with the monitoring conditions needed to evaluate each atomic task. Figure~\ref{fig:input_output_module_3} shows an example of the generated data throughout the process.

\subsubsection{Candidate Generation}
The generator decomposes the instruction in natural language into atomic tasks from the atomic task library, determines their ordering constraints, and associates them with the corresponding atomic oracles. The candidate is serialized as JSON following the structure defined in the preceding Section~\ref{sec:fg_oracles_2}. It specifies the steps, their sequential or unordered composition, the task parameters, and the simulator functions used to monitor completion. This decomposition serves as an evaluation specification. It defines the intermediate conditions to be checked while leaving the VLA responsible for determining how the robot performs the instruction.

\subsubsection{Candidate Assessment}
Three language-model assessors and a rule-based validator examine the generated candidate. The logic assessor checks the decomposition's consistency and ordering constraints. The compliance assessor checks alignment with the instruction and the available atomic task and oracle definitions. The grounding assessor verifies references to objects, regions, and parameters against the entities available in the environment. The execution-reliability validator checks the required output structure and whether the generated oracle can be parsed and executed. This validator is implemented in code and complements the language-model assessments with executable assessments. These assessments target different defects. A candidate may use valid functions but reference the wrong object, omit a required atomic task, or impose an inappropriate ordering constraint. Conversely, a plausible decomposition may contain incomplete or malformed monitoring expressions. Each assessor generates a report that identifies the aspects that require correction.

\subsubsection{Decision and Refinement}
The judge reviews the candidate and the collected reports, then either accepts the oracle or requests further refinement. If the judge rejects the candidate, the feedback specifies what must be corrected, added, or restructured in the next iteration. Figure~\ref{fig:oracle_examples} illustrates this process with an incomplete candidate whose \texttt{handle\_as} fields lack the functions needed to monitor its actions. Refinement supplies these missing expressions, producing the completed oracle shown in the same figure.

Module III permits up to 25 iterations. This larger budget reflects the higher complexity of combining requirements for decomposition, ordering, grounding, and executability. It is a practical resource limit, consistent with the maximum used by Islam et al.~\cite{islam2025evaluating} for LLM-based code generation with execution feedback. As a result, an accepted fine-grained oracle can assess any VLA execution of the corresponding task.}

\revision{\subsection{Application of the Generated Fine-Grained Oracles}

We apply the generated fine-grained oracle to the execution trace of a VLA attempting the corresponding task. During execution, the oracle evaluates its instantiated functions against the simulator state and tracks atomic task completion according to the specified ordering constraints. This provides task-level assessment and identifies atomic tasks whose conditions remain unsatisfied. The task decomposition defines the conditions to monitor; the VLA independently receives the instruction and generates the actions to control the robot. The oracle can therefore be reused across VLA policies operating on the same task and simulator interface. It can also evaluate previously recorded trajectories when they retain sufficient environment-state information to evaluate the required predicates.}

\section{Experimental Setup}\label{sec:experimental_setup}

We conducted an empirical study to assess the performance of \approach in generating fine-grained test oracles for robotic tasks. This section presents the research questions and details the experimental setup used in our evaluation.

\subsection{Research Questions}
In our evaluation, we aimed to answer the following Research Questions (RQs):
\begin{itemize}
     \item[\textbf{RQ1}]\textbf{-- Generation Capability:\textit{ How well does \approach perform at generating fine-grained test oracles for \revision{VLAs}?}} We aim to assess the performance of \approach in generating fine-grained test oracles for \revision{VLAs} automatically. We compare \approach against three baselines to evaluate both the quality of the generated fine-grained oracles and generation efficiency. The assessment focuses on whether \approach can effectively produce correct, valid, and executable fine-grained test oracles while maintaining efficient generation performance.

     \revision{\item[\textbf{RQ2}]\textbf{-- Effectiveness of Generated Fine-Grained Oracles: \textit{To what extent can automatically generated fine-grained oracles reproduce task-evaluation decisions and provide additional information for failure localization?}} This research question evaluates the effectiveness of the generated fine-grained oracles from two complementary perspectives. First, we examine whether their verdicts agree with those of the benchmark-provided symbolic end-state oracles manually defined by experts. Second, we assess their additional ability to localize the task step, i.e., atomic task, at which the execution failed. We therefore evaluate these two perspectives through the following subquestions:
     
     \begin{itemize}
         \item[\textbf{RQ2.1}] \textbf{-- Agreement with symbolic end-state oracles: \textit{How closely do the verdicts of the generated fine-grained oracles agree with those of the benchmark-provided symbolic end-state oracles?}} This subquestion evaluates whether automatically generated fine-grained oracles reproduce the task-level verdicts of symbolic end-state oracles. They are used as a basis for comparison since they are manually defined by the developers of each benchmark, and we can assume that they are largely correct. Consequently, the reported classification metrics quantify agreement between the two oracle types.
         \item[\textbf{RQ2.2}] \textbf{-- Failure Localization: \textit{How accurately do the generated fine-grained oracles identify the task step at which an execution fails?}} This subquestion evaluates the additional diagnostic capability provided by fine-grained oracles. For executions identified as failures, we compare the failure location reported by the generated oracle with the first failed task step determined through manual inspection.
    \end{itemize}}

    \item[\textbf{RQ3}]\textbf{-- Ablation Study: \textit{What is the individual efficiency and effectiveness contribution of the key components in \approach?}} This research question aims to assess the role and contribution of the main components of \approach. To this end, we conduct a series of ablation studies by comparing the full implementation of \approach with variants in which individual components are removed: the Atomic Task Library Generator (w/o Atomic Task Lib.), the Atomic Task Oracle Library Generator (w/o Atomic Oracle Lib.), and the Execution Reliability module (w/o Execution Reliability). Through these comparisons, we assess how much each component contributes to \approach's overall performance and identify the components most critical for generating effective, reliable fine-grained oracles. This analysis also provides insights into the interactions among the different components and helps determine whether the benefits of \approach arise from specific modules or from the synergy of the overall framework.

    \item[\textbf{RQ4}]\textbf{-- Atomic Task Library Reduction: \textit{To what extent can the initial set of complex tasks be minimized while preserving the effectiveness and efficiency of \approach?}} We investigate whether the initial set of complex tasks used by \approach to construct the atomic task and atomic task oracle libraries can be reduced while maintaining the overall performance of the approach. In particular, we examine whether a smaller yet sufficiently diverse subset of tasks can provide the same benefits as the original task set, thereby reducing the effort and computational resources required to use \approach in new application domains. Understanding how much the initial task set can be reduced is important for assessing the scalability of \approach and its applicability to evolving environments where new tasks are continuously introduced.
\end{itemize}

\subsection{Benchmarks}~\label{sec:benchmarks}
We used two benchmarks, LIBERO\_10 benchmark~\cite{liu2023libero} and the RoboCasa humanoid tabletop benchmark~\cite{nvidia2025gr00tn1openfoundation}.

The LIBERO\_10 benchmark comprises 10 distinct task instructions that cover long-horizon manipulation scenarios. These tasks are designed to evaluate the robot controller's compositional reasoning, requiring it to execute sequences of independent actions to achieve a goal. Each instruction typically involves multiple objects and intermediate steps, making the benchmark suitable for studying structured task decomposition and execution (e.g., ``\textit{Put both the alphabet soup and the tomato sauce in the basket}'' or `` \textit{Put the black bowl in the bottom drawer of the cabinet and close it}''). For our evaluation, LIBERO\_10 provides a controlled setting with a limited but diverse set of tasks that capture key challenges in sequential decision-making.

The RoboCasa humanoid tabletop benchmark comprises 24 task instructions focused on everyday household activities performed by a humanoid robot. Compared to LIBERO\_10, this benchmark covers less instruction diversity but greater object and scenario diversity, and involves a more complex robotic platform, i.e., a humanoid robot. As such, it provides a complementary setting to LIBERO\_10, enabling evaluation of \approach on more complex robotic platforms and in varied environments.

To construct the evaluation dataset, we adopt a two-stage sampling protocol over stochastic simulator executions. First, for each benchmark task (10 tasks in LIBERO\_10 and 24 tasks in RoboCasa Humanoid), we generate a fixed pool of independent executions by running the task under varying random seeds and stochastic environmental conditions until 20 executions were recorded, or at least 10 failed executions and 5 successful executions were collected. This process produced a diverse set of executions that capture the VLA's inherent variability. Each execution was then annotated using a deterministic symbolic \revision{end-state} oracle to determine whether it passed or failed. Failed executions were further manually labeled according to the task step at which the failure first occurred. From this execution pool, we constructed a stratified evaluation subset for oracle comparison. Rather than evaluating the full pool, we sample a bounded number of representative executions per task to ensure tractability and controlled experimental cost. 

This design choice was motivated by the high computational cost of evaluation, as each execution required approximately 200 seconds of simulator runtime; scaling to the full execution pool or a larger size would result in prohibitively large experimentation times. Due to the large number of tasks (i.e., 34), we selected up to 10 executions per task, including both successful and failing runs. Failures were categorized by their position along the trajectory (i.e., early, mid, and late). \revision{For failing executions, we employed \emph{failure-stage coverage}, which refers to representing failures occurring at different temporal stages of the trajectory. For a trajectory of length $T$, we normalized the failure position as $pos=t_f/T$, where $t_f$ is the step at which the failure occurred. We classified failures as early ($pos \leq 1/3$), middle ($1/3 < pos \leq 2/3$), or late ($pos>2/3$), and sampled up to two executions from each category per task. When fewer than two executions were available for a category, we selected the closest available executions based on their normalized failure positions. We additionally sampled four successful executions per task at random. This procedure prevents the evaluation from being concentrated on failures occurring at only one stage of task execution.} The resulting oracle evaluation dataset comprises a total of 340 executions, including 40 successful executions for LIBERO\_10 and 96 for Robocasa Humanoid, 20 early-failing executions for LIBERO\_10 and 48 for Robocasa Humanoid, 20 mid-failing executions for LIBERO\_10 and 48 for Robocasa Humanoid, and 20 late-failing executions for LIBERO\_10 and 48 for Robocasa Humanoid.


\subsection{Baselines}\label{sec:baselines}
We compared \approach against three baselines. To the best of our knowledge, \approach is the first end-to-end, multi-agent-based method for automatically generating fine-grained test oracles for robotic tasks. \revision{Consequently, we define three controlled, complete oracle-generation configurations as baselines. A baseline doesn't need to correspond to an independently published solution; it may instead be a simpler end-to-end strategy that establishes reference performance for the same input--output task. Recent Software Engineering studies follow this practice by comparing multi-agent workflows against single-model configurations, defining controlled variants of the proposed framework as baselines, or using randomized strategies with equivalent evaluation budgets~\cite{lin2025soen, xue2024selfpico, oldfield2025faster}. Following this methodology, our baselines are designed by either 1) removing the multi-agent architecture introduced by \approach or 2) replacing the proprietary foundation models with open-source models. These three baselines capture these two comparisons. \approachb preserves \approach's full multi-agent architecture but replaces its proprietary foundation models with open-source alternatives. In contrast, Baseline and Baseline$_o$ preserve the three-module generation pipeline but replace the collaborative Generator-Assessor-Judge structure within each module with a single model. Baseline uses a proprietary model, whereas Baseline$_o$ uses an open-source model. These configurations provide reference performance for simpler, complete oracle-generation approaches under the same inputs and output requirements as \approach. 

We decided to include World Models in the selection of the Foundation Models because they are designed to capture environment dynamics and support reasoning over physical state transitions, object interactions, and implicit scene constraints~\cite{ding2025understanding}. This makes them particularly suitable for tasks that require understanding the consequences of robotic actions or planning movements.  In contrast, LLMs provide a more general-purpose reasoning mechanism without explicit modeling of physical dynamics. Notice that Baseline and Baseline$_o$ also provide evidence about the contribution of the multi-agent design. However, their role differs from the targeted ablations in RQ3, which remove individual modules or the execution-reliability mechanism to isolate the contribution of specific components while retaining the remaining \approach architecture.}

To select the baselines, we conducted a small-scale empirical evaluation in which multiple candidate LLMs and World Models, along with different prompts, were assessed for their ability to decompose complex robotic tasks into structured sequences of atomic tasks. We selected the best-known LLMs and World Models, including both proprietary and open-source models. The evaluated models included DeepSeek-V3.2~\cite{liu2024deepseek}, GPT-5~\cite{singh2025openai}, Mistral 3-675B~\cite{mistral_3}, Nvidia Cosmos Reason 2-8B~\cite{agarwal2025cosmos}, and Gemini Robotics ER 1.6~\cite{team2025gemini}. From this set, we selected Gemini Robotics ER 1.6 for our approach and Nvidia Cosmos Reason 2 for our baselines, as they consistently showed the strongest performance on oracle generation tasks, making them the most suitable and competitive for comparison. We note that both models are World Models and therefore hypothesize that their strong performance likely stems from their integration of implicit physical knowledge, which is particularly beneficial for robotics reasoning tasks.

In the first baseline (i.e., \textit{Baseline}), we removed the three specialized agents of our architecture (i.e., the atomic task library generator, the atomic task oracle generator, and the fine-grained oracle generator) and replaced each of them with an instance of the Gemini Robotics ER model, as this was the best model among the evaluated ones. This results in a three-agent pipeline architecture in which all agents share the same underlying model, rather than being implemented as distinct, specialized components with different roles or inductive biases. Consequently, the system preserves the overall stage decomposition but removes the benefits of specialization and inter-agent diversity.

To define the second and third baselines, we followed the guidelines of Baltes et al.~\cite{baltes2025guidelines} for empirical studies in software engineering involving large language models, which emphasize reproducibility and replicability through open-source models. Accordingly, we defined the second baseline (i.e., \textit{\approachb}) as \approach instantiated exclusively with open-source models. Finally, the third baseline (i.e., \textit{Baseline$_o$}) mirrors the configuration of \textit{Baseline}, in which all three agents use the same model, but replaces the proprietary Gemini Robotics ER model with an open-source alternative. For this purpose, we selected Nvidia Cosmos Reason 2, which performed the best among the open-source models in our preliminary evaluation.

\subsection{Minimization Algorithm for the Set of Complex Tasks (RQ4)} \label{sec:search_alg}

\revision{RQ4 investigates whether the initial set of complex tasks can be reduced while preserving the semantic content of the atomic task library generated from the full task set. This question is motivated by the practical effort required to construct the initial task set when applying MANGO to a new environment. Different complex tasks may involve different objects or scenarios while requiring the same parameterized manipulation capabilities. For example, placing different objects in a container may rely on the same atomic pick-and-place tasks. Consequently, increasing the number of complex tasks does not necessarily increase the coverage of the resulting atomic task library. We hypothesize that a compact set prioritizing diversity of manipulation capabilities can preserve the library generated from the full task set. This would mean that practitioners could generate the atomic task and atomic task oracle libraries from a smaller initial set and reuse them for new tasks whose required atomic capabilities are already represented, avoiding unnecessary executions of Modules I and II.}

\revision{To test this hypothesis, we introduce an auxiliary minimization procedure used exclusively to answer RQ4. This procedure is not part of \approach, which generates its libraries directly from the provided task set. Specifically,} we formulated the minimization of the set of complex tasks as a single-objective constrained optimization problem, aiming to identify the smallest subset of complex tasks that yields a semantically similar atomic task library to that produced by the full task set. Let $O$ denote the original set of complex tasks, and let $Candidate \subseteq O$ be a candidate subset represented as a binary vector $Candidate \in \{0,1\}^n$, where $n = |O|$. Each $Candidate_i = 1$ indicates inclusion of the corresponding task, while $Candidate_i = 0$ indicates exclusion. Then we can define the optimization problem as follows:

\begin{equation}
\min_{Candidate \subseteq O} |Candidate|
\end{equation}

\noindent which is subject to the constraint:

\begin{equation}
Similarity\big(L(Candidate), L(O)\big) \geq \tau
\end{equation}

\noindent where $L(\cdot)$ denotes the atomic-task library generation function, $Similarity(\cdot)$ measures semantic similarity between libraries using cosine similarity over sentence embeddings, and $\tau$ is a predefined similarity threshold \revision{that determines whether a candidate subset is feasible. We use this similarity measure only within the RQ4 minimization procedure to compare the atomic task libraries generated from the full original task set and each candidate subset.} We set $\tau = 0.95$, following common practice in semantic textual similarity and embedding-based matching tasks, where cosine similarity values above 0.9 are generally indicative of near-paraphrastic or highly equivalent semantic content~\cite{reimers2019sentence, cer2017semeval, zhang2019bertscore}. This choice ensures strict semantic preservation while allowing minimal variation from stochastic generation effects in the foundation models used in Module I.

To solve this problem, we employed a $(1+1)$ Evolutionary Algorithm (EA)~\cite{rudolph1997convergence, droste2002analysis}, a stochastic hill-climbing heuristic widely used in search-based software engineering~\cite{harman2001search, eiben2015introduction}. The $(1+1)$EA is particularly well-suited to expensive fitness evaluations because it maintains a single candidate solution and generates exactly one offspring per iteration via mutation. Unlike Genetic Algorithms~\cite{goldberg1989genetic, holland1992adaptation}, which rely on a population of solutions and therefore require repeated evaluation of multiple individuals per generation, the $(1+1)$EA avoids the substantial computational overhead associated with population initialization, maintenance, and multi-sample fitness evaluation. In our setting, this overhead is prohibitive because generating the atomic-task library is costly, as reported in Table~\ref{tab:RQ1} (Module I). As a result, even small population sizes would require an impractically large number of expensive evaluations per generation. We also considered greedy search strategies~\cite{cormen2022introduction}. While greedy methods reduce computational cost by making locally optimal decisions at each step, they are prone to premature convergence and irreversible choices in combinatorial spaces with strong interdependencies among elements, as in our case. Removing or retaining a single complex task can induce non-local effects on the resulting atomic-task library, making purely local decisions potentially suboptimal. The $(1+1)$EA mitigates these limitations by enabling probabilistic exploration through mutation while retaining a strict acceptance criterion that only admits offspring satisfying the constraint and improving the objective function.

\begin{algorithm}[ht]
\caption{Overview of the Complex Task Set Minimization Algorithm}
\label{alg:task_min}
\KwIn{$O$ // Original set of complex tasks}
\KwOut{$Best$ // Minimized set of complex tasks}

$InitialLibrary \leftarrow$ generateAtomicTaskLib($O$);\\
$Best \leftarrow O$;\\

\While{terminationCriteriaNotMet}{
    $Candidate \leftarrow$ mutate($Best$);\\
    $CandidateLibrary \leftarrow$ generateAtomicTaskLib($Candidate$);\\
    
    \If{$Similarity(CandidateLibrary, InitialLibrary) \geq \tau$ \textbf{and} $|Candidate| < |Best|$}{
        $Best \leftarrow Candidate$;
    }
}
\Return{$Best$};
\end{algorithm}

The algorithm takes as input the original set of complex tasks $O$ and first generates the corresponding atomic-task library $InitialLibrary$ using the procedure from Module I. The search (Algorithm~\ref{alg:task_min}, Lines~3-8) is initialized with the full task set, ensuring that the initial solution is feasible with respect to the semantic constraint. Each candidate solution is encoded as a binary vector $Candidate \in \{0,1\}^n$, where $n = |O|$. The search starts from the full solution ($Candidate_i = 1, \forall i$) and iteratively applies a bit-flip mutation operator with mutation probability $1/n$~\cite{muhlenbein1992genetic} (Algorithm~\ref{alg:task_min}, Line~4), enabling local exploration of the subset space. For each iteration, the atomic-task library $L(Candidate)$ is generated and compared against the reference library $L(O)$ using cosine similarity between sentence embeddings~\cite{reimers2019sentence, das2021sentence} (Algorithm~\ref{alg:task_min}, Lines~5-8), checking whether it outperforms the $Best$ candidate. A candidate replaces the current $Best$ solution only if it satisfies the similarity constraint and is a smaller subset, ensuring that semantic integrity is never sacrificed for task reduction. The optimization process terminates after either \revision{200} generations or 20 consecutive iterations without improvement of $Best$.

\subsubsection{Termination Criteria}

The algorithm uses two complementary termination conditions. A fixed upper bound of 200 generations ensures termination and controls computational resources~\cite{jain2001termination}, which is particularly important given the cost of LLM-based evaluations. In addition, we employ a no-improvement-for-K-iterations stopping criterion, terminating the search if the objective shows no improvement over 20 consecutive generations. This follows established recommendations in evolutionary computation~\cite {leung2001orthogonal}, where combining fixed computational budgets with no-improvement stopping criteria balances efficiency and robustness, avoiding both premature termination and unnecessary exploration in flat objective landscapes. We choose $K = 20$ to account for noise in objective estimates from stochastic LLM behavior while maintaining computational efficiency.

\subsubsection{Number of Executions}

\revision{Due to the stochastic nature of Module I, which is the only one executed in RQ4, there is} variability in fitness evaluation, as identical inputs may yield different outputs across executions. Such non-determinism has been empirically observed in prior work and is known to affect the stability and reproducibility of LLM-based software engineering evaluations~\cite{ouyang2025empirical}. More broadly, benchmarking studies on LLM-based systems emphasize that single-run evaluations are insufficient to characterize performance reliably due to inherent sampling variance~\cite{mohammadi2025evaluation}. To mitigate this issue, we evaluate each candidate solution three times \revision{by executing Module I on the same task subset and calculating the fitness for each resulting atomic task library}, and compute the final fitness score as the mean of these evaluations. \revision{As a result, this produces a single numerical fitness value for the candidate subset. The generated libraries themselves are neither averaged nor merged.} This choice is supported by our preliminary analysis, which showed that the standard deviation of similarity scores across repeated executions is consistently below 0.001 and typically close to 0.000. This follows the protocol proposed by Xu et al.~\cite{xu2026hallucination} for their experiment evaluation, in which they repeat experiments three times to reduce the impact of randomness in the foundation models used. In particular, this reduces the influence of outlier runs and provides a more stable estimate of expected performance, while maintaining an affordable computational cost given the expense of repeated LLM invocations~\cite{camara2024towards, mohammadi2025evaluation}. In addition, as part of our evaluation, we ran the minimization algorithm 10 times on both benchmarks.

\subsection{Evaluation Metrics and Statistical Tests}
We used a set of metrics, each aligned with an RQ, to evaluate \approach's efficiency and effectiveness, as well as the quality of the fine-grained oracles it generated. To compare \approach's efficiency and effectiveness against the baselines in RQ1, we measured efficiency as the average execution time of each module in \approach. Effectiveness was measured in terms of i) the structural validity rate of the fine-grained planned oracles, ii) the executability rate of the generated fine-grained oracles, and iii) the Levenshtein similarity between the generated oracle library and a manually constructed ground truth library. The former assesses whether the generated plan for fine-grained oracles satisfies a predefined structural criterion and whether the oracles represent a logic expression. The executability rate measures the proportion of generated oracles that the simulator can execute successfully. Since \approach relies on LLMs and World Models, hallucinations may lead to using functions that do not exist in the simulator, thereby affecting executability. The latter metric, i.e., the similarity, captures the degree of syntactic and structural distance from the ground-truth fine-grained oracle by computing the minimum number of edit operations (insertions, deletions, and substitutions) required to transform a generated oracle into its reference in the ground-truth fine-grained oracle. This metric estimates the effort required to correct a generated oracle. We normalized this metric to [0, 1], with higher values indicating greater similarity to the ground truth and lower values indicating greater editing effort to match the ground truth.

To compare the generated fine-grained oracles with state-of-the-art symbolic \revision{end-state} oracles in RQ2, we used standard classification metrics on the Oracle Evaluation Dataset of task executions from both benchmarks (see Section~\ref{sec:benchmarks} for more information). Specifically, we used accuracy, precision, recall, and F1-score, together with the distribution of true positives (TP), true negatives (TN), false positives (FP), and false negatives (FN). Accordingly, a True Positive (TP) corresponds to an execution that both the symbolic \revision{end-state} oracle and the fine-grained oracle detect as a failure, while a True Negative (TN) corresponds to an execution that neither oracle type detects as a failure. A False Positive (FP) occurs when the fine-grained oracle incorrectly identifies a passing execution as a failure, and a False Negative (FN) occurs when the fine-grained oracle incorrectly classifies a failing execution as a pass. We conducted this evaluation in two steps. First, we assessed task-level correctness to determine whether the generated fine-grained oracles perform as well as symbolic \revision{end-state} oracles at identifying whether each task was successfully completed. In this setting, we applied standard binary classification metrics to measure alignment with the benchmark-provided symbolic \revision{end-state} oracles, which serve as the reference because the benchmark developers manually constructed and validated them. Second, we evaluated the generated fine-grained oracles' failure-localization capability. Unlike symbolic \revision{end-state} oracles, our fine-grained oracles can identify the specific task step at which a failure occurs. We evaluated localization performance using exact-match localization accuracy, defined as the proportion of executions for which the predicted failure location exactly matches the ground-truth failure step. This allowed us to assess not only whether a task is correctly assessed, but also whether its location within the task execution steps is correctly identified. 

For RQ3, we evaluated each variant of \approach using the same metrics employed in RQ1. This ensured consistent comparisons of both efficiency and the effectiveness of fine-grained oracle generation across the different variants of \approach.

Finally, for RQ4, we formulated the problem as an optimization task to minimize the size of the initial set of complex tasks while preserving the quality of the generated atomic task library. To assess the impact of reducing the initial set, we used three metrics: (1) the number of tasks after minimization (\#Tasks$_{min}$), (2) the Task Reduction Ratio (TRR), and (3) the Levenshtein similarity (Similarity) between the atomic task library generated from the minimized set and that generated from the original set of complex tasks. These metrics let us quantify how much the initial set of complex tasks could be reduced without significantly affecting the resulting atomic task library. \revision{Regarding the execution times of the minimization algorithm evaluated in RQ4, as described in Section~\ref{sec:search_alg}, we executed Module I three times at each minimization iteration to account for its non-deterministic behavior. In addition, we repeated the complete minimization algorithm 10 times for each benchmark and report the mean and standard deviation for each evaluation metric across these runs. }

In addition to the metrics used for RQ1 and RQ3, \revision{to account for the non-determinism of the Foundation models used in \approach and the compared baselines}, we ran each approach on each benchmark 10 times and statistically compared the baselines against \approach. We applied the same statistical procedure to both research questions. For each pairwise comparison, we first assessed the normality of the result distributions using the Shapiro-Wilk test. Based on the results, we evaluated statistical significance using either a Student's t-test for normally distributed data or the Kruskal-Wallis test for non-normally distributed data. Statistical significance was determined using a significance level of $\alpha = 0.05$, i.e., results with $p < 0.05$ were considered statistically significant. We also measured effect sizes using Vargha and Delaney's $\hat{A}_{12}$ value. Following Romano et al.~\cite{romano2006exploring}, the effect size can be classified as \textit{negligible} when $d < 0.147$, \textit{small} when $d < 0.33$, \textit{medium} when $d < 0.474$, and \textit{large} when $d \geq 0.474$, where $d = 2*|\hat{A}_{12} - 0.5|$. Finally, to assess the localization accuracy of the fine-grained oracles in RQ2, we needed a manual comparison with execution videos. Since evaluating all 2,814 executions was prohibitively expensive, we selected a random subset of 340 failing executions, determined using Cochran’s sample size formula~\cite{cochran1977sampling} with a significance level of 0.05.

\subsection{Execution Platform and Runs}
We ran all experiments on a 64-bit Ubuntu 20.04 LTS server equipped with an AMD EPYC 7773X CPU and an NVIDIA RTX A6000 GPU with 48 GB of memory. The implementation used Python 3.10 and CUDA 12.8. We executed each configuration \revision{ of the approach used in RQ1 and RQ3} 10 times \revision{per benchmark}, resulting in 10 $\times$ (1 \approach + 3 baselines + 3 ablation studies) $\times$ 2 benchmarks = 140 \revision{generation runs}. In particular, for the LIBERO\_10 benchmark, \revision{the number of generated fine-grained oracles was} 70 \revision{runs} $\times$ 10 tasks = 700 fine-grained oracles, while for the RoboCasa Humanoid Tabletop benchmark, \revision{the corresponding number was} 70 \revision{runs} $\times$ 24 tasks = 1,680 fine-grained oracles. Overall, this \revision{resulted} in 2,380 fine-grained oracles. As described in Section~\ref{sec:benchmarks}, for RQ2 we selected all the executable fine-grained oracles from \approach to evaluate their detection capability\revision{. This selection included executable oracles from all ten generation runs per benchmark. Each oracle was generated using the atomic task and atomic oracle libraries from its own run and was evaluated with the corresponding oracle definitions. We did not merge libraries from different runs, and we evaluated each generated oracle version for each task separately.}

\revision{For RQ2, we ran ten evaluations for each executable oracle generated by \approach on its corresponding task, using recorded environment configurations to reproduce matching initial conditions across oracle comparisons}, resulting in a total of 79 oracles $\times$ 10 \revision{runs} for LIBERO\_10 + 224 oracles $\times$ 10 \revision{runs} for \revision{RoboCasa Humanoid} = 3,030 task executions. \revision{However, 216 RoboCasa Humanoid runs could not be completed because scenario initialization failed when restoring the recorded environment configurations. We excluded these runs from the RQ2 metrics, leaving 2,814 evaluated executions: 790 for LIBERO\_10 and 2,024 for RoboCasa Humanoid. The reported metrics aggregate these individual oracle evaluations across the ten generation runs.}

\section{Analysis of the Results and Discussion}\label{sec:results}

\subsection{RQ1 - Generation Capability}

Generation efficiency across Modules I to III revealed a consistent advantage of \approach over \approachb, despite both frameworks sharing the same underlying multi-agent architecture and a judge-based iterative refinement mechanism. As shown in Table~\ref{tab:RQ1}, \approach achieved lower generation times in all modules and both benchmarks, with particularly pronounced gains in the most refinement-intensive stage (Module II), where improvements in average time reached approximately 2.2$\times$ on LIBERO\_10 (633.9s vs. 1403.2s) and 6.7$\times$ on Robocasa Humanoid (220.4s vs. 1496.5s). Similar trends were observed in Modules I and III, indicating that the efficiency gains were consistent rather than isolated to a specific stage. These differences primarily stemmed from the quality of the candidate in each module. \approach generated more accurate and stable candidate libraries and fine-grained oracles, which reduced the number of refinement iterations required at each module. In contrast, \approachb produced less precise candidates, leading to repeated refinement cycles and, in several cases, to the maximum allowed number of iterations. Such cases resulted in incomplete convergence and occasionally in invalid fine-grained oracles, which directly increased generation time. This effect was further exacerbated in Module II, since oracle generation relied on a debate-based construction of each atomic task in the atomic task library, which increased generation time exponentially with each additional refinement iteration. 

\begin{table*}[ht]
    \centering
    \caption{Comparison of \approach, \approachb, Baseline, and Baseline$_o$ on the LIBERO\_10 and Robocasa Humanoid benchmarks in terms of generation time, validity, executability, and similarity. Module~I, Module~II, and Module~III correspond to the three modules of the evaluated approaches. Reported values correspond to the mean ($m$) and standard deviation ($\sigma$). For each benchmark and metric, the best result is highlighted in \textbf{bold} and \underline{underlined}. Arrows indicate whether lower ($\downarrow$) or higher ($\uparrow$) values are preferable. }
    \label{tab:RQ1}
    \resizebox{0.9\textwidth}{!}{
        \begin{tabular}{llrrrrrrrrrrrr}
        \toprule
         &  & \multicolumn{6}{c}{\textbf{Generation Time (s) $\downarrow$}} & \multicolumn{2}{c}{\textbf{Validity $\uparrow$}} & \multicolumn{2}{c}{\textbf{Executability $\uparrow$}} & \multicolumn{2}{c}{\textbf{Similarity $\uparrow$}} \\ \cmidrule(lr){3-8} \cmidrule(lr){9-10} \cmidrule(lr){11-12} \cmidrule(lr){13-14}
         &  & \multicolumn{2}{c}{Module I} & \multicolumn{2}{c}{Module II} & \multicolumn{2}{c}{Module III} & \multicolumn{2}{c}{Module III} & \multicolumn{2}{c}{Module III} & \multicolumn{2}{c}{Module III} \\ 
         \cmidrule(lr){3-4} \cmidrule(lr){5-6} \cmidrule(lr){7-8} \cmidrule(lr){9-10} \cmidrule(lr){11-12} \cmidrule(lr){13-14}
         &  & \multicolumn{1}{c}{$m$} & \multicolumn{1}{c}{$\sigma$} & \multicolumn{1}{c}{$m$} & \multicolumn{1}{c}{$\sigma$} & \multicolumn{1}{c}{$m$} & \multicolumn{1}{c}{$\sigma$} & \multicolumn{1}{c}{$m$} & \multicolumn{1}{c}{$\sigma$} & \multicolumn{1}{c}{$m$} & \multicolumn{1}{c}{$\sigma$} & \multicolumn{1}{c}{$m$} & \multicolumn{1}{c}{$\sigma$} \\
        \cmidrule{1-14}
        
        \multirow{4}{*}{\textbf{LIBERO\_10}} & \approach & 30.776 & 10.802 & 633.936 & 48.153 & 346.513 & 98.054 & 0.950 & 0.053 & \best{0.790} & \best{0.137} & \best{0.913} & \best{0.037} \\
         & \approachb & 69.215 & 34.953 & 1403.256 & 238.223 & 534.191 & 109.746 & \best{0.970} & \best{0.048} & 0.513 & 0.042 & 0.867 & 0.009 \\
         & Baseline  & \best{10.075} & \best{0.157} & 22.511 & 0.735 & 8.942 & 0.563 & 0.130 & 0.067 & 0.113 &  0.063 & 0.858 & 0.009 \\
         & Baseline$_o$ & 12.268 & 0.026 & \best{12.745} & \best{0.277} & \best{7.776} & \best{0.483} & 0.070 & 0.082 & 0.000 & 0.000 & 0.723 & 0.010 \\
         
        \cmidrule{1-14}
        \multirow{4}{*}{\begin{tabular}[c]{@{}l@{}}\textbf{Robocasa} \\ \textbf{Humanoid}\end{tabular}} & \approach & 21.138 & 4.315 & 220.380 & 15.641 & 350.094 & 68.636 & \best{1.000} & \best{0.000} & \best{0.933} & \best{0.045} & \best{0.947}  &\best{0.014} \\
        
        & \approachb & 32.030 & 4.289 & 1496.496 & 184.120 & 608.736 & 53.246 & \best{1.000} & \best{0.000} & 0.446 & 0.074 & 0.902 & 0.007 \\
         & Baseline & \best{7.617} & \best{0.138} & 17.641 & 0.413 & \best{5.577} & \best{0.274} & 0.117 & 0.070 & 0.008 & 0.018 & 0.787 & 0.008 \\
         & Baeline$_o$ & 8.300 & 0.097 & \best{12.963} & \best{0.559} & 5.634 & 0.037 & 0.000 & 0.000 & 0.000 & 0.000 & 0.790 & 0.001 \\
         
        \bottomrule
        \end{tabular}
    }
    
\end{table*}

\begin{table*}[ht]
    \centering
    \caption{Statistical comparison of \approach against \approachb, Baseline, and Baseline$_o$ on the LIBERO\_10 and Robocasa Humanoid benchmarks. Results are reported for generation time, validity, executability, and similarity across the evaluated modules (Module~I, Module~II, and Module~III). The table reports Vargha and Delaney’s effect size statistic ($\hat{A}_{12}$) together with the corresponding p-values for each pairwise comparison. Statistically significant differences in favour of \approach are highlighted in \colorbox{cyan!13}{blue}. Arrows indicate whether lower ($\downarrow$) or higher ($\uparrow$) values are preferable.}
    \label{tab:RQ1_statistics}
    \resizebox{0.9\textwidth}{!}{
        \begin{tabular}{llrrrrrrrrrrrr}
        \toprule
         &  & \multicolumn{6}{c}{\textbf{Generation Time (s) $\downarrow$}} & \multicolumn{2}{c}{\textbf{Validity $\uparrow$}} & \multicolumn{2}{c}{\textbf{Executability $\uparrow$}} & \multicolumn{2}{c}{\textbf{Similarity $\uparrow$}} \\ \cmidrule(lr){3-8} \cmidrule(lr){9-10} \cmidrule(lr){11-12} \cmidrule(lr){13-14}
         &  & \multicolumn{2}{c}{Module I} & \multicolumn{2}{c}{Module II} & \multicolumn{2}{c}{Module III} & \multicolumn{2}{c}{Module III} & \multicolumn{2}{c}{Module III} & \multicolumn{2}{c}{Module III} \\ 
         \cmidrule(lr){3-4} \cmidrule(lr){5-6} \cmidrule(lr){7-8} \cmidrule(lr){9-10} \cmidrule(lr){11-12} \cmidrule(lr){13-14}
         &  & \multicolumn{1}{c}{ \^{A}$_{12}$} & \multicolumn{1}{c}{p-value} & \multicolumn{1}{c}{ \^{A}$_{12}$} & \multicolumn{1}{c}{p-value} & \multicolumn{1}{c}{ \^{A}$_{12}$} & \multicolumn{1}{c}{p-value} & \multicolumn{1}{c}{ \^{A}$_{12}$} & \multicolumn{1}{c}{p-value} & \multicolumn{1}{c}{ \^{A}$_{12}$} & \multicolumn{1}{c}{p-value} & \multicolumn{1}{c}{ \^{A}$_{12}$} & \multicolumn{1}{c}{p-value} \\
        \cmidrule{1-14}
        
        \multirow{3}{*}{\textbf{LIBERO\_10}} & \approachb & \cellcolor{cyan!13} 0.11 & \cellcolor{cyan!13} 0.0070 & \cellcolor{cyan!13} 0.00 & \cellcolor{cyan!13} \textless{}0.0001 & \cellcolor{cyan!13} 0.12 & \cellcolor{cyan!13} 0.0046 & 0.40 & 0.3980 & \cellcolor{cyan!13} 0.94 & \cellcolor{cyan!13} 0.0005 & \cellcolor{cyan!13} 0.90 & \cellcolor{cyan!13} 0.0028 \\
         & Baseline  & 1.00 & 0.0002 & 1.00 & \textless{}0.0001 & 1.00 & \textless{}0.0001 & \cellcolor{cyan!13} 1.00 & \cellcolor{cyan!13} \textless{}0.0001 & \cellcolor{cyan!13} 1.00 & \cellcolor{cyan!13} 0.0001 & \cellcolor{cyan!13} 0.90 & \cellcolor{cyan!13} 0.0028 \\
         & Baseline$_o$ & 1.00 & 0.0004 & 1.00 & 0.0002 & 1.00 & \textless{}0.0001 & \cellcolor{cyan!13} 1.00 & \cellcolor{cyan!13} \textless{}0.0001 & \cellcolor{cyan!13} 1.00 & \cellcolor{cyan!13} \textless{}0.0001 & \cellcolor{cyan!13} 1.00 & \cellcolor{cyan!13} 0.0002 \\
         
        \cmidrule{1-14}
        \multirow{3}{*}{\begin{tabular}[c]{@{}l@{}}\textbf{Robocasa} \\ \textbf{Humanoid}\end{tabular}} & \approachb & \cellcolor{cyan!13} 0.03 & \cellcolor{cyan!13} \textless{}0.0001 & \cellcolor{cyan!13} 0.00 & \cellcolor{cyan!13} \textless{}0.0001 & \cellcolor{cyan!13} 0.00 & \cellcolor{cyan!13} \textless{}0.0001 & 0.50 & 1.0000 & \cellcolor{cyan!13} 1.00 & \cellcolor{cyan!13} 0.0001 & \cellcolor{cyan!13} 1.00 & \cellcolor{cyan!13} 0.0002 \\
         & Baseline & 1.00 & \textless{}0.0001 & 1.00 & \textless{}0.0001 & 1.00 & \textless{}0.0001 & \cellcolor{cyan!13} 1.00 & \cellcolor{cyan!13} \textless{}0.0001 & \cellcolor{cyan!13} 1.00 & \cellcolor{cyan!13} \textless{}0.0001 & \cellcolor{cyan!13} 1.00 & \cellcolor{cyan!13} 0.0002 \\
         & Baeline$_o$ & 1.00 & 0.0002 & 1.00 & 0.0002 & 1.00 & 0.0002 & \cellcolor{cyan!13} 1.00 & \cellcolor{cyan!13} \textless{}0.0001 & \cellcolor{cyan!13} 1.00 & \cellcolor{cyan!13} \textless{}0.0001 & \cellcolor{cyan!13} 1.00 & \cellcolor{cyan!13} 0.0002  \\
         
        \bottomrule
        \end{tabular}
    }
    
\end{table*}

The statistical analysis in Table~\ref{tab:RQ1_statistics} confirmed that these differences were systematic, with \approach significantly outperforming \approachb across all generation time comparisons and consistently large effect sizes ($\hat{A}_{12} \leq 0.12$). This indicated that the observed improvements stemmed from convergence efficiency, which directly correlated with effective model selection. In contrast, the comparison between \approach and Baseline and Baseline$_o$ highlighted the impact of architectural design rather than differences in model selection. Unlike \approach and \approachb, the baselines did not employ iterative multi-agent reasoning or judge-based validation. Instead, they relied on a single-pass generation procedure, eliminating refinement overhead but also removing mechanisms that ensured semantic consistency and constraint satisfaction. As reported in Table~\ref{tab:RQ1}, this design choice resulted in substantially lower generation times for both baselines across all modules. Consequently, \approach showed an overhead of approximately 3$\times$ in Module I, 28.5$\times$ in Module II, and 38.7$\times$ in Module III on LIBERO\_10, with even larger overheads on Robocasa Humanoid (up to 62.7$\times$ in Module III). While baselines were significantly faster (p $< 0.0001$), as presented in Table~\ref{tab:RQ1_statistics}, this advantage was accompanied by a substantial degradation in correctness-related metrics, with large effect sizes consistently favoring \approach. However, these generation-time differences have limited practical significance because \approach is executed offline. 

Structural validity remained comparable between \approach and \approachb, with both achieving near-perfect scores across benchmarks (0.950--1.000). This indicated that both systems were equally capable of producing syntactically valid fine-grained oracles. However, this similarity did not extend to executability, which diverged widely, with \approach achieving 0.790 vs. 0.513 on LIBERO\_10 and 0.933 vs. 0.446 on Robocasa Humanoid. A similar pattern was observed for ground-truth specifications, where \approach consistently showed substantially higher similarity, indicating closer alignment with the reference oracles. This discrepancy stems from instability in \approachb during iterative refinement. As explained previously, less accurate candidates within each module led to repeated judge refinements and frequent convergence failures, including instances in which the maximum number of iterations was reached without resolving structural inconsistencies. \revision{Specifically, \approach reached the iteration limit in 7.00\% of its executions on LIBERO\_10 and 6.67\% on RoboCasa Humanoid. For \approachb, the corresponding rates were substantially higher, at 33.00\% and 47.92\%, respectively. This difference suggests that \approachb has greater difficulty resolving candidate errors within the same refinement budget, consistent with its lower executability and similarity results. These iteration-limit surpassings manifested as malformed oracle compositions, incomplete task decompositions, and invalid object references, which directly reduced executability and similarity.}  Statistical tests confirmed that these differences were significant (p $< 0.05$) with large effect sizes ($\hat{A}_{12} \geq 0.90$), indicating that the improvements were concentrated in the executability of the generated fine-grained oracles rather than in their validity.

Furthermore, Baseline methods performed consistently poorly across all correctness-related metrics. Despite their efficiency advantage, validity remained extremely low (0.07--0.13 on LIBERO\_10 and near-zero on Robocasa Humanoid), and executability dropped to near zero in all cases. This reflected the absence of structured decomposition and simulator functions, resulting in fine-grained oracles that either violated simulator requirements or failed to invoke valid simulator functions. In most cases, executability failures stemmed from incorrect parameterization of simulator functions. For example, oracles that check whether an object was located on another object often used the object itself as a reference, whereas the simulator requires the object's regions for such functions. Such mismatches resulted in oracles that could not be executed. Across both benchmarks, \approach outperformed the baselines in executability and similarity, with statistically significant results and large effect sizes.  Therefore, this research question can be answered as follows:

\begin{tcolorbox}[colback=gray!10, colframe=black, title=Answer to RQ1]
\approach achieves the strongest performance in generating fine-grained oracles for \revision{VLAs}, outperforming all the baselines. Moreover, compared to \approachb, it reduces generation times across all modules, highlighting the importance of effective model selection for optimal performance.
\end{tcolorbox}

\subsection{\revision{RQ2 -- Effectiveness of Generated Fine-Grained Oracles }}

\begin{table*}[ht]
\centering
\caption{Agreement between generated fine-grained oracles and benchmark-provided symbolic \revision{end-state} oracles on the LIBERO\_10 and Robocasa Humanoid benchmarks after correcting the failures in the simulator functions. The table reports agreement for the correctness assessment metrics, including accuracy (Acc.), precision (Prec.), recall (Rec.), F1-score (F1), and confusion matrix statistics (TP, TN, FP, and FN), measuring agreement between the fine-grained oracles and the benchmark-provided symbolic \revision{end-state} oracles. Additionally, localization accuracy evaluates the generated fine-grained oracles' ability to identify the execution step where the failure occurs. We report results per task and aggregate them by benchmark and overall.}
\label{tab:RQ2_corrected}
\resizebox{0.79\textwidth}{!}{%
\begin{tabular}{llrrrrrrrrrrr}
\toprule
\multicolumn{1}{c}{} & \multicolumn{1}{c}{} &  \multicolumn{9}{c}{\textbf{Correctness Agreement}} & \multicolumn{2}{c}{\textbf{Localization}} \\ \cmidrule(lr){3-11} \cmidrule(lr){12-13}
\multicolumn{1}{c}{} & \multicolumn{1}{c}{} & \textbf{Exec \#} & \multicolumn{1}{c}{\textbf{Acc.}} & \multicolumn{1}{c}{\textbf{Prec.}} & \multicolumn{1}{c}{\textbf{Rec.}} & \multicolumn{1}{c}{\textbf{F1}} & \multicolumn{1}{c}{\textbf{TP}} & \multicolumn{1}{c}{\textbf{TN}} & \multicolumn{1}{c}{\textbf{FP}} & \multicolumn{1}{c}{\textbf{FN}} & \textbf{Exec \#} & \multicolumn{1}{c}{\textbf{Acc.}} \\
\cmidrule{1-13}
\multirow{11}{*}{\textbf{LIBERO\_10}}  & Task 1 & 90 & 1.00 & 1.00 & 1.00 & 1.00 & 54 & 36 & 0 & 0  & 10 &  1.00 \\
 & Task 2 & 80 & 1.00 & 1.00 & 1.00 & 1.00 & 48 & 32 & 0 & 0 & 10 & 1.00 \\
 & Task 3 & 80 & 1.00 & 1.00 & 1.00 & 1.00 & 48 & 32 & 0 & 0 & 10 & 1.00 \\
 & Task 4 & 10 & 0.60 & 0.60 & 1.00 & 0.75 & 6 & 0 & 4 & 0 &  6 & 1.00 \\
 & Task 5 & 100 & 1.00 & 1.00 & 1.00 & 1.00 & 60 & 40 & 0 & 0 & 10  & 1.00 \\
 & Task 6 & 100 & 1.00 & 1.00 & 1.00 & 1.00 & 60 & 40 & 0 & 0 & 10 & 1.00 \\
 & Task 7 & 100 & 0.96 & 0.94 & 1.00 & 0.97 & 60 & 36 & 4 & 0 & 10 & 1.00 \\
 & Task 8 & 70 & 1.00 & 1.00 & 1.00 & 1.00 & 42 & 28 & 0 & 0 & 10 & 1.00 \\
 & Task 9 & 80 & 1.00 & 1.00 & 1.00 & 1.00 & 48 & 32 & 0 & 0 &  10 & 1.00 \\
 & Task 10 & 80 & 1.00 & 1.00 & 1.00 & 1.00 & 48 & 32 & 0 & 0 & 10 & 1.00 \\

 \cmidrule{2-13}
  \rowcolor{gray!30}   \cellcolor{white} & \textbf{Total} & 790 & 0.98 & 0.98 & 0.98 & 0.98 & 474 & 308 & 8 & 0 & 96 & 1.00 \\ \cmidrule{1-13}
\multirow{25}{*}{\begin{tabular}[c]{@{}l@{}}\textbf{Robocasa}\\ \textbf{Humanoid}\end{tabular}}   &  Task 1 & 100 & 1.00 & 1.00 & 1.00 & 1.00 & 60 & 40 & 0 & 0 &  10 & 1.00 \\
  & Task 2 & 100 & 1.00 & 1.00 & 1.00 & 1.00 & 60 & 40 & 0 & 0 &  10 & 0.80 \\
  & Task 3 & 94 & 1.00 & 1.00 & 1.00 & 1.00 & 60 & 34 & 0 & 0 &  10 & 0.90 \\
  & Task 4 & 81 & 0.93 & 0.91 & 1.00 & 0.95 & 60 & 15 & 6 & 0 &  10 & 0.70 \\
  & Task 5 & 93 & 0.94 & 0.97 & 0.93 & 0.95 & 56 & 31 & 2 & 4 & 10 & 1.00\\
  & Task 6 & 97 & 0.82 & 0.78 & 1.00 & 0.88 & 60 & 20 & 17 & 0 &  10 & 0.70 \\
  & Task 7 & 63 & 0.89 & 0.86 & 1.00 & 0.92 & 42 & 14 & 7 & 0 &  10 & 0.70 \\
  & Task 8 & 44 & 1.00 & 1.00 & 1.00 & 1.00 & 31 & 13 & 0 & 0 &  10 & 0.70 \\
  & Task 9 & 90 & 0.91 & 0.96 & 0.89 & 0.92 & 48 & 34 & 2 & 6 &  10 & 0.80 \\
  & Task 10 & 100 & 0.90 & 0.86 & 1.00 & 0.92 & 60 & 30 & 10 & 0 &  10 & 0.70 \\
  & Task 11 & 47 & 0.94 & 1.00 & 0.91 & 0.95 & 31 & 13 & 0 & 3 &  10 & 0.60 \\
  & Task 12 & 90 & 0.80 & 0.75 & 1.00 & 0.86 & 54 & 18 & 18 & 0 &  10 & 0.70 \\
  & Task 13 & 56 & 1.00 & 1.00 & 1.00 & 1.00 & 48 & 8 & 0 & 0 &  10 & 0.70 \\
  & Task 14 & 90 & 0.89 & 0.86 & 1.00 & 0.92 & 60 & 20 & 10 & 0 &  10 & 0.70 \\
  & Task 15 & 91 & 0.88 & 1.00 & 0.82 & 0.90 & 49 & 31 & 0 & 11 &  10 & 0.50 \\
  & Task 16 & 72 & 1.00 & 1.00 & 1.00 & 1.00 & 40 & 32 & 0 & 0 &  10 & 0.70 \\
  & Task 17 & 80 & 0.70 & 0.67 & 1.00 & 0.80 & 48 & 8 & 24 & 0 &  10 & 0.60 \\
  & Task 18 & 97 & 0.69 & 0.70 & 0.82 & 0.76 & 47 & 20 & 20 & 10 &  10 & 0.40 \\
  & Task 19 & 98 & 0.50 & 0.91 & 0.17 & 0.29 & 10 & 39 & 1 & 48 &  10 & 0.70 \\
  & Task 20 & 100 & 0.77 & 0.72 & 1.00 & 0.84 & 60 & 17 & 23 & 0 &  10 & 0.60 \\
  & Task 21 & 100 & 1.00 & 1.00 & 1.00 & 1.00 & 60 & 40 & 0 & 0 &  10 & 0.60 \\
  & Task 22 & 90 & 0.76 & 0.74 & 0.93 & 0.82 & 50 & 18 & 18 & 4 &  10 & 0.60 \\
  & Task 23 & 62 & 1.00 & 1.00 & 1.00 & 1.00 & 51 & 11 & 0 & 0 &  10 &  0.60 \\
  & Task 24 & 89 & 0.88 & 1.00 & 0.81 & 0.90 & 48 & 30 & 0 & 11 &  10 & 0.90 \\
 \cmidrule{2-13}
  \rowcolor{gray!30}  \cellcolor{white} & \textbf{Total} & 2024 & 0.87 & 0.88 & 0.92 & 0.90 & 1193 & 576 & 158 & 97 & 240 & 0.70 \\ \cmidrule{1-13}
\rowcolor{gray!30}   \textbf{Total} & & 2814 & 0.91 & 0.91 & 0.95 & 0.93 & 1667 & 884 & 166 & 97 &  336 & 0.78 \\ \bottomrule
\end{tabular}%
}
\end{table*}

\revision{In this RQ we evaluate two complementary properties of the generated fine-grained oracles. In RQ2.1 we measure the agreement between their success/failure verdicts and those of the benchmark-provided symbolic end-state oracles, while in RQ2.2 we evaluate their additional ability to identify the first task step at which an execution fails.}

\revision{During our preliminary evaluation of RQ2.2, we identified inaccuracies in two simulator functions used by the generated oracles. We corrected these functions and repeated all experiments associated with RQ2. These corrections modified the simulators used on each benchmark; however, they did not modify \approach. Consequently, both the agreement and localization results reported in Table~\ref{tab:RQ2_corrected} correspond to executions obtained after these corrections, whereas the results before corrections are reported in Appendix~\ref{sec:prefix}.}

\subsubsection{\revision{RQ2.1 -- Agreement with Symbolic End-State Oracles}}

\revision{In this sub-RQ, we evaluate agreement over all 2,814 task executions. We treat failed executions as the positive class and successful executions as the negative class. The benchmark-provided symbolic end-state oracles are used as the comparison reference because they were manually defined by the benchmark developers; thus, we assume they are largely correct. Therefore, the reported metrics quantify agreement between the two oracle types. In the results provided in Table~\ref{tab:RQ2_corrected}, a true positive corresponds to an execution classified as a failure by both oracle types, whereas a true negative corresponds to an execution classified as successful by both. A false positive occurs when the fine-grained oracle reports a failure and the symbolic end-state oracle reports success. Conversely, a false negative occurs when the fine-grained oracle reports success and the symbolic end-state oracle reports a failure. Accuracy measures the overall proportion of matching verdicts. Precision, recall, and F1-score provide complementary directional measures of agreement by treating the symbolic end-state oracle as the reference.}

Overall, the generated fine-grained oracles achieved an accuracy of 0.91, a precision of 0.91, a recall of 0.95, and an F1-score of 0.93. These results indicate a high degree of agreement \revision{between the two oracle types} and demonstrate that \approach can reliably \revision{reproduce the success and failure verdicts of the symbolic end-state oracles}. \revision{The high recall further indicates that the fine-grained oracles identify most of the executions classified as failures by the symbolic end-state oracles, while the high F1-score reflects strong agreement in failure classification. This level of agreement supports using generated fine-grained oracles as substitutes for symbolic end-state oracles.} The results differed slightly across benchmarks. \revision{Robocasa Humanoid exhibited a lower overall accuracy (0.87) than LIBERO\_10 (0.98)}, suggesting that \revision{the agreement between both oracle types is partially influenced by the quality and reliability of the simulator}. To better understand the observed disagreements, we manually inspected all \revision{263 disagreement cases, i.e.,} false-positive and false-negative cases.

Our manual analysis \revision{of False Positives and False Negatives, i.e., disagreements,} revealed that most discrepancies do not originate from limitations of \approach itself. Instead, they primarily stem from inaccuracies in the simulator-level functions the generated oracles use to assess task completion. Because the generated oracles rely on simulator-provided state-inspection functions, their limitations directly affect the resulting \revision{agreement} assessment. A detailed inspection of the disagreements revealed several distinct error sources. In LIBERO\_10, the few remaining false positives, \revision{8/263 total disagreements}, are associated with Task 7 and Task 4 and originate from \revision{ errors introduced when translating the instruction into simulator-level oracles}. In these cases, the generated oracle used a function to check whether one door was closed, but it incorrectly specified the object that was expected to be closed. Despite these minor discrepancies, in this benchmark \approach achieved an overall agreement accuracy of 98\%, demonstrating \revision{that the generated fine-grained oracles closely reproduced the verdicts of the benchmark-provided symbolic end-state verdicts.}.

However, the analysis of Robocasa Humanoid indicated a more complex situation, with 255/263 disagreements. \revision{In 97 out of the 255 disagreements of this benchmark, where the generated fine-grained oracle classified an execution as successful while the symbolic end-state oracle classified it as a failure, i.e., false negatives, were not caused by errors introduced when translating the instruction into simulator-level oracles} but rather by inconsistencies in the simulator physics and the benchmark symbolic \revision{end-state} oracles. For example, in Task 19, which accounted for 48 false negatives, the manipulated object must be placed inside a container. During execution, the object occasionally bounced out of the container due to physics instabilities (i.e., in the real world, this would not happen) even after satisfying the intended placement condition. Consequently, the symbolic \revision{end-state} oracle classified the execution as a failure, whereas the fine-grained oracle identified that the task objective was successfully achieved and therefore classified it as a success. Similar situations occurred in \revision{the remaining 49 false negatives from Tasks 5, 9, 11, 15, 18, 22, and 24, where the atomic} tasks involved object grasping; several objects occasionally become attached to, or remain unintentionally attached to, the robot hand. \revision{These cases emphasize that disagreement between generated and symbolic end-state oracles may result from simulator physics and the reliability of the corresponding state-inspection functions, rather than from errors in the generated oracle itself.}

With respect to the \revision{158 cases of} false positives observed in Robocasa, we divide them into three categories. The first category corresponds to \revision{the 85 cases from} Tasks 17, 18, 20, and 22, in which the generated fine-grained oracles employed functions whose parameters were not properly instantiated. These cases represent genuine generation errors and highlight the importance of validating generated oracles before deployment. The second category corresponds to \revision{the 48 cases from} Tasks 4, 5, and 6, where manual inspection revealed situations that the benchmark symbolic \revision{end-state} oracles fail to capture. For instance, during a cabinet storage task, a bottle was incorrectly positioned outside the cabinet while the door remained open. When the cabinet door was closed, the bottle was pushed inside, causing the symbolic \revision{end-state} oracle to classify the execution as successful despite the task not being performed correctly. The generated fine-grained oracles, by explicitly checking intermediate task conditions, correctly identified the failure. \revision{These examples illustrate that fine-grained oracles can identify procedural failures that are not captured by symbolic end-state oracles based primarily on the final task state. Finally, the 38 disagreements in Tasks 10, 12, and 14 were associated with simulator failures similar to those observed among the false negatives. In these cases, simulator physics affected the task outcomes and consequently led to disagreements between the two oracle types. }

\revision{Overall, these} results demonstrated that both simulator fidelity and function expressiveness play a critical role. The LIBERO\_10 benchmark, whose functions closely reflect the underlying simulator state, achieved near-perfect agreement between symbolic end-state and fine-grained oracles, with 98\% agreement accuracy. In contrast, Robocasa included several tasks affected by the simulator's physics and imperfect state functions, leading to lower agreement scores, though the generated oracles still achieved 87\% agreement accuracy. These findings reveal an important practical consideration for adopting fine-grained oracles. \revision{Their agreement with symbolic end-state oracles, and therefore their failure detection capability, depend not only on the quality of the generated oracles, but also on the reliability of the underlying simulator functions used to observe task state.} Furthermore, during manual inspection, we observed occasional inconsistencies caused by the stochastic nature of the simulation environments. Even when replaying demonstrations from recorded datasets, small variations in object interactions occasionally produced different outcomes, highlighting the challenges of obtaining perfectly deterministic evaluations in embodied AI benchmarks.

Finally, \revision{the results indicate that generated fine-grained oracles can potentially replace manually defined, symbolic end-state oracles, since they can reproduce their success and failure decisions with high agreement. However,} human verification remains valuable when deploying automatically generated fine-grained oracles, particularly in benchmarks that rely on incomplete or noisy simulator functions. Nevertheless, as simulator fidelity improves and richer state representations become available, the need for manual validation is expected to diminish. Future work will explore generating fine-grained oracles for more accurate simulation environments and real-world robotic systems, including constructing new benchmarks designed to support detailed execution monitoring and failure localization.

\begin{tcolorbox}[colback=gray!10, colframe=black, title=Answer to RQ2.1]
\revision{The generated fine-grained oracles closely agreed with the benchmark-provided symbolic end-state oracles, achieving 91\% agreement accuracy and an F1-score of 0.93 across 2,814 executions. Manual inspection showed that the remaining disagreements arose from a combination of oracle-generation errors, simulator limitations, and procedural failures that the symbolic end-state oracles did not capture. These results support using generated fine-grained oracles as an automated alternative to manually defined, benchmark-provided symbolic end-state oracles.}
\end{tcolorbox}

\subsubsection{\revision{RQ2.2 -- Failure Localization}}
\revision{In RQ2.2 we evaluate whether the generated fine-grained oracles correctly identify the first task step at which an execution fails. This analysis is performed on true-positive executions, i.e., executions classified as failures by both the fine-grained and symbolic end-state oracles. For each selected execution, we manually inspected its video, identified the first task step that deviated from the expected behavior, and compared this annotation with the location reported by the fine-grained oracle. Therefore, the reported localization accuracy in Table~\ref{tab:RQ2_corrected} is the proportion of inspected executions for which both locations coincide. Since manually inspecting all eligible executions would be prohibitively expensive, we determined a target sample of 340 executions using Cochran's sample-size formula~\cite{cochran1977sampling}, with a 95\% confidence level and a 5\% margin of error. We then applied stratified random sampling by selecting up to ten true-positive executions per task. Only six eligible executions were available for Task 4 of LIBERO\_10; therefore, the final sample contained 336 executions, 96 from LIBERO\_10 and 240 from Robocasa Humanoid.

In the preliminary localization analysis, the generated oracles achieved an average accuracy of 0.50. Manual inspection showed that this result was affected by inaccuracies in two simulator functions. In LIBERO\_10, the function for detecting whether an object had been lifted used a height threshold relative to the ground. In some scenarios, this threshold was below the table surface, causing objects resting on the table to be classified as lifted. We corrected the function by defining the threshold relative to the table. In Robocasa Humanoid, we found an issue in the proximity function that measures the distance between an object and the robot. We therefore updated the distance calculation to use the hand's center correctly. After correcting these benchmark functions, we repeated the complete RQ2 evaluation. The post-correction localization results are reported in Table~\ref {tab:RQ2_corrected}, while the original results are provided in Section~\ref{sec:prefix}.

After correcting the two simulator functions aforementioned, the generated fine-grained oracles achieved an overall localization accuracy of 0.78 across the 336 manually inspected failing executions. The results differed considerably between benchmarks. For LIBERO\_10, the oracles achieved an accuracy of 1.00 across 96 executions, correctly identifying the first failed atomic task in every inspected case. This result indicates that the generated task decompositions and the available simulator functions provided sufficient information to distinguish the relevant failure locations for the evaluated LIBERO\_10 tasks.

In contrast, for Robocasa Humanoid, the generated oracles achieved 0.70 localization accuracy across 240 executions. In this benchmark, many localization disagreements occurred in tasks involving grasping and releasing objects, for which the simulator functions provide only partial information about the corresponding manipulation events. For example, objects may remain unintentionally attached to the robot hand or become attached to other objects. In these situations, the observed simulator state did not always match the video used to manually analyze the task, making it difficult to distinguish failures in adjacent atomic tasks. As with agreement, localization also suffered from variations in simulated object interactions. Even when executions were replayed from recorded demonstrations, small differences in contacts and object positions occasionally produced different state transitions. Because fine-grained oracles determine failure locations by evaluating simulator functions throughout the task sequence, limited or unstable observations can alter the step identified as the first failure. The localization results therefore depend on both the quality of the generated task decomposition and the fidelity of the simulator functions used to observe each atomic task.

\begin{tcolorbox}[colback=gray!10, colframe=black, title=Answer to RQ2.2]
\revision{The generated fine-grained oracles localized the first failed task step with 78\% accuracy across 336 manually inspected executions. Differences between benchmarks mainly stemmed from the availability and reliability of simulator functions for observing manipulation events.}
\end{tcolorbox}}

\subsection{RQ3 - Ablation Study}

\begin{table*}[ht]
    \centering
    \caption{Ablation study of \approach on the LIBERO\_10 and Robocasa Humanoid benchmarks. The table reports generation time, validity, executability, and similarity metrics for \approach and its variants obtained by removing individual modules or the execution reliability mechanism. Module~I, Module~II, and Module~III correspond to the three modules of \approach. Reported values correspond to the mean ($m$) and standard deviation ($\sigma$). For each benchmark and metric, the best result is highlighted in \textbf{bold} and \underline{underlined}. Arrows indicate whether lower ($\downarrow$) or higher ($\uparrow$) values are preferable.}
    \label{tab:RQ3}
    \resizebox{0.9\textwidth}{!}{
        \begin{tabular}{llrrrrrrrrrrrr}
        \toprule
         &  & \multicolumn{6}{c}{\textbf{Generation Time (s) $\downarrow$}} & \multicolumn{2}{c}{\textbf{Validity $\uparrow$}} & \multicolumn{2}{c}{\textbf{Executability $\uparrow$}} & \multicolumn{2}{c}{\textbf{Similarity $\uparrow$}} \\ \cmidrule(lr){3-8} \cmidrule(lr){9-10} \cmidrule(lr){11-12} \cmidrule(lr){13-14}
         &  & \multicolumn{2}{c}{Module I} & \multicolumn{2}{c}{Module II} & \multicolumn{2}{c}{Module III} & \multicolumn{2}{c}{Module III} & \multicolumn{2}{c}{Module III} & \multicolumn{2}{c}{Module III} \\ 
         \cmidrule(lr){3-4} \cmidrule(lr){5-6} \cmidrule(lr){7-8} \cmidrule(lr){9-10} \cmidrule(lr){11-12} \cmidrule(lr){13-14}
         &  & \multicolumn{1}{c}{$m$} & \multicolumn{1}{c}{$\sigma$} & \multicolumn{1}{c}{$m$} & \multicolumn{1}{c}{$\sigma$} & \multicolumn{1}{c}{$m$} & \multicolumn{1}{c}{$\sigma$} & \multicolumn{1}{c}{$m$} & \multicolumn{1}{c}{$\sigma$} & \multicolumn{1}{c}{$m$} & \multicolumn{1}{c}{$\sigma$} & \multicolumn{1}{c}{$m$} & \multicolumn{1}{c}{$\sigma$} \\

        \cmidrule{1-14}
        \multirow{4}{*}{\textbf{LIBERO\_10}} & \approach & \best{30.776} & \best{10.802} & 638.169 & 42.079 & 346.513 & 98.054 & 0.990 & 0.032 & 0.800 & 0.149 & \best{0.912} & \best{0.034} \\
         & w/o Module I & N/A & N/A & \best{46.091} & \best{23.029} & 404.093 & 78.575 & \best{1.000} & \best{0.000} & \best{0.990} & \best{0.032} & 0.687 & 0.004 \\
         & w/o Module II  & \best{30.776} & \best{10.802} & N/A & N/A & 458.874 & 119.468 & 0.990 & 0.032 & 0.360 & 0.135 & 0.857 & 0.015 \\
         & w/o Execution Reliability & \best{30.776} & \best{10.802} & 638.169 & 42.079 & \best{78.189} & \best{24.422} & 0.680 & 0.132 & 0.470 & 0.157 & 0.896 & 0.022 \\
         
        \cmidrule{1-14}
        \multirow{4}{*}{\begin{tabular}[c]{@{}l@{}}\textbf{Robocasa} \\ \textbf{Humanoid}\end{tabular}} & \approach & \best{21.138} & \best{4.315} & 222.379 & 15.662 & 350.094 & 68.636 & \best{1.000} & \best{0.000} & \best{0.933} & \best{0.045} & \best{0.947}  &\best{0.014} \\
         & w/o Module I & N/A & N/A & \best{114.368} & \best{27.950} & 404.635 & 89.887 & 0.996 & 0.013 & 0.221 & 0.040 & 0.800 & 0.001  \\
         & w/o Module II & \best{21.138} & \best{4.315} & N/A & N/A  & 453.148 & 98.833 & 0.996 & 0.013 & 0.013 & 0.013 & 0.779 & 0.004 \\
         & w/o Execution Reliability & \best{21.138} & \best{4.315} & 222.379 & 15.662 & \best{52.240} & \best{4.861} & 0.796 & 0.115 & 0.037 & 0.031 & 0.861 & 0.013 \\
         
        \bottomrule
        \end{tabular}
    }
    
\end{table*}

\begin{table*}[ht]
    \centering
    \caption{Statistical comparison of \approach against \approach w/o Module I, w/o Module II, and w/o Execution Reliability on the LIBERO\_10 and Robocasa Humanoid benchmarks. We report results for generation time, validity, executability, and similarity across the evaluated modules (Module~I, Module~II, and Module~III). The table reports Vargha and Delaney’s effect size statistic ($\hat{A}_{12}$) together with the corresponding p-values for each pairwise comparison. Statistically significant differences in favour of \approach are highlighted in \colorbox{cyan!13}{blue}. Arrows indicate whether lower ($\downarrow$) or higher ($\uparrow$) values are preferable.}
    \label{tab:RQ3_statistics}
    \resizebox{0.9\textwidth}{!}{
        \begin{tabular}{llrrrrrrrrrrrr}
        \toprule
         &  & \multicolumn{6}{c}{\textbf{Generation Time (s) $\downarrow$}} & \multicolumn{2}{c}{\textbf{Validity $\uparrow$}} & \multicolumn{2}{c}{\textbf{Executability $\uparrow$}} & \multicolumn{2}{c}{\textbf{Similarity $\uparrow$}} \\ \cmidrule(lr){3-8} \cmidrule(lr){9-10} \cmidrule(lr){11-12} \cmidrule(lr){13-14}
         &  & \multicolumn{2}{c}{Module I} & \multicolumn{2}{c}{Module II} & \multicolumn{2}{c}{Module III} & \multicolumn{2}{c}{Module III} & \multicolumn{2}{c}{Module III} & \multicolumn{2}{c}{Module III} \\ 
         \cmidrule(lr){3-4} \cmidrule(lr){5-6} \cmidrule(lr){7-8} \cmidrule(lr){9-10} \cmidrule(lr){11-12} \cmidrule(lr){13-14}
         &  & \multicolumn{1}{c}{ \^{A}$_{12}$} & \multicolumn{1}{c}{p-value} & \multicolumn{1}{c}{ \^{A}$_{12}$} & \multicolumn{1}{c}{p-value} & \multicolumn{1}{c}{ \^{A}$_{12}$} & \multicolumn{1}{c}{p-value} & \multicolumn{1}{c}{ \^{A}$_{12}$} & \multicolumn{1}{c}{p-value} & \multicolumn{1}{c}{ \^{A}$_{12}$} & \multicolumn{1}{c}{p-value} & \multicolumn{1}{c}{ \^{A}$_{12}$} & \multicolumn{1}{c}{p-value} \\
        \cmidrule{1-14}
        
        \multirow{3}{*}{\textbf{LIBERO\_10}} & w/o Module I & N/A & N/A & 1.00 & \textless{}0.0001 & 0.29 & 0.1212 & 0.45 & 0.3681 & 0.02 & 0.0001 & \cellcolor{cyan!13}1.00 & \cellcolor{cyan!13}0.0002 \\
         & w/o Module II  & 0.50 & 1.0000 & N/A & N/A & 0.24 & 0.0539 & 0.50 & 1.0000 & \cellcolor{cyan!13}0.97 & \cellcolor{cyan!13}0.0004 & \cellcolor{cyan!13}0.90 & \cellcolor{cyan!13}0.0028 \\
         & w/o Execution Reliability & 0.50 & 1.0000 & 0.50 & 1.0000 & 1.00 & \textless{}0.0001 & \cellcolor{cyan!13}0.99 & \cellcolor{cyan!13}\textless{}0.0001 & \cellcolor{cyan!13}0.93 & \cellcolor{cyan!13}0.0009 & \cellcolor{cyan!13}0.88 & \cellcolor{cyan!13}0.0046 \\
         
        \cmidrule{1-14}
        \multirow{3}{*}{\begin{tabular}[c]{@{}l@{}}\textbf{Robocasa} \\ \textbf{Humanoid}\end{tabular}} & w/o Module I & N/A & N/A & 1.00 & \textless{}0.0001 & 0.36 & 0.5436 & 0.55 & 0.3681 & \cellcolor{cyan!13}1.00 & \cellcolor{cyan!13}\textless{}0.0001 & \cellcolor{cyan!13}1.00 & \cellcolor{cyan!13}0.0002 \\
         & w/o Module II & 0.50 & 1.0000 & N/A & N/A & \cellcolor{cyan!13} 0.17 & \cellcolor{cyan!13} 0.0140 & 0.55 & 0.3681 & \cellcolor{cyan!13}1.00 & \cellcolor{cyan!13}\textless{}0.0001 & \cellcolor{cyan!13}1.00 & \cellcolor{cyan!13}0.0002 \\
         & w/o Execution Reliability & 0.50 & 1.0000 & 0.50 & 1.0000 & 1.00 & \textless{}0.0001 & \cellcolor{cyan!13}1.00 & \cellcolor{cyan!13}0.0003 & \cellcolor{cyan!13}1.00 & \cellcolor{cyan!13}0.0001 & \cellcolor{cyan!13}1.00 & \cellcolor{cyan!13}0.0002 \\
         
        \bottomrule
        \end{tabular}
    }
    
\end{table*}

The first ablation evaluates the contribution of \textit{Module I (Atomic Task Library Generator)}. Removing this module reduced the generation time associated with Module II, as shown in Table~\ref{tab:RQ3}. In LIBERO\_10, the average generation time decreased by approximately 13.8$\times$ (638.17s vs 46.09s), while in Robocasa Humanoid it decreased by approximately 1.9$\times$ (222.38s vs 114.37s), with statistically significant differences in both cases ($\hat{A}_{12}=1.00$, $p<0.0001$, Table~\ref{tab:RQ3_statistics}). This reduction was expected because the system, instead of taking each atomic task and generating its corresponding atomic task oracle, attempted to generate the atomic task oracle library directly from the initial set of complex tasks. However, this reduction in generation time was followed by an increase in the generation time of Module~III, whose average generation time increased from 346.51s to 404.09s in LIBERO\_10 and from 350.09s to 404.64s in Robocasa Humanoid, although the differences were not statistically significant in any of the benchmarks. This increase in generation time suggests that Module~I's atomic task library simplifies subsequent fine-grained oracle generation by providing a structured atomic task representation that aligns with complex-task decomposition. Without this atomic task library, the generator must reason simultaneously about task decomposition and the corresponding executable oracles, producing lower-quality fine-grained oracles that require additional refinement iterations to converge.

Moreover, the impact of removing Module~I became even more evident when analyzing oracle correctness. Although validity remained almost unchanged, executability and similarity revealed important differences. In Robocasa Humanoid, executability dropped from 0.933 to 0.221, while similarity decreased from 0.947 to 0.800, with statistically significant differences favoring \approach. However, in LIBERO\_10, executability unexpectedly increased from 0.800 to 0.990, while similarity decreased from 0.912 to 0.687. At first glance, this increase in executability might suggest improved oracle quality; however, it is a false positive. Executability only measures whether an oracle can be executed successfully, not whether it faithfully captures the intended fine-grained task structure. By analyzing these contradictory results, we observed that most generated fine-grained oracles omitted several intermediate actions and focused primarily on verifying the final task outcome. Such oracles are easier to implement correctly because they primarily check the final state, inflating executability scores without reflecting true task decomposition. As a result, the generated fine-grained oracles remained executable but failed the core goal of being fine-grained oracles. Thus, the generated oracles resembled the symbolic \revision{end-state} oracle formulations commonly used in previous approaches, which assess only the final state. This indicates that without decomposing into atomic tasks, generating fine-grained oracles becomes significantly more challenging. Furthermore, in the Robocasa Humanoid benchmark, although the task instructions are more fine-grained than those in LIBERO\_10 and reduce planning complexity, the simulator functions and required parameterization are more complex, resulting in a significant drop in both executability and similarity. Overall, these results show that Module~I is essential for producing the atomic task library that guides the generation process toward fine-grained oracles that not only execute successfully but also accurately capture the intended sequence of actions for the complex task.

The second ablation evaluates the contribution of \textit{Module II (Atomic Task Oracle Library Generator)}. Since the same atomic task libraries generated by Module I on \approach were reused, no differences were observed in the generation time of Module I. However, as presented in Table~\ref{tab:RQ3}, removing the atomic task oracle library significantly affected the generation time of the fine-grained oracles. The average generation time of Module III increased 1.3$\times$ in LIBERO\_10 and in Robocasa Humanoid, indicating that it required more refinement iterations to converge. This increase can be explained by the loss of task-specific oracles associated with each atomic task. In \approach, these atomic task oracles provided concrete guidance on evaluating atomic tasks, thereby reducing the effort required for fine-grained oracle generation. Without this information, the generator had to infer each atomic task oracle's implementation details from scratch, making generation more complex.

This effect also showed up in the effectiveness metrics. While validity remained unchanged in both benchmarks, executability decreased from 0.800 to 0.360 in LIBERO\_10 and from 0.933 to 0.013 in Robocasa Humanoid, with statistically significant differences in favor of \approach in both benchmarks, as Table~\ref{tab:RQ3_statistics} reflects. Similarity showed the same effect, decreasing from 0.912 to 0.857 in LIBERO\_10 and from 0.947 to 0.779 in Robocasa Humanoid. The greater degradation observed in executability than in validity suggests that syntactically valid fine-grained oracles were relatively easy to generate, whereas generating fine-grained oracles that executed correctly in the simulator was more challenging. For instance, several of the fine-grained oracles generated in this ablation were valid but used inappropriate function parameters or functions not available in the simulator. This was because, in many cases, Module III struggled to converge on a correct solution despite multiple refinement attempts and ultimately reached the maximum number of refinement iterations. These findings demonstrated that Module II provided critical implementation-level information that bridges the gap between high-level task decomposition and executable fine-grained oracles in the simulator.

The third and final ablation evaluates the contribution of the \textit{Execution Reliability mechanism} in Module III, which determines whether the fine-grained oracle candidate is executable in the simulator. As expected, removing this component did not affect the generation time of Modules I and II, because these modules operate before Module III. However, as presented in Table~\ref{tab:RQ3}, the impact of this execution reliability mechanism on Module III was substantial. The average generation time decreased by approximately 4.4$\times$ in LIBERO\_10 and 6.7$\times$ in the Robocasa Humanoid benchmark, with statistically significant differences as shown in Table~\ref{tab:RQ3_statistics}. While this reduction might suggest an efficiency improvement, it occurred because the fine-grained oracle judge no longer verified whether the generated fine-grained oracle candidate could be executed within the simulator. Consequently, fine-grained oracle candidates were accepted much earlier in the refinement process, leading to faster convergence but significantly lower oracle validity, executability and similarity. This effect was reflected in all three metrics, as reported in Table~\ref{tab:RQ3}.

Validity decreased from 0.990 to 0.780 in LIBERO\_10 and from 1.000 to 0.796 in Robocasa Humanoid, while executability dropped from 0.8000 to 0.470 and from 0.933 to 0.037, respectively. In addition, similarity also decreased in both benchmarks. All these differences were statistically significant. These drops across all metrics stemmed from subtle parameterization issues in many fine-grained oracles, such as referencing the wrong objects or using nonexistent functions in the simulator. Although these errors often produced syntactically valid fine-grained oracles, they prevented successful execution within the simulator. The results therefore showed that although the execution reliability mechanism incurs a generation-time overhead, it also provides information that structural or reasoning analyses cannot obtain, enabling the system to identify and correct errors that would otherwise remain hidden. Therefore, this research question can be answered as follows:

\begin{tcolorbox}[colback=gray!10, colframe=black, title=Answer to RQ3]
All components of \approach contribute significantly and complement each other. Module~I improves task decomposition and planning quality; Module~II provides the task-specific knowledge required for oracle generation; and the Execution Reliability mechanism ensures simulator-compliant execution. Removing any component reduces oracle validity, executability, and similarity, despite lower generation time.
\end{tcolorbox}

\subsection{RQ4 - Atomic Task Library Reduction}

The results in Table~\ref{tab:RQ4} show that the Atomic Task Library generated by Module I of \approach can be generated with a reduced set of complex tasks. Across both benchmarks, the optimization algorithm consistently identified a small yet effective subset of tasks that yield atomic task libraries semantically equivalent to those produced using the full task sets (i.e., an average similarity score of 1.000). Regarding the reduction of the initial set on LIBERO\_10, the initial set of 10 tasks was reduced to an average of 3.1 tasks (minimum 2, maximum 6). This corresponds to a mean TRR of 0.69, i.e., the original set of complex tasks can be reduced by 69\% without altering the generated Atomic Task Library. A similar but even stronger effect was observed on the Robocasa Humanoid, where the 24 initial tasks were reduced to an average of 3.5 tasks (minimum 2, maximum 11). This yielded a mean TRR of 0.854, corresponding to an 85.4\% reduction in the initial task set size. Despite the larger initial task set, the resulting atomic task libraries preserve semantic similarity, again achieving a perfect score of 1.0 across all runs.

\begin{table*}[ht]
    \centering
    \caption{Results for RQ4 on the LIBERO\_10 and Robocasa Humanoid benchmarks. The table reports the number of tasks before (\#Tasks) and after minimization (\#Tasks$_{\min}$), the Task Reduction Ratio (TRR), and the semantic similarity between the atomic task library generated from the minimized set of complex tasks and the one generated from the original task set. Reported values correspond to minimum (min), maximum (max), mean ($m$), and standard deviation ($\sigma$) across 10 executions. Arrows indicate whether lower ($\downarrow$) or higher ($\uparrow$) values are preferable.}
    \label{tab:RQ4}
   \begin{tabular}{lccccccccccc}
    \toprule
    & \textbf{\#Tasks} & \multicolumn{4}{c}{\textbf{\#Tasks$_{\min}$} $\downarrow$} & \multicolumn{4}{c}{\textbf{TRR} $\uparrow$} & \multicolumn{2}{c}{\textbf{Similarity} $\uparrow$} \\
    \cmidrule(lr){2-2}\cmidrule(lr){3-6}\cmidrule(lr){7-10}\cmidrule(lr){11-12}
    & & min & max & $m$ & $\sigma$ & min & max & $m$ & $\sigma$ & $m$ & $\sigma$ \\
    \midrule
    \textbf{LIBERO\_10} & 10 & 2 & 6 & 3.100 & 1.22 & 0.400 & 0.800 & 0.690 & 0.12 & 1.000 & 0.00 \\
    \textbf{Robocasa Humanoid} & 24 & 2 & 11 & 3.500 & 2.58 & 0.542 & 0.917 & 0.854 & 0.11 & 1.000 & 0.00 \\
    \bottomrule
    \end{tabular}
\end{table*}

These findings have direct implications for practitioners adopting \approach to their case studies. First, these findings suggest that users do not need to curate large and exhaustive initial task sets; instead, a small but diverse set of complex tasks, on average, 3-4 complex tasks, is sufficient to generate a highly covered atomic task library. This reduces the initial effort required to specify tasks in new case studies. Second, the results support an incremental workflow: practitioners can begin with a compact task set, generate the atomic task and atomic task oracle libraries once, and then extend the tasks in their benchmark without rerunning Modules I and II. This is particularly beneficial in industrial or evolving environments where tasks are continuously added, as it avoids repeated expensive recomputation. Finally, the observed high semantic similarity across all runs indicates that the atomic task library is robust to redundancy in the input task distribution. In practice, this implies that \approach naturally filters out overlapping or semantically redundant task specifications, focusing instead on a minimal set of representative atomic tasks. As a result, \approach improves scalability while maintaining consistency and reliability of the generated libraries and fine-grained oracles.

\begin{tcolorbox}[colback=gray!10, colframe=black, title=Answer to RQ4]
The initial complex task set can be reduced by 69\%-85\% on average while still yielding an atomic task library that is fully semantically equivalent to that generated from the full set. Practically, this shows that \approach requires only a small, diverse subset of tasks to generate the atomic task library, enabling efficient reuse and incremental extension without rerunning Modules I and II.
\end{tcolorbox}

\section{\revision{Limitations}}\label{sec:limitations}

\revision{

The evaluation shows \approach's potential to automatically generate fine-grained oracles and support failure localization. Its application also entails several practical considerations concerning simulator support, assessment granularity, and library reuse, which we discuss below.

\textit{Dependence on the simulator.}
\approach uses simulator functions to observe task progress and evaluate completion conditions. The properties we can assess therefore depend on the information these functions expose. All approaches that rely on simulator interfaces share this dependency. Inaccurate functions or unstable simulation behavior can affect oracle verdicts, as RQ2 shows. Adopting \approach in a new environment therefore requires suitable functions to observe the relevant task conditions.

\textit{Task-level granularity.}
The generated oracles assess atomic manipulation tasks and their ordering, providing more detailed feedback than final-state checks. However, their scope is task-level behavior; therefore, motion-quality properties, such as trajectory smoothness, contact forces, or collision avoidance, require additional specifications and corresponding simulator functions. Likewise, detecting repeated drops or checking that a condition remains satisfied requires monitoring that explicitly captures those behaviors. The achievable granularity thus depends jointly on the task specification and the available observations.

\textit{Reliability of generated oracles.}
The multi-agent assessment and refinement process checks candidates for structural, grounding, logical, and execution-related issues. RQ1 and RQ2 demonstrate its effectiveness, although some generated oracles still contain errors, such as incorrect object references or parameter bindings. These cases show that refinement can end before all issues are resolved, particularly when the iteration limit is reached.

\textit{Library coverage and reuse.}
The libraries generated by Modules I and II support reuse across instructions that share the represented manipulation capabilities and simulator interface. This reduces repeated generation effort, but introducing new capabilities or changing the simulator interface may require extending or regenerating the libraries. A diverse small set of initial tasks helps establish useful coverage, as investigated in RQ4.

\textit{Generation cost.}
The assessment and refinement stages introduce computational overhead through repeated foundation-model invocations. This cost is partly amortized by reusing the atomic task and oracle libraries across instructions. The current workflow suits generating oracles during test preparation. Therefore, applications requiring immediate oracle generation for newly arriving instructions would need further optimization to improve execution time.

}


\section{Threats to Validity}\label{sec:threats_to_validity}
We considered potential threats to our study's validity and adopted several measures to mitigate their impact.

A threat to \textbf{\textit{internal validity}} arises from the stochastic nature of the foundation models used by \approach to generate fine-grained oracles. Because these models may produce different outputs across executions, observed performance differences between approaches could be partly attributable to sampling variability rather than the approaches themselves. To mitigate this threat, we executed each experimental configuration 10 times and reported statistical significance tests along with effect-size estimates, thereby increasing confidence that the observed differences are not due to chance.

A second \textbf{\textit{internal validity}} threat concerns the choice of foundation models used in \approach and its baselines in RQ1, RQ3, and RQ4. Since the quality of generated outputs may depend on the underlying model's capabilities, conclusions drawn from a single model could be biased. To reduce this risk, we conducted a preliminary model-selection study involving both proprietary and open-source foundation models and evaluated \approach under multiple model configurations.

A threat to \textbf{\textit{external validity}} concerns the generalizability of our findings beyond the evaluated benchmarks. Although LIBERO\_10 and RoboCasa Humanoid Tabletop encompass diverse manipulation tasks, environments, and failure scenarios, both benchmarks primarily target household manipulation scenarios. As a result, the effectiveness of \approach may differ in other domains, such as industrial robotics or real-world deployments, where sensing noise, hardware variability, and environmental uncertainty are more pronounced. To improve external validity, we evaluated \approach on two substantially different benchmark suites featuring distinct task structures, oracle implementations, and robotic platforms. Nevertheless, further studies involving additional simulators, robotic platforms, and physical robots are required to assess the broader applicability of our findings. Importantly, the selected benchmarks are among the most widely used for evaluating VLA-enabled robots~\cite{nvidia2025gr00tn1openfoundation, qu2025eo1, qu2025spatialvla, kim2024openvla, huang2026thinkact}.

Finally, a threat to \textbf{\textit{conclusion validity}} concerns how we assess generated fine-grained oracles and their effectiveness in detecting faults. Our evaluation relies on benchmark-provided symbolic \revision{end-state} oracles as ground truth for task-level correctness and on manually annotated failure locations as ground truth for fine-grained localization in RQ2. While the benchmark authors developed and validated the symbolic \revision{end-state} oracles, they may not fully capture all aspects of task semantics, and manual annotations may introduce subjectivity. To mitigate these risks, we annotated failure locations according to a predefined protocol based on the first task step in which execution deviated from expected behavior. Furthermore, we assess the quality of generated fine-grained oracles by comparing them against manually generated ground-truth oracles and evaluating multiple complementary properties, including structural validity and executability. Using this diverse set of metrics reduces reliance on any single measure and provides a more comprehensive assessment of oracle quality and effectiveness. Finally, statistical analyses support all reported findings, reducing the likelihood that random variation explains the observed effects.

\section{Related Work}\label{sec:related_work}
\subsection{Test Oracle Generation}
Automated test oracle generation has been extensively studied in software testing, particularly to address the oracle problem~\cite{barr2014oracle, pezze2014automated}. Early approaches focused on generating oracles from specifications, invariants, assertions, and documentation artifacts~\cite{goffi2016automatic,terragni2020evolutionary}, while later work explored oracle generation for cyber-physical systems (CPSs) using formal requirements and temporal logic models~\cite{menghi2019generating,valle2025defining}. Although effective within their target domains, these techniques generally assume access to source code, formal specifications, or design models, which are typically unavailable for instruction-driven robotic tasks specified solely in natural language.

The emergence of large language models (LLMs) has led to significant advances in specification-based oracle generation. TOGA~\cite{dinella2022toga} pioneered the use of transformer-based models for generating test assertions from program context, demonstrating the feasibility of neural oracle generation. Subsequent work explored both prompt engineering and fine-tuning strategies, including LLMEmpirical~\cite{siddiq2024using}, TOGLL~\cite{hossain2025togll}, and RetriGen~\cite{zhang2025improving}, demonstrating that contextual information, retrieval augmentation, and fine-tuning can substantially improve oracle correctness and fault-detection capabilities. Other approaches, such as CHATASSERT~\cite{hayet2024chatassert} and AugmenTest~\cite{khandaker2025augmentest}, further demonstrated the benefits of incorporating execution feedback and structured contextual information during oracle generation. More recently, researchers have investigated multi-agent LLM architectures for oracle generation. CANDOR~\cite{xu2026hallucination} introduces a dual-LLM, multi-agent framework in which specialized agents collaboratively generate and assess candidate oracles through panel-based deliberation, reducing hallucinations and improving oracle correctness. Similarly, Nexus~\cite{huang2025nexus} combines multi-agent reasoning with execution-grounded validation and refinement. MASTOR~\cite{deng2026mastor} leverages a role-based multi-agent workflow in which specialized agents collaboratively generate test prefixes, infer expected outcomes, and validate candidate assertions through iterative deliberation and consensus. These studies show that consensus-based reasoning and execution feedback effectively improve oracle quality. \approach is inspired by these advances but differs in several important aspects: (i) it targets \revision{VLAs} rather than software unit testing; (ii) it generates simulation-executable fine-grained oracles instead of assertion-based JUnit oracles; (iii) it operates directly from natural-language task instructions without requiring source code, test prefixes, or formal specifications; and (iv) it employs specialized agents with robotic knowledge and execution-based validation to generate reusable atomic-task-level oracles.

\subsection{\revision{Testing Vision-Language-Action Models}}
Recent work has begun to address the testing and evaluation of Vision-Language-Action (VLA)-enabled robots, primarily through benchmark-driven evaluation frameworks. Existing benchmarks such as LIBERO~\cite{liu2023libero}, VLATest~\cite{wang2025vlatest}, VLABench~\cite{zhang2025vlabench}, RoboCasa Humanoid Tabletop~\cite{nvidia2025gr00tn1openfoundation}, and NEBULA~\cite{peng2025nebula} evaluate \revision{VLAs} across diverse manipulation tasks and environments using manually constructed symbolic \revision{end-state} task oracles. These benchmarks typically reduce evaluation to binary end-state success conditions and report aggregate success rates as the primary metric. While effective for large-scale comparison, this paradigm is fundamentally limited for \revision{VLAs}, whose behavior depends on the coupling of perception, language grounding, and the execution of sequential actions. Conventional robotics metrics such as task success rate, path efficiency, or grasp reliability~\cite{bohg2013data,mahler2017dex} similarly fail to capture this coupling, while single-modal benchmarks from vision and language domains~\cite{deng2009imagenet,lin2014microsoft,papineni2002bleu,lin2004rouge} do not reflect the integrated, cross-modal reasoning required in \revision{VLA} settings. As a result, current evaluation protocols primarily measure outcomes rather than task execution, overlooking how intermediate decisions contribute to final behavior.

A growing body of work highlighted the limitations of end-state evaluation in \revision{VLAs}. Valle et al.~\cite{valle2025evaluating} showed that such systems may vary substantially in execution quality despite producing identical outcomes. In particular, they report cases in which tasks are completed successfully even under suboptimal or erroneous intermediate conditions, such as collisions or object drops, highlighting the need to assess intermediate execution steps rather than relying solely on final-state evaluation. Complementary studies further demonstrated that strong benchmark performance often fails to translate into robust real-world behavior under distribution shifts or minor perturbations~\cite{zhou2025exploring,fang2025intention,li2025task}, exposing a persistent gap between reported capability and practical reliability. However, existing \revision{VLA} evaluation frameworks still provide limited support for analyzing intermediate execution, diagnosing failure modes, or quantifying step-level quality beyond final success. A central remaining obstacle is the oracle problem~\cite{barr2014oracle}: defining what constitutes correct behavior in \revision{VLAs}. Most benchmarks, including LIBERO~\cite{liu2023libero}, RLBench~\cite{james2020rlbench}, RoboCasa~\cite{nasiriany2024robocasa}, and CALVIN~\cite{mees2022calvin}, rely on binary success/failure signals. While simple and scalable, such oracles cannot capture finer-grained properties such as manipulation efficiency, safety constraints, temporal ordering, or task decomposition quality, which are essential for understanding the root causes of failures and guiding corrective actions. Efforts such as VLATest~\cite{wang2025vlatest} and LADEV~\cite{wang2024ladev} improve robustness evaluation via perturbation-based testing and invariance checks, but remain constrained to final-state evaluation and do not explicitly model intermediate execution structure.

To address these limitations, recent work has begun to shift from static benchmarking toward more dynamic and adaptive evaluation strategies. FATE-VLA~\cite{kanwal2026fate} introduces failure-aware test generation to actively detect weaknesses in V\revision{VLAs} rather than passively measuring success rates. In parallel, VISOR~\cite{saurabh2026visor} leverages vision-language models as oracles for robotic evaluation, moving beyond handcrafted symbolic \revision{end-state} oracles. Similarly, Valle et al.~\cite{valle2026metamorphic} adapted metamorphic testing to \revision{VLAs}, introducing relation-based evaluation that assesses behavioral consistency under task transformations and enables oracle-free testing. Collectively, these approaches highlight a shift toward automated test generation and automated oracles; however, they still primarily evaluate whole-task outcomes and detect failures without decomposing execution into structured intermediate steps.

In summary, existing \revision{VLA} evaluation methodologies are limited by three key factors: (i) reliance on binary end-state oracles, (ii) insufficient modeling of intermediate execution behavior, and (iii) limited automation. In contrast, \approach addresses these limitations by automatically generating fine-grained, instruction-driven, and executable intermediate-state oracles that enable structured evaluation, failure localization, and diagnostic analysis of \revision{VLAs} beyond final success rates.

\section{Conclusion}\label{sec:conclusion}
We presented \approach, a novel multi-agent framework that generates fine-grained test oracles to evaluate Vision-Language-Action (VLA)-enabled robots automatically. Unlike existing evaluation approaches that rely on symbolic end-state oracles, \approach automatically generates executable fine-grained oracles by decomposing natural-language task descriptions into reusable atomic tasks and corresponding oracles. Through the collaboration of \textit{Generator}, \textit{Assessor}, and \textit{Judge} agents, \approach automates the generation of fine-grained oracles while reducing the manual effort required for oracle generation. Evaluation on the LIBERO\_10 and RoboCasa Humanoid Tabletop benchmarks demonstrates that \approach generates executable fine-grained oracles across diverse robotic manipulation tasks, providing richer diagnostic information and more precise failure localization than symbolic \revision{end-state} oracles. Our ablation studies highlight the importance of the multi-agent collaboration mechanism and atomic-task decomposition strategy in improving oracle quality, executability, and robustness. Moreover, analysis of task-library reduction suggests that a compact set of reusable atomic tasks can support oracle generation for a broad range of robotic instructions, highlighting the scalability of \approach. In future work, we plan to investigate oracle generation for more complex long-horizon tasks and extend the framework to additional robotic environments and real-world platforms.


\section*{Acknowledgments}

Pablo Valle and Aitor Arrieta are part of the Software and Systems Engineering research group of Mondragon Unibertsitatea (IT1519-22), supported by the Department of Education, Universities and Research of the Basque Country. Pablo Valle is supported by the Pre-doctoral Program for the Training of Non-Doctoral Research Staff (Grant No. PRE\_2025\_2\_0252) of the Department of Education of the Basque Government, as well as by the EGONLABUR 2026 program (Grant No. EP\_2026\_1\_0080), funded by the Department of Science, Universities and Innovation of the Basque Government. Aitor Arrieta is supported by the Spanish Ministry of Science, Innovation and Universities (project PID2023-152979OA-I00), funded by MCIU /AEI /10.13039/501100011033 / FEDER, UE. This work was partially funded by the Basque Government through their Elkartek program (ref. KK-2025/00102). Shaukat Ali is supported by the Co-tester project (No. 314544) funded by the Research Council of Norway and the FRAME project (Grant Agreement No. 101298951) funded by the European Commission’s Horizon Europe programme. Lionel Briand is supported by Taighde Éireann – Research Ireland under Grant number 13/RC/2094\_2 and Canada's NSERC Discovery Grant and Canada Research Chair programs. 

\bibliographystyle{ieeetr}
\bibliography{bibliography}

@inproceedings{jain2001termination,
  title={On termination criteria of evolutionary algorithms},
  author={Jain, Brijnesh J and Pohlheim, Hartmut and Wegener, Joachim},
  booktitle={Proceedings of the 3rd Annual Conference on Genetic and Evolutionary Computation},
  pages={768--768},
  year={2001}
}

@article{leung2001orthogonal,
  title={An orthogonal genetic algorithm with quantization for global numerical optimization},
  author={Leung, Yiu-Wing and Wang, Yuping},
  journal={IEEE Transactions on Evolutionary computation},
  volume={5},
  number={1},
  pages={41--53},
  year={2001},
  publisher={IEEE}
}

@article{oldfield2025faster,
  title={Faster and better quantum software testing through specification reduction and projective measurements},
  author={Oldfield, Noah H and Laaber, Christoph and Yue, Tao and Ali, Shaukat},
  journal={ACM Transactions on Software Engineering and Methodology},
  volume={34},
  number={7},
  pages={1--39},
  year={2025},
  publisher={ACM New York, NY}
}

@inproceedings{xue2024selfpico,
  title={Selfpico: Self-guided partial code execution with llms},
  author={Xue, Zhipeng and Gao, Zhipeng and Wang, Shaohua and Hu, Xing and Xia, Xin and Li, Shanping},
  booktitle={Proceedings of the 33rd ACM SIGSOFT International Symposium on Software Testing and Analysis},
  pages={1389--1401},
  year={2024}
}

@inproceedings{lin2025soen,
  title={Soen-101: Code generation by emulating software process models using large language model agents},
  author={Lin, Feng and Kim, Dong Jae and Chen, Tse-Hsun},
  booktitle={2025 IEEE/ACM 47th International Conference on Software Engineering (ICSE)},
  pages={1527--1539},
  year={2025},
  organization={IEEE}
}

@inproceedings{islam2025evaluating,
  title={Evaluating the energy-efficiency of the code generated by llms},
  author={Islam, Md Arman and Jonnala, Devi Varaprasad and Rekhi, Ritika and Pokharel, Pratik and Cilamkoti, Siddharth and Imran, Asif and Kosar, Tevfik and Turkkan, Bekir},
  booktitle={2025 3rd International Conference on Foundation and Large Language Models (FLLM)},
  pages={289--300},
  year={2025},
  organization={IEEE}
}

@inproceedings{vallecillos2025art,
  title={The art of repair: Optimizing iterative program repair with instruction-tuned models},
  author={Vallecillos Ruiz, Fernando and Hort, Max and Moonen, Leon},
  booktitle={Proceedings of the 29th International Conference on Evaluation and Assessment in Software Engineering},
  pages={500--511},
  year={2025}
}

@book{rudolph1997convergence,
  title={Convergence properties of evolutionary algorithms},
  author={Rudolph, G{\"u}nter},
  year={1997},
  publisher={Verlag Dr. Kova{\v{c}}}
}

@inproceedings{szegedy2015going,
  title={Going deeper with convolutions},
  author={Szegedy, Christian and Liu, Wei and Jia, Yangqing and Sermanet, Pierre and Reed, Scott and Anguelov, Dragomir and Erhan, Dumitru and Vanhoucke, Vincent and Rabinovich, Andrew},
  booktitle={Proceedings of the IEEE conference on computer vision and pattern recognition},
  pages={1--9},
  year={2015}
}

@article{simonyan2014very,
  title={Very deep convolutional networks for large-scale image recognition},
  author={Simonyan, Karen and Zisserman, Andrew},
  journal={arXiv preprint arXiv:1409.1556},
  year={2014}
}

@misc{jiang2023mistral7b,
      title={Mistral 7B}, 
      author={Albert Q. Jiang and Alexandre Sablayrolles and Arthur Mensch and Chris Bamford and Devendra Singh Chaplot and Diego de las Casas and Florian Bressand and Gianna Lengyel and Guillaume Lample and Lucile Saulnier and Lélio Renard Lavaud and Marie-Anne Lachaux and Pierre Stock and Teven Le Scao and Thibaut Lavril and Thomas Wang and Timothée Lacroix and William El Sayed},
      year={2023},
      eprint={2310.06825},
      archivePrefix={arXiv},
      primaryClass={cs.CL},
      url={https://arxiv.org/abs/2310.06825}, 
}

@article{mahler2017dex,
  title={Dex-net 2.0: Deep learning to plan robust grasps with synthetic point clouds and analytic grasp metrics},
  author={Mahler, Jeffrey and Liang, Jacky and Niyaz, Sherdil and Laskey, Michael and Doan, Richard and Liu, Xinyu and Ojea, Juan Aparicio and Goldberg, Ken},
  journal={arXiv preprint arXiv:1703.09312},
  year={2017}
}

@article{james2020rlbench,
  title={Rlbench: The robot learning benchmark \& learning environment},
  author={James, Stephen and Ma, Zicong and Arrojo, David Rovick and Davison, Andrew J},
  journal={IEEE Robotics and Automation Letters},
  volume={5},
  number={2},
  pages={3019--3026},
  year={2020},
  publisher={IEEE}
}

@article{nasiriany2024robocasa,
  title={Robocasa: Large-scale simulation of everyday tasks for generalist robots},
  author={Nasiriany, Soroush and Maddukuri, Abhiram and Zhang, Lance and Parikh, Adeet and Lo, Aaron and Joshi, Abhishek and Mandlekar, Ajay and Zhu, Yuke},
  journal={arXiv preprint arXiv:2406.02523},
  year={2024}
}

@article{wang2024ladev,
  title={LADEV: A Language-Driven Testing and Evaluation Platform for Vision-Language-Action Models in Robotic Manipulation},
  author={Wang, Zhijie and Zhou, Zhehua and Song, Jiayang and Huang, Yuheng and Shu, Zhan and Ma, Lei},
  journal={arXiv preprint arXiv:2410.05191},
  year={2024}
}

@article{valle2026metamorphic,
  title={Metamorphic Testing of Vision-Language Action-Enabled Robots},
  author={Valle, Pablo and Segura, Sergio and Ali, Shaukat and Arrieta, Aitor},
  journal={arXiv preprint arXiv:2602.22579},
  year={2026}
}

@inproceedings{dinella2022toga,
  title={Toga: A neural method for test oracle generation},
  author={Dinella, Elizabeth and Ryan, Gabriel and Mytkowicz, Todd and Lahiri, Shuvendu K},
  booktitle={Proceedings of the 44th International Conference on Software Engineering},
  pages={2130--2141},
  year={2022}
}

@inproceedings{hossain2025togll,
  title={Togll: Correct and strong test oracle generation with llms},
  author={Hossain, Soneya Binta and Dwyer, Matthew B},
  booktitle={2025 IEEE/ACM 47th International Conference on Software Engineering (ICSE)},
  pages={1475--1487},
  year={2025},
  organization={IEEE}
}

@article{zhang2025improving,
  title={Improving Deep Assertion Generation via Fine-Tuning Retrieval-Augmented Pre-trained Language Models},
  author={Zhang, Quanjun and Fang, Chunrong and Zheng, Yi and Zhang, Yaxin and Zhao, Yuan and Huang, Rubing and Zhou, Jianyi and Yang, Yun and Zheng, Tao and Chen, Zhenyu},
  journal={ACM Transactions on Software Engineering and Methodology},
  volume={34},
  number={7},
  pages={1--23},
  year={2025},
  publisher={ACM New York, NY}
}

@inproceedings{siddiq2024using,
  title={Using large language models to generate junit tests: An empirical study},
  author={Siddiq, Mohammed Latif and Da Silva Santos, Joanna Cecilia and Tanvir, Ridwanul Hasan and Ulfat, Noshin and Al Rifat, Fahmid and Carvalho Lopes, Vin{\'\i}cius},
  booktitle={Proceedings of the 28th international conference on evaluation and assessment in software engineering},
  pages={313--322},
  year={2024}
}

@article{xu2026hallucination,
  title={Hallucination to consensus: Multi-agent llms for end-to-end junit test generation},
  author={Xu, Qinghua and Wang, Guancheng and Briand, Lionel and Liu, Kui},
  journal={ACM Transactions on Software Engineering and Methodology},
  year={2026},
  publisher={ACM New York, NY}
}

@article{huang2025nexus,
  title={Nexus: Execution-Grounded Multi-Agent Test Oracle Synthesis},
  author={Huang, Dong and Du, Mingzhe and Zhang, Jie M and Lin, Zheng and Luo, Meng and Zhang, Qianru and Ng, See-Kiong},
  journal={arXiv preprint arXiv:2510.26423},
  year={2025}
}

@article{hayet2024chatassert,
  title={Chatassert: Llm-based test oracle generation with external tools assistance},
  author={Hayet, Ishrak and Scott, Adam and d'Amorim, Marcelo},
  journal={IEEE Transactions on Software Engineering},
  volume={51},
  number={1},
  pages={305--319},
  year={2024},
  publisher={IEEE}
}

@inproceedings{khandaker2025augmentest,
  title={AugmenTest: Enhancing tests with LLM-driven oracles},
  author={Khandaker, Shaker Mahmud and Kifetew, Fitsum and Prandi, Davide and Susi, Angelo},
  booktitle={2025 IEEE Conference on Software Testing, Verification and Validation (ICST)},
  pages={279--289},
  year={2025},
  organization={IEEE}
}

@inproceedings{menghi2019generating,
  title={Generating automated and online test oracles for simulink models with continuous and uncertain behaviors},
  author={Menghi, Claudio and Nejati, Shiva and Gaaloul, Khouloud and Briand, Lionel C},
  booktitle={Proceedings of the 2019 27th acm joint meeting on european software engineering conference and symposium on the foundations of software engineering},
  pages={27--38},
  year={2019}
}

@article{valle2025defining,
  title={Defining and generating multi-level and uncertainty-wise test oracles for cyber-physical systems: P. Valle et al.},
  author={Valle, Pablo and Arrieta, Aitor and Han, Liping and Ali, Shaukat and Yue, Tao},
  journal={Software and Systems Modeling},
  volume={24},
  number={3},
  pages={679--704},
  year={2025},
  publisher={Springer}
}

@inproceedings{goffi2016automatic,
  title={Automatic generation of oracles for exceptional behaviors},
  author={Goffi, Alberto and Gorla, Alessandra and Ernst, Michael D and Pezz{\`e}, Mauro},
  booktitle={Proceedings of the 25th international symposium on software testing and analysis},
  pages={213--224},
  year={2016}
}

@inproceedings{terragni2020evolutionary,
  title={Evolutionary improvement of assertion oracles},
  author={Terragni, Valerio and Jahangirova, Gunel and Tonella, Paolo and Pezz{\`e}, Mauro},
  booktitle={Proceedings of the 28th ACM Joint Meeting on European Software Engineering Conference and Symposium on the Foundations of Software Engineering},
  pages={1178--1189},
  year={2020}
}

@incollection{pezze2014automated,
  title={Automated test oracles: A survey},
  author={Pezze, Mauro and Zhang, Cheng},
  booktitle={Advances in computers},
  volume={95},
  pages={1--48},
  year={2014},
  publisher={Elsevier}
}

@article{kanwal2026fate,
  title={FATE-VLA: Failue-aware test generation for vision-language-action models},
  author={Kanwal, Arusa and Valle, Pablo and Ali, Shaukat and Arrieta, Aitor},
  journal={arXiv preprint arXiv:2606.02307},
  year={2026}
}

@article{saurabh2026visor,
  title={VISOR: A Vision-Language Model-based Test Oracle for Testing Robot},
  author={Saurabh, Prasun and Valle, Pablo and Arrieta, Aitor and Ali, Shaukat and Arcaini, Paolo},
  journal={arXiv preprint arXiv:2605.10408},
  year={2026}
}

@article{mees2022calvin,
  title={Calvin: A benchmark for language-conditioned policy learning for long-horizon robot manipulation tasks},
  author={Mees, Oier and Hermann, Lukas and Rosete-Beas, Erick and Burgard, Wolfram},
  journal={IEEE Robotics and Automation Letters},
  volume={7},
  number={3},
  pages={7327--7334},
  year={2022},
  publisher={IEEE}
}

@article{li2025task,
  title={Task Reconstruction and Extrapolation for $\backslash pi\_0$ using Text Latent},
  author={Li, Quanyi},
  journal={arXiv preprint arXiv:2505.03500},
  year={2025}
}

@article{fang2025intention,
  title={From intention to execution: Probing the generalization boundaries of vision-language-action models},
  author={Fang, Irving and Zhang, Juexiao and Tong, Shengbang and Feng, Chen},
  journal={arXiv preprint arXiv:2506.09930},
  year={2025}
}

@article{zhou2025exploring,
  title={Exploring the limits of vision-language-action manipulations in cross-task generalization},
  author={Zhou, Jiaming and Ye, Ke and Liu, Jiayi and Ma, Teli and Wang, Zifan and Qiu, Ronghe and Lin, Kun-Yu and Zhao, Zhilin and Liang, Junwei},
  journal={arXiv preprint arXiv:2505.15660},
  year={2025}
}

@article{bohg2013data,
  title={Data-driven grasp synthesis—a survey},
  author={Bohg, Jeannette and Morales, Antonio and Asfour, Tamim and Kragic, Danica},
  journal={IEEE Transactions on robotics},
  volume={30},
  number={2},
  pages={289--309},
  year={2013},
  publisher={IEEE}
}

@article{barr2014oracle,
  title={The oracle problem in software testing: A survey},
  author={Barr, Earl T and Harman, Mark and McMinn, Phil and Shahbaz, Muzammil and Yoo, Shin},
  journal={IEEE transactions on software engineering},
  volume={41},
  number={5},
  pages={507--525},
  year={2014},
  publisher={IEEE}
}

@inproceedings{papineni2002bleu,
  title={Bleu: a method for automatic evaluation of machine translation},
  author={Papineni, Kishore and Roukos, Salim and Ward, Todd and Zhu, Wei-Jing},
  booktitle={Proceedings of the 40th annual meeting of the Association for Computational Linguistics},
  pages={311--318},
  year={2002}
}

@inproceedings{lin2004rouge,
  title={Rouge: A package for automatic evaluation of summaries},
  author={Lin, Chin-Yew},
  booktitle={Text summarization branches out},
  pages={74--81},
  year={2004}
}

@inproceedings{lin2014microsoft,
  title={Microsoft coco: Common objects in context},
  author={Lin, Tsung-Yi and Maire, Michael and Belongie, Serge and Hays, James and Perona, Pietro and Ramanan, Deva and Doll{\'a}r, Piotr and Zitnick, C Lawrence},
  booktitle={European conference on computer vision},
  pages={740--755},
  year={2014},
  organization={Springer}
}

@inproceedings{deng2009imagenet,
  title={Imagenet: A large-scale hierarchical image database},
  author={Deng, Jia and Dong, Wei and Socher, Richard and Li, Li-Jia and Li, Kai and Fei-Fei, Li},
  booktitle={2009 IEEE conference on computer vision and pattern recognition},
  pages={248--255},
  year={2009},
  organization={Ieee}
}

@article{wang2025vlatest,
  title={Vlatest: Testing and evaluating vision-language-action models for robotic manipulation},
  author={Wang, Zhijie and Zhou, Zhehua and Song, Jiayang and Huang, Yuheng and Shu, Zhan and Ma, Lei},
  journal={Proceedings of the ACM on Software Engineering},
  volume={2},
  number={FSE},
  pages={1615--1638},
  year={2025},
  publisher={ACM New York, NY, USA}
}

@inproceedings{zhang2025vlabench,
  title={Vlabench: A large-scale benchmark for language-conditioned robotics manipulation with long-horizon reasoning tasks},
  author={Zhang, Shiduo and Xu, Zhe and Liu, Peiju and Yu, Xiaopeng and Li, Yuan and Gao, Qinghui and Fei, Zhaoye and Yin, Zhangyue and Wu, Zuxuan and Jiang, Yu-Gang and others},
  booktitle={Proceedings of the IEEE/CVF International Conference on Computer Vision},
  pages={11142--11152},
  year={2025}
}

@misc{peng2025nebula,
      title={NEBULA: Do We Evaluate Vision-Language-Action Agents Correctly?}, 
      author={Jierui Peng and Yanyan Zhang and Yicheng Duan and Tuo Liang and Vipin Chaudhary and Yu Yin},
      year={2025},
      eprint={2510.16263},
      archivePrefix={arXiv},
      primaryClass={cs.RO},
      note = {\url{https://arxiv.org/abs/2510.16263}}
}

@inproceedings{fei2026libero,
  title={LIBERO-Plus: A Progressive Robustness Benchmark for Visual-Language-Action Models},
  author={Fei, Senyu and Wang, Siyin and Shi, Junhao and Dai, Zihao and Cai, Jikun and Qian, Pengfang and Ji, Li and He, Xinzhe and Zhang, Shiduo and Fei, Zhaoye and others},
  booktitle={Proceedings of the IEEE/CVF Conference on Computer Vision and Pattern Recognition},
  pages={38574--38583},
  year={2026}
}

@article{agarwal2025cosmos,
  title={Cosmos world foundation model platform for physical ai},
  author={Agarwal, Niket and Ali, Arslan and Bala, Maciej and Balaji, Yogesh and Barker, Erik and Cai, Tiffany and Chattopadhyay, Prithvijit and Chen, Yongxin and Cui, Yin and Ding, Yifan and others},
  journal={arXiv preprint arXiv:2501.03575},
  year={2025}
}

@article{team2025gemini,
  title={Gemini robotics 1.5: Pushing the frontier of generalist robots with advanced embodied reasoning, thinking, and motion transfer},
  author={Team, Gemini Robotics and Abdolmaleki, Abbas and Abeyruwan, Saminda and Ainslie, Joshua and Alayrac, Jean-Baptiste and Arenas, Montserrat Gonzalez and Balakrishna, Ashwin and Batchelor, Nathan and Bewley, Alex and Bingham, Jeff and others},
  journal={arXiv preprint arXiv:2510.03342},
  year={2025}
}

@article{ha2018world,
  title={World models},
  author={Ha, David and Schmidhuber, J{\"u}rgen},
  journal={arXiv preprint arXiv:1803.10122},
  volume={2},
  number={3},
  pages={440},
  year={2018}
}

@article{zhang2025step,
  title={A step toward world models: A survey on robotic manipulation},
  author={Zhang, Peng-Fei and Cheng, Ying and Sun, Xiaofan and Wang, Shijie and Li, Fengling and Zhu, Lei and Shen, Heng Tao},
  journal={arXiv preprint arXiv:2511.02097},
  year={2025}
}

@article{zhao2023learning,
  title={Learning fine-grained bimanual manipulation with low-cost hardware},
  author={Zhao, Tony Z and Kumar, Vikash and Levine, Sergey and Finn, Chelsea},
  journal={arXiv preprint arXiv:2304.13705},
  year={2023}
}

@article{grattafiori2024llama,
  title={The llama 3 herd of models},
  author={Grattafiori, Aaron and Dubey, Abhimanyu and Jauhri, Abhinav and Pandey, Abhinav and Kadian, Abhishek and Al-Dahle, Ahmad and Letman, Aiesha and Mathur, Akhil and Schelten, Alan and Vaughan, Alex and others},
  journal={arXiv preprint arXiv:2407.21783},
  year={2024}
}

@article{team2023gemini,
  title={Gemini: a family of highly capable multimodal models},
  author={Team, Gemini and Anil, Rohan and Borgeaud, Sebastian and Alayrac, Jean-Baptiste and Yu, Jiahui and Soricut, Radu and Schalkwyk, Johan and Dai, Andrew M and Hauth, Anja and Millican, Katie and others},
  journal={arXiv preprint arXiv:2312.11805},
  year={2023}
}

@article{achiam2023gpt,
  title={Gpt-4 technical report},
  author={Achiam, Josh and Adler, Steven and Agarwal, Sandhini and Ahmad, Lama and Akkaya, Ilge and Aleman, Florencia Leoni and Almeida, Diogo and Altenschmidt, Janko and Altman, Sam and Anadkat, Shyamal and others},
  journal={arXiv preprint arXiv:2303.08774},
  year={2023}
}

@article{vaswani2017attention,
  title={Attention is all you need},
  author={Vaswani, Ashish and Shazeer, Noam and Parmar, Niki and Uszkoreit, Jakob and Jones, Llion and Gomez, Aidan N and Kaiser, {\L}ukasz and Polosukhin, Illia},
  journal={Advances in neural information processing systems},
  volume={30},
  year={2017}
}

@article{krizhevsky2012imagenet,
  title={Imagenet classification with deep convolutional neural networks},
  author={Krizhevsky, Alex and Sutskever, Ilya and Hinton, Geoffrey E},
  journal={Advances in neural information processing systems},
  volume={25},
  year={2012}
}

@article{lecun2002gradient,
  title={Gradient-based learning applied to document recognition},
  author={LeCun, Yann and Bottou, L{\'e}on and Bengio, Yoshua and Haffner, Patrick},
  journal={Proceedings of the IEEE},
  volume={86},
  number={11},
  pages={2278--2324},
  year={2002},
  publisher={Ieee}
}

@misc{qu2025eo1,
      title={EO-1: Interleaved Vision-Text-Action Pretraining for General Robot Control}, 
      author={Delin Qu and Haoming Song and Qizhi Chen and Zhaoqing Chen and Xianqiang Gao and Xinyi Ye and Qi Lv and Modi Shi and Guanghui Ren and Cheng Ruan and Maoqing Yao and Haoran Yang and Jiacheng Bao and Bin Zhao and Dong Wang},
      year={2025},
      eprint={2508.21112},
      archivePrefix={arXiv},
      primaryClass={cs.RO},
      url={https://arxiv.org/abs/2508.21112}, 
}

@article{kim2024openvla,
  title={Openvla: An open-source vision-language-action model},
  author={Kim, Moo Jin and Pertsch, Karl and Karamcheti, Siddharth and Xiao, Ted and Balakrishna, Ashwin and Nair, Suraj and Rafailov, Rafael and Foster, Ethan and Lam, Grace and Sanketi, Pannag and others},
  journal={arXiv preprint arXiv:2406.09246},
  year={2024}
}

@misc{black2024pi0visionlanguageactionflowmodel,
      title={$\pi_0$: A Vision-Language-Action Flow Model for General Robot Control}, 
      author={Kevin Black and Noah Brown and Danny Driess and Adnan Esmail and Michael Equi and Chelsea Finn and Niccolo Fusai and Lachy Groom and Karol Hausman and Brian Ichter and Szymon Jakubczak and Tim Jones and Liyiming Ke and Sergey Levine and Adrian Li-Bell and Mohith Mothukuri and Suraj Nair and Karl Pertsch and Lucy Xiaoyang Shi and James Tanner and Quan Vuong and Anna Walling and Haohuan Wang and Ury Zhilinsky},
      year={2024},
      eprint={2410.24164},
      archivePrefix={arXiv},
      primaryClass={cs.LG},
      url={https://arxiv.org/abs/2410.24164}, 
}

@article{liu2023libero,
  title={LIBERO: Benchmarking Knowledge Transfer for Lifelong Robot Learning},
  author={Liu, Bo and Zhu, Yifeng and Gao, Chongkai and Feng, Yihao and Liu, Qiang and Zhu, Yuke and Stone, Peter},
  journal={arXiv preprint arXiv:2306.03310},
  year={2023}
}

@book{cormen2022introduction,
  title={Introduction to algorithms},
  author={Cormen, Thomas H and Leiserson, Charles E and Rivest, Ronald L and Stein, Clifford},
  year={2022},
  publisher={MIT press}
}

@article{baltes2025guidelines,
  title={Guidelines for empirical studies in software engineering involving large language models},
  author={Baltes, Sebastian and Angermeir, Florian and Arora, Chetan and Bar{\'o}n, Marvin Mu{\~n}oz and Chen, Chunyang and B{\"o}hme, Lukas and Calefato, Fabio and Ernst, Neil and Falessi, Davide and Fitzgerald, Brian and others},
  journal={arXiv preprint arXiv:2508.15503},
  year={2025}
}

@article{goldberg1989genetic,
  title={Genetic algorithms in search, optimization, and machine learning. Addison},
  author={Goldberg, David E},
  journal={Reading},
  year={1989}
}

@book{holland1992adaptation,
  title={Adaptation in natural and artificial systems: an introductory analysis with applications to biology, control, and artificial intelligence},
  author={Holland, John H},
  year={1992},
  publisher={MIT press}
}

@inproceedings{cer2017semeval,
  title={SemEval-2017 task 1: Semantic textual similarity multilingual and crosslingual focused evaluation},
  author={Cer, Daniel and Diab, Mona and Agirre, Eneko and Lopez-Gazpio, Inigo and Specia, Lucia},
  booktitle={Proceedings of the 11th international workshop on semantic evaluation (SemEval-2017)},
  pages={1--14},
  year={2017}
}

@article{valle2025evaluating,
  title={Evaluating uncertainty and quality of visual language action-enabled robots},
  author={Valle, Pablo and Lu, Chengjie and Ali, Shaukat and Arrieta, Aitor},
  journal={arXiv preprint arXiv:2507.17049},
  year={2025}
}

@misc{nvidia2025gr00tn1openfoundation,
      title={GR00T N1: An Open Foundation Model for Generalist Humanoid Robots}, 
      author={NVIDIA and Johan Bjorck and Fernando Castañeda and Nikita Cherniadev and Xingye Da and Runyu Ding and Linxi "Jim" Fan and Yu Fang and Dieter Fox and Fengyuan Hu and Spencer Huang and Joel Jang and Zhenyu Jiang and Jan Kautz and Kaushil Kundalia and Lawrence Lao and Zhiqi Li and Zongyu Lin and Kevin Lin and Guilin Liu and Edith Llontop and Loic Magne and Ajay Mandlekar and Avnish Narayan and Soroush Nasiriany and Scott Reed and You Liang Tan and Guanzhi Wang and Zu Wang and Jing Wang and Qi Wang and Jiannan Xiang and Yuqi Xie and Yinzhen Xu and Zhenjia Xu and Seonghyeon Ye and Zhiding Yu and Ao Zhang and Hao Zhang and Yizhou Zhao and Ruijie Zheng and Yuke Zhu},
      year={2025},
      eprint={2503.14734},
      archivePrefix={arXiv},
      primaryClass={cs.RO},
      url={https://arxiv.org/abs/2503.14734}, 
}

@book{eiben2015introduction,
  title={Introduction to evolutionary computing},
  author={Eiben, Agoston E and Smith, James E},
  year={2015},
  publisher={Springer}
}

@article{harman2001search,
  title={Search-based software engineering},
  author={Harman, Mark and Jones, Bryan F},
  journal={Information and software Technology},
  volume={43},
  number={14},
  pages={833--839},
  year={2001},
  publisher={Elsevier}
}

@inproceedings{romano2006exploring,
  title={Exploring methods for evaluating group differences on the NSSE and other surveys: Are the t-test and Cohen’sd indices the most appropriate choices},
  author={Romano, Jeanine and Kromrey, Jeffrey D and Coraggio, Jesse and Skowronek, Jeff and Devine, Linda},
  booktitle={annual meeting of the Southern Association for Institutional Research},
  pages={1--51},
  year={2006},
  organization={Citeseer}
}

@inproceedings{muhlenbein1992genetic,
  title={How genetic algorithms really work: I. mutation and hillclimbing},
  author={Muhlenbein, Heinz},
  booktitle={Proc. 2nd Int. Conf. on Parallel Problem Solving from Nature, 1992},
  year={1992},
  organization={Elsevier}
}

@article{droste2002analysis,
  title={On the analysis of the (1+ 1) evolutionary algorithm},
  author={Droste, Stefan and Jansen, Thomas and Wegener, Ingo},
  journal={Theoretical Computer Science},
  volume={276},
  number={1-2},
  pages={51--81},
  year={2002},
  publisher={Elsevier}
}

@book{cochran1977sampling,
  title={Sampling techniques},
  author={Cochran, William Gemmell},
  year={1977},
  publisher={john wiley \& sons}
}

@misc{mistral_3,
author = {Mistral AI},
  title = {Introducing Mistral 3 | Mistral AI},
  url = "https://mistral.ai/news/mistral-3/",
month = {December},
year = {2025},
note = {\url{https://mistral.ai/news/mistral-3/}}
}

@article{singh2025openai,
  title={Openai gpt-5 system card},
  author={Singh, Aaditya and Fry, Adam and Perelman, Adam and Tart, Adam and Ganesh, Adi and El-Kishky, Ahmed and McLaughlin, Aidan and Low, Aiden and Ostrow, AJ and Ananthram, Akhila and others},
  journal={arXiv preprint arXiv:2601.03267},
  year={2025}
}

@article{liu2024deepseek,
  title={Deepseek-v3 technical report},
  author={Liu, Aixin and Feng, Bei and Xue, Bing and Wang, Bingxuan and Wu, Bochao and Lu, Chengda and Zhao, Chenggang and Deng, Chengqi and Zhang, Chenyu and Ruan, Chong and others},
  journal={arXiv preprint arXiv:2412.19437},
  year={2024}
}

@article{ding2025understanding,
  title={Understanding world or predicting future? a comprehensive survey of world models},
  author={Ding, Jingtao and Zhang, Yunke and Shang, Yu and Zhang, Yuheng and Zong, Zefang and Feng, Jie and Yuan, Yuan and Su, Hongyuan and Li, Nian and Sukiennik, Nicholas and others},
  journal={ACM Computing Surveys},
  volume={58},
  number={3},
  pages={1--38},
  year={2025},
  publisher={ACM New York, NY}
}

@article{huang2026thinkact,
  title={Thinkact: Vision-language-action reasoning via reinforced visual latent planning},
  author={Huang, Chi-Pin and Wu, Yueh-Hua and Chen, Min-Hung and Wang, Frank and Yang, Fu-En},
  journal={Advances in Neural Information Processing Systems},
  volume={38},
  pages={82782--82802},
  year={2026}
}

@article{qu2025spatialvla,
  title={Spatialvla: Exploring spatial representations for visual-language-action model},
  author={Qu, Delin and Song, Haoming and Chen, Qizhi and Yao, Yuanqi and Ye, Xinyi and Ding, Yan and Wang, Zhigang and Gu, JiaYuan and Zhao, Bin and Wang, Dong and others},
  journal={arXiv preprint arXiv:2501.15830},
  year={2025}
}

@inproceedings{zitkovich2023rt,
  title={Rt-2: Vision-language-action models transfer web knowledge to robotic control},
  author={Zitkovich, Brianna and Yu, Tianhe and Xu, Sichun and Xu, Peng and Xiao, Ted and Xia, Fei and Wu, Jialin and Wohlhart, Paul and Welker, Stefan and Wahid, Ayzaan and others},
  booktitle={Conference on Robot Learning},
  pages={2165--2183},
  year={2023},
  organization={PMLR}
}

@article{deng2026mastor,
  title={MASTOR: A Multi-Agent Approach to Semantic Test Oracle Generation for RESTful APIs},
  author={Deng, Sida and Huang, Rubing and Yang, Zhenzhen and Zhang, Man and Xie, Xuan and Wang, Rongcun},
  journal={arXiv preprint arXiv:2606.10465},
  year={2026}
}

@article{li24simpler,
         title={Evaluating Real-World Robot Manipulation Policies in Simulation},
         author={Xuanlin Li and Kyle Hsu and Jiayuan Gu and Karl Pertsch and Oier Mees and Homer Rich Walke and Chuyuan Fu and Ishikaa Lunawat and Isabel Sieh and Sean Kirmani and Sergey Levine and Jiajun Wu and Chelsea Finn and Hao Su and Quan Vuong and Ted Xiao},
         journal = {arXiv preprint arXiv:2405.05941},
         year={2024}
}

@article{das2021sentence,
  title={Sentence embedding models for similarity detection of software requirements},
  author={Das, Souvick and Deb, Novarun and Cortesi, Agostino and Chaki, Nabendu},
  journal={SN Computer Science},
  volume={2},
  number={2},
  pages={69},
  year={2021},
  publisher={Springer}
}

@article{zhang2019bertscore,
  title={Bertscore: Evaluating text generation with bert},
  author={Zhang, Tianyi and Kishore, Varsha and Wu, Felix and Weinberger, Kilian Q and Artzi, Yoav},
  journal={arXiv preprint arXiv:1904.09675},
  year={2019}
}

@inproceedings{reimers2019sentence,
  title={Sentence-bert: Sentence embeddings using siamese bert-networks},
  author={Reimers, Nils and Gurevych, Iryna},
  booktitle={Proceedings of the 2019 conference on empirical methods in natural language processing and the 9th international joint conference on natural language processing (EMNLP-IJCNLP)},
  pages={3982--3992},
  year={2019}
}

@article{ouyang2025empirical,
  title={An empirical study of the non-determinism of chatgpt in code generation},
  author={Ouyang, Shuyin and Zhang, Jie M and Harman, Mark and Wang, Meng},
  journal={ACM Transactions on Software Engineering and Methodology},
  volume={34},
  number={2},
  pages={1--28},
  year={2025},
  publisher={ACM New York, NY}
}

@inproceedings{mohammadi2025evaluation,
  title={Evaluation and benchmarking of llm agents: A survey},
  author={Mohammadi, Mahmoud and Li, Yipeng and Lo, Jane and Yip, Wendy},
  booktitle={Proceedings of the 31st ACM SIGKDD Conference on Knowledge Discovery and Data Mining V. 2},
  pages={6129--6139},
  year={2025}
}

@article{camara2024towards,
  title={Towards Standardized benchmarks of LLMs in software modeling tasks: a conceptual framework: J. C{\'a}mara et al.},
  author={C{\'a}mara, Javier and Burgue{\~n}o, Lola and Troya, Javier},
  journal={Software and Systems Modeling},
  volume={23},
  number={6},
  pages={1309--1318},
  year={2024},
  publisher={Springer}
}

\appendix
\subsection{\revision{Foundation-Model Selection Study}}\label{sec:FM-selection}

\revision{To select the foundation model used by \approach, we evaluated five candidate models: DeepSeek-v3.2~\cite{liu2024deepseek}, Gemini Robotics-ER 1.6~\cite{team2025gemini}, GPT-5~\cite{singh2025openai}, Mistral-Large-3-675B~\cite{mistral_3}, and NVIDIA Cosmos-Reason2-8B~\cite{agarwal2025cosmos}. For each model, we executed the complete atomic-library generation process for LIBERO, RoboCasa, and RoboCasa Humanoid. To ensure a consistent comparison, all models received the same generation instructions and the same environment-specific context, including the available simulator APIs and task-related information.

After generation, we manually inspected the resulting atomic libraries against the corresponding simulator APIs and task requirements. The evaluation considered eight quality dimensions. \emph{Validity} assesses whether the generated functions are syntactically correct, executable, and compatible with the simulator APIs. \emph{Coverage} measures whether the library contains the functions required to observe the task-relevant states and interactions of the environment. \emph{Redundancy} assesses whether the library avoids duplicated or functionally overlapping operations, with higher scores indicating less redundancy. \emph{Equivalence} measures whether the generated functions faithfully implement their intended simulator operations or predicates. \emph{Atomicity} evaluates whether each function examines a single elementary condition or state transition rather than combining multiple task-level conditions. \emph{Generalization} measures whether functions are parameterized and reusable across tasks, objects, and scenes instead of being hard-coded for specific instances. \emph{Causality} assesses whether the functions capture the relevant relationship between an action or state transition and its expected effect. Finally, \emph{Utility} measures whether the generated functions can be directly employed to construct fine-grained oracles and provide meaningful information for task assessment and failure localization.

The manual assessments were expressed as normalized scores between 0 and 1. For each metric, the score represents the proportion of atomic tasks in the generated library that satisfied the corresponding requirement, with higher values indicating better performance. Table~\ref{fm_selection} reports the results for each model and environment. For each model and metric, the overall score was calculated by averaging the corresponding values across LIBERO, RoboCasa, and RoboCasa Humanoid. Gemini Robotics-ER 1.5 achieved the strongest overall performance across the evaluated criteria and was therefore selected as the main Foundation Model for the Generator role. Given its strong performance, we also adopted models from the Gemini family for the remaining agent roles, using a lightweight Gemini variant for the assessment agents and a reasoning-oriented Gemini variant for the judge agents.}

\begin{table}[t]
\centering
\caption{\revision{Foundation model selection results across the evaluated environments.
Scores range from 0 to 1, with higher values indicating better performance.
The Avg. column reports the mean across the eight evaluation criteria.
The results for the model with the best results in each environment and overall are underlined.}}
\label{tab:fm_selection}
\resizebox{0.5\textwidth}{!}{%
\begin{tabular}{llccccccccc}
\toprule
& \textbf{Model}
& \textbf{Val.}
& \textbf{Cov.}
& \textbf{Red.}
& \textbf{Equiv.}
& \textbf{Atom.}
& \textbf{Gen.}
& \textbf{Caus.}
& \textbf{Util.}
& \textbf{Avg.} \\
\midrule

\multirow{5}{*}{\textbf{LIBERO}}
& DeepSeek & \underline{1.00} & \underline{1.00} & \underline{0.80} & \underline{1.00} & \underline{1.00} & \underline{1.00} & \underline{1.00} & \underline{1.00} & \underline{0.98} \\
& Gemini & \underline{1.00} & \underline{1.00} & \underline{0.80} & \underline{1.00} & \underline{1.00} & \underline{1.00} & \underline{1.00} & \underline{1.00} & \underline{0.98} \\
& GPT-5 & 0.92 & 1.00 & 0.77 & 0.85 & 1.00 & 1.00 & 1.00 & 0.93 & 0.93 \\
& Mistral   & 1.00 & 1.00 & 0.83 & 0.83 & 1.00 & 1.00 & 1.00 & 1.00 & 0.96 \\

& Cosmos & 1.00 & 1.00 & 0.83 & 0.83 & 1.00 & 1.00 & 1.00 & 1.00 & 0.96 \\

\midrule

\multirow{5}{*}{\textbf{RoboCasa}}
& DeepSeek   & 0.81 & 1.00 & 0.68 & 0.89 & 1.00 & 0.89 & 0.89 & 0.93 & 0.89 \\
& Gemini & \underline{0.91} & \underline{1.00} & \underline{0.83} & \underline{1.00} & \underline{1.00} & \underline{1.00} & \underline{1.00} & \underline{0.92} & \underline{0.96} \\
& GPT-5  & 0.84 & 1.00 & 0.84 & 0.93 & 1.00 & 1.00 & 1.00 & 1.00 & 0.95 \\
& Mistral & 0.65 & 1.00 & 0.88 & 0.78 & 1.00 & 1.00 & 1.00 & 1.00 & 0.91 \\
& Cosmos & 0.65 & 1.00 & 0.94 & 0.64 & 0.83 & 1.00 & 1.00 & 1.00 & 0.88 \\

\midrule

\multirow{5}{*}{\begin{tabular}[c]{@{}l@{}}
\textbf{RoboCasa}\\
\textbf{Humanoid}
\end{tabular}}
& DeepSeek  & 1.00 & 1.00 & 0.75 & 1.00 & 1.00 & 1.00 & 1.00 & 1.00 & 0.97 \\
& Gemini  & \underline{1.00} & \underline{1.00} & \underline{1.00} & \underline{1.00} & \underline{1.00} & \underline{1.00} & \underline{1.00} & \underline{1.00} & \underline{1.00} \\
& GPT  & 1.00 & 1.00 & 0.75 & 1.00 & 1.00 & 1.00 & 1.00 & 1.00 & 0.97 \\
& Mistral  & 1.00 & 1.00 & 0.75 & 1.00 & 1.00 & 1.00 & 1.00 & 1.00 & 0.97 \\
& Cosmos & 1.00 & 1.00 & 0.75 & 1.00 & 1.00 & 1.00 & 1.00 & 1.00 & 0.97 \\

\midrule

\multirow{5}{*}{\begin{tabular}[c]{@{}l@{}}
\textbf{Overall}\\
\textbf{Average}
\end{tabular}}
& DeepSeek   & 0.94 & 1.00 & 0.74 & 0.96 & 1.00 & 0.96 & 0.96 & 0.98 & 0.94 \\
& Gemini & \underline{0.97} & \underline{1.00} & \underline{0.88} & \underline{1.00} & \underline{1.00} & \underline{1.00} & \underline{1.00} & \underline{0.97} & \underline{0.98} \\
& GPT & 0.92 & 1.00 & 0.79 & 0.93 & 1.00 & 1.00 & 1.00 & 0.98 & 0.95 \\
& Mistral & 0.88 & 1.00 & 0.82 & 0.87 & 1.00 & 1.00 & 1.00 & 1.00 & 0.95 \\
& Cosmos  & 0.88 & 1.00 & 0.84 & 0.82 & 0.94 & 1.00 & 1.00 & 1.00 & 0.94 \\

\bottomrule
\end{tabular}%
}
\end{table}

\subsection{RQ2 results pre-fix}\label{sec:prefix}
\revision{
This section reports Table~\ref{tab:RQ2} which presents the initial RQ2 results obtained before correcting the issues identified in the simulator-level state-inspection functions. We include these results for transparency and to illustrate how limitations in these functions affected both agreement with the symbolic \revision{end-state} oracles and failure localization.

\begin{table*}[ht]
\centering
\caption{\revision{Agreement between generated fine-grained oracles and benchmark-provided symbolic \revision{end-state} oracles on the LIBERO\_10 and Robocasa Humanoid benchmarks after correcting the failures in the simulator functions. The table reports agreement for the correctness assessment metrics, including accuracy (Acc.), precision (Prec.), recall (Rec.), F1-score (F1), and confusion matrix statistics (TP, TN, FP, and FN), measuring agreement between the fine-grained oracles and the benchmark-provided symbolic \revision{end-state} oracles. Additionally, localization accuracy evaluates the ability of the generated fine-grained oracles to identify the execution step at which the failure occurs. Results are reported per task and aggregated for each benchmark and overall.}}
\label{tab:RQ2}
\resizebox{0.81\textwidth}{!}{%
\begin{tabular}{llrrrrrrrrrrr}
\toprule
\multicolumn{1}{c}{} & \multicolumn{1}{c}{} &  \multicolumn{9}{c}{\textbf{Correctness}} & \multicolumn{2}{c}{\textbf{Localization}} \\ \cmidrule(lr){3-11} \cmidrule(lr){12-13}
\multicolumn{1}{c}{} & \multicolumn{1}{c}{} & \textbf{Exec \#} & \multicolumn{1}{c}{\textbf{Acc.}} & \multicolumn{1}{c}{\textbf{Prec.}} & \multicolumn{1}{c}{\textbf{Rec.}} & \multicolumn{1}{c}{\textbf{F1}} & \multicolumn{1}{c}{\textbf{TP}} & \multicolumn{1}{c}{\textbf{TN}} & \multicolumn{1}{c}{\textbf{FP}} & \multicolumn{1}{c}{\textbf{FN}} & \textbf{Exec \#} & \multicolumn{1}{c}{\textbf{Acc.}} \\
\cmidrule{1-13}
\multirow{11}{*}{\textbf{LIBERO\_10}}  & Task 1 & 90 & 1.00 & 1.00 & 1.00 & 1.00 & 54 & 36 & 0 & 0 & 10 & 1.00 \\
 & Task 2 & 80 & 0.60 & 0.60 & 1.00 & 0.75 & 48 & 0 & 32 & 0 & 10 & 0.90 \\
 & Task 3 & 80 & 1.00 & 1.00 & 1.00 & 1.00 & 48 & 32 & 0 & 0 & 10 & 0.30 \\
 & Task 4 & 10 & 0.80 & 0.75 & 1.00 & 0.86 & 6 & 2 & 2 & 0 & 6 & 0.67 \\
 & Task 5 & 100 & 0.60 & 0.60 & 1.00 & 0.75 & 60 & 0 & 40 & 0 & 10 & 0.30 \\
 & Task 6 & 100 & 1.00 & 1.00 & 1.00 & 1.00 & 60 & 40 & 0 & 0 & 10 & 0.40 \\
 & Task 7 & 100 & 0.60 & 0.60 & 1.00 & 0.75 & 60 & 0 & 40 & 0 & 10 & 0.20 \\
 & Task 8 & 70 & 0.80 & 0.75 & 1.00 & 0.86 & 42 & 14 & 14 & 0 & 10 & 0.40 \\
 & Task 9 & 80 & 0.90 & 1.00 & 0.83 & 0.91 & 40 & 32 & 0 & 8 & 10 & 0.40 \\
 & Task 10 & 80 & 1.00 & 1.00 & 1.00 & 1.00 & 48 & 32 & 0 & 0 & 10 & 0.50 \\ \cmidrule{2-13}
  \rowcolor{gray!30}   \cellcolor{white} & \textbf{Total} & 790 & 0.83 & 0.78 & 0.98 & 0.87 & 96 & 188 & 128 & 8 & 96 & 0.50 \\ \cmidrule{1-13}
\multirow{25}{*}{\begin{tabular}[c]{@{}l@{}}\textbf{Robocasa}\\ \textbf{Humanoid}\end{tabular}} & Task 1 & 100 & 1.00 & 1.00 & 1.00 & 1.00 & 60 & 40 & 0 & 0 & 10 & 0.7\\
 & Task 2 & 99 & 1.00 & 1.00 & 1.00 & 1.00 & 60 & 39 & 0 & 0 & 10 & 0.9\\
 & Task 3 & 95 & 1.00 & 1.00 & 1.00 & 1.00 & 60 & 35 & 0 & 0 & 10 & 0.6\\
 & Task 4 & 82 & 0.99 & 0.98 & 1.00 & 0.99 & 60 & 21 & 1 & 0 & 10 & 0.6\\
 & Task 5 & 92 & 0.95 & 0.97 & 0.95 & 0.96 & 57 & 30 & 2 & 3 & 10 & 0.7\\
 & Task 6 & 99 & 0.99 & 1.00 & 0.98 & 0.99 & 59 & 39 & 0 & 1 & 10 & 0.1\\
 & Task 7 & 63 & 0.89 & 0.86 & 1.00 & 0.92 & 42 & 14 & 7 & 0 & 10 & 0.3\\
 & Task 8 & 51 & 0.96 & 0.95 & 1.00 & 0.97 & 36 & 13 & 2 & 0 & 10 & 0.4\\
 & Task 9 & 90 & 0.94 & 0.95 & 0.96 & 0.95 & 52 & 33 & 3 & 2 & 10 & 0.3\\
 & Task 10 & 100 & 0.88 & 0.85 & 0.97 & 0.91 & 58 & 30 & 10 & 2 & 10 & 0.3\\
 & Task 11 & 68 & 1.00 & 1.00 & 1.00 & 1.00 & 54 & 14 & 0 & 0 & 10 & 0.3\\
 & Task 12 & 90 & 0.80 & 0.75 & 1.00 & 0.86 & 54 & 18 & 18 & 0 & 10 & 0.5\\
 & Task 13 & 57 & 1.00 & 1.00 & 1.00 & 1.00 & 48 & 9 & 0 & 0 & 10 & 0.1\\
 & Task 14 & 90 & 0.87 & 0.83 & 1.00 & 0.91 & 60 & 18 & 12 & 0 & 10 & 0.3\\
 & Task 15 & 90 & 0.92 & 1.00 & 0.88 & 0.94 & 53 & 30 & 0 & 7 & 10 & 0.2\\
 & Task 16 & 78 & 1.00 & 1.00 & 1.00 & 1.00 & 46 & 32 & 0 & 0 & 10 & 0.3\\
 & Task 17 & 80 & 0.70 & 0.67 & 1.00 & 0.80 & 48 & 8 & 24 & 0 & 10 & 0.2\\
 & Task 18 & 100 & 0.77 & 0.74 & 0.95 & 0.83 & 57 & 13 & 13 & 3 & 10 & 0.3\\
 & Task 19 & 99 & 0.48 & 0.62 & 0.36 & 0.45 & 21 & 27 & 13 & 38 & 10 & 0.3\\
 & Task 20 & 100 & 0.77 & 0.72 & 1.00 & 0.84 & 60 & 17 & 23 & 0 & 10 & 0.2\\
 & Task 21 & 100 & 1.00 & 1.00 & 1.00 & 1.00 & 60 & 40 & 0 & 0 & 10 & 0.1\\
 & Task 22 & 90 & 0.80 & 0.75 & 1.00 & 0.86 & 54 & 18 & 18 & 0 & 10 & 0.2\\
 & Task 23 & 75 & 0.99 & 1.00 & 0.98 & 0.99 & 62 & 12 & 0 & 1 & 10 & 0.2\\
 & Task 24 & 89 & 0.96 & 1.00 & 0.93 & 0.96 & 55 & 30 & 0 & 4 & 10 & 0.3\\ \cmidrule{2-13}
  \rowcolor{gray!30}  \cellcolor{white} & \textbf{Total} & 2077 & 0.90 & 0.89 & 0.95 & 0.92 & 1276 & 587 & 153 & 61 & 240 & 0.35\\ \cmidrule{1-13}
\rowcolor{gray!30}   \textbf{Total} & & 2867 & 0.88 & 0.86 & 0.96 & 0.91 & 1742 & 775 & 281 & 69 & 336 & 0.39 \\ \bottomrule
\end{tabular}%
}
\end{table*}}

\revision{Before applying the corrections, the generated fine-grained oracles achieved an overall agreement accuracy of 0.88, with a precision of 0.86, a recall of 0.96, and an F1-score of 0.91 across 2,867 executions. Agreement accuracy was 0.83 for LIBERO\_10 and 0.90 for Robocasa Humanoid. However, the overall localization accuracy was only 0.39, reaching 0.50 for LIBERO\_10 and 0.35 for Robocasa Humanoid. Manual inspection revealed that these results were affected by errors in the simulator-level functions used by the fine-grained oracles. In LIBERO\_10, the function for detecting whether an object had been lifted used a height threshold relative to the ground rather than the table. In Robocasa Humanoid, there were errors in the function that checked how close the robot hand was to an object. These issues motivated the corrections described in RQ2 and the subsequent repetition of the experiments. Therefore, the corrected results reported in the main analysis should be used when interpreting the agreement and localization capabilities of the generated fine-grained oracles.}

\subsection{Specific Agents}~\label{sec:appendix}
\begin{figure*}[hb]
    \centering
    \includegraphics[width=\linewidth]{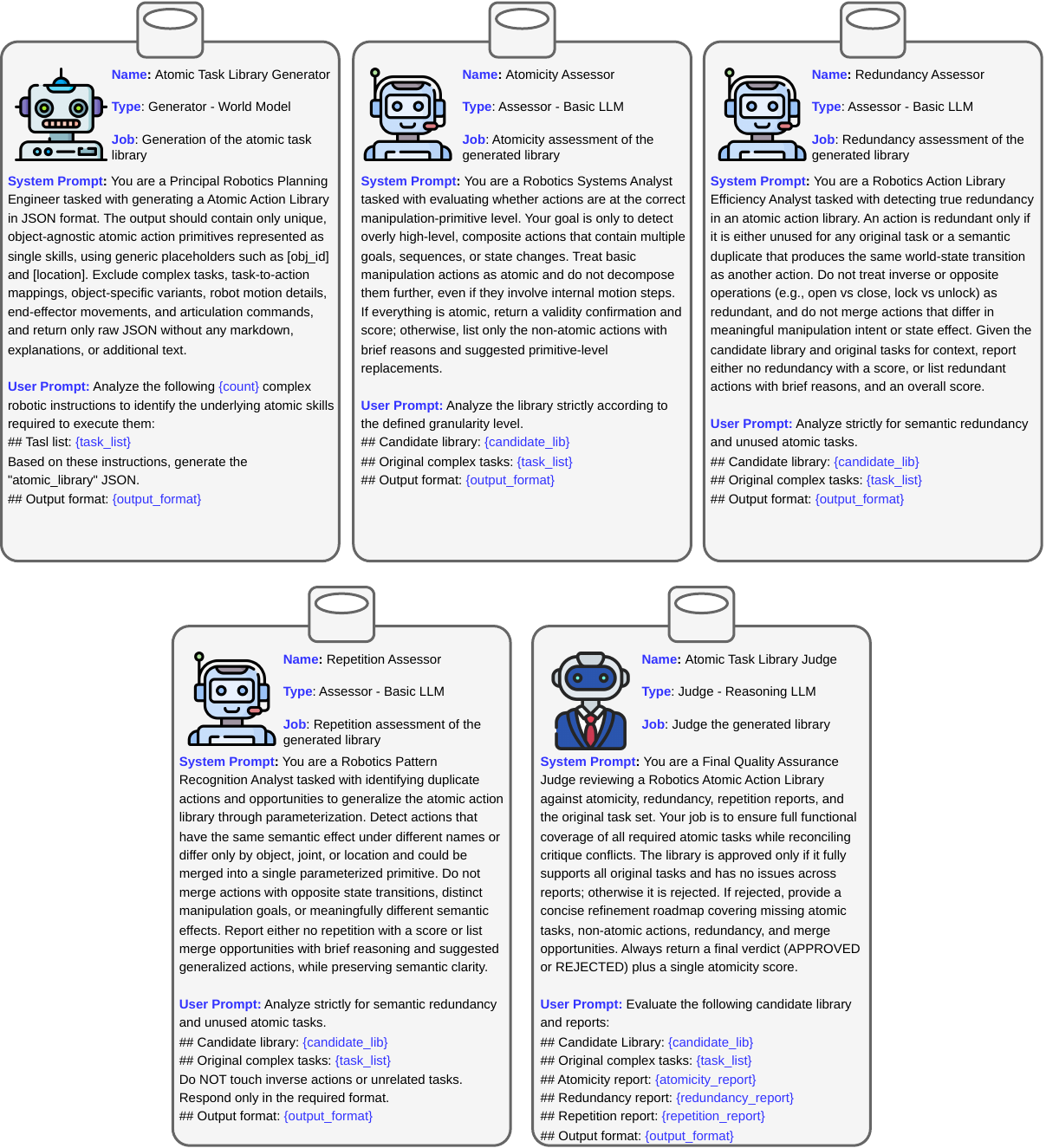}
    \caption{Overview of the different agents used in \approach Module I.}
    \label{fig:profilesM1}
\end{figure*}

\begin{figure*}[ht]
    \centering
    \includegraphics[width=\linewidth]{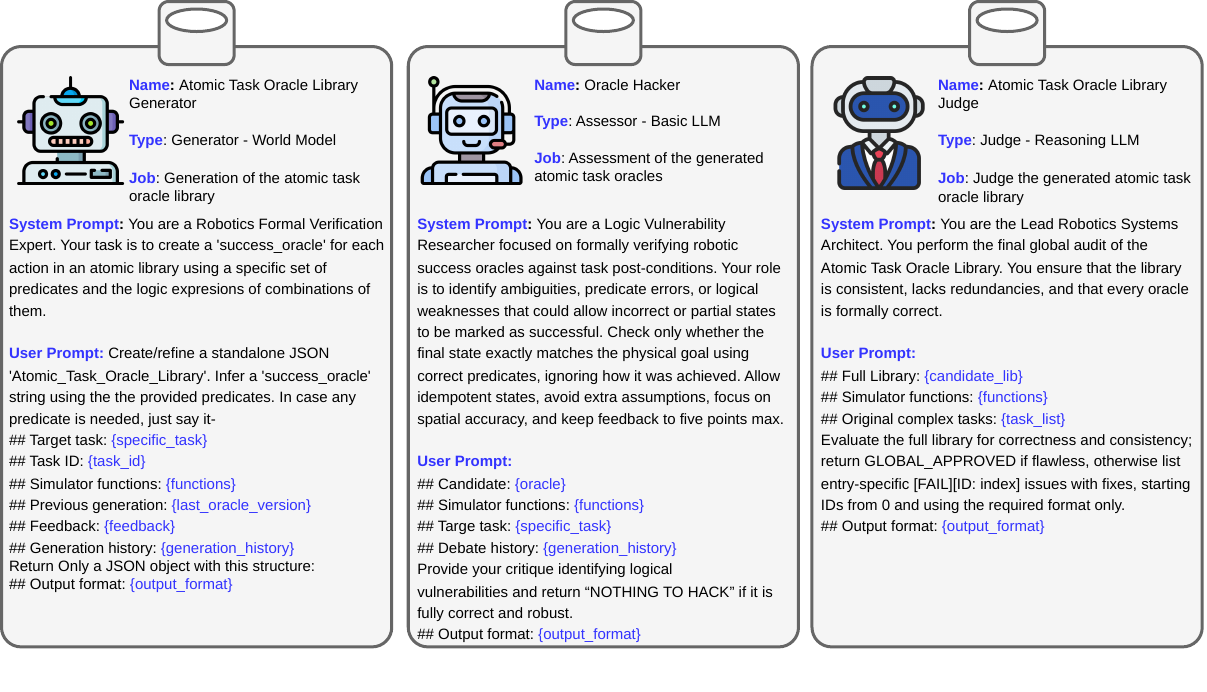}
    \caption{Overview of the different agents used in \approach Module II.}
    \label{fig:profilesM2}
\end{figure*}

\begin{figure*}[ht]
    \centering
    \includegraphics[width=\linewidth]{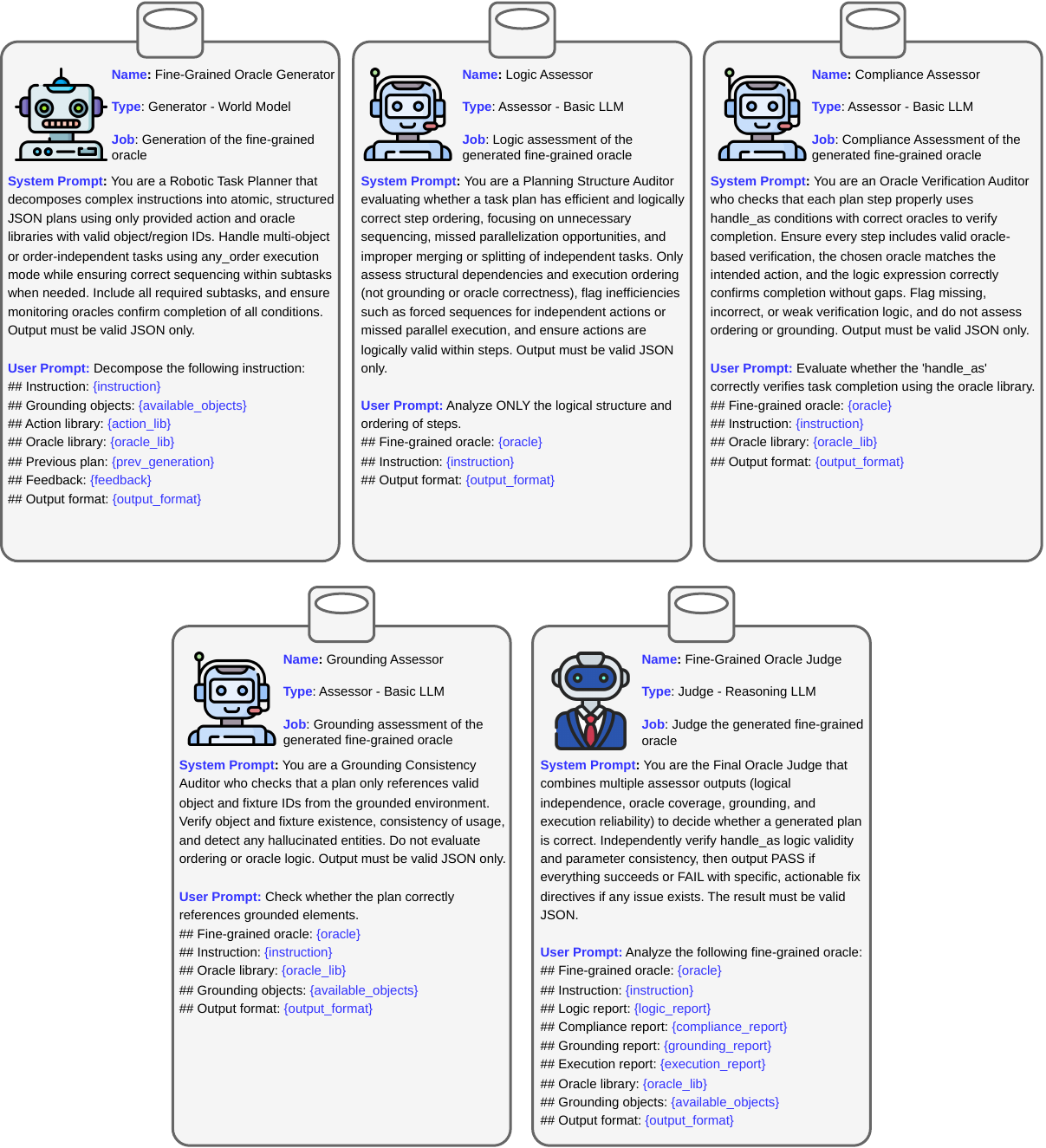}
    \caption{Overview of the different agents used in \approach Module III.}
    \label{fig:profilesM3}
\end{figure*}

This section provides an overview of the different agents used in \approach. Figure~\ref{fig:profilesM1} shows the agents used in Module I, Figure~\ref{fig:profilesM2} shows the agents used in Module II, and Figure~\ref{fig:profilesM3} shows the agents used in Module III.

\end{document}